%% file: main_final.tex
\documentclass[12pt,english]{article}
\usepackage{lscape}
\usepackage[english]{babel}

\usepackage{eurosym} 

\usepackage{soul}

\usepackage[a4paper,
left=1.25in,
right=1.25in,
top=1.25in,
bottom=1.25in]{geometry}

\usepackage{amsmath}
\usepackage{graphicx}
\usepackage[colorlinks=true, allcolors=blue,hyperfootnotes=false]{hyperref}
\usepackage{tabularx}
\usepackage{booktabs}
\usepackage{adjustbox}
\usepackage{amssymb}
\usepackage{amsfonts}
\usepackage{placeins}
\usepackage{dsfont}
\usepackage{xcolor}
\usepackage{soul}
\usepackage[comma,authoryear]{natbib}
\usepackage{multibib}
\newcites{sec}{References in the Online Appendix}

\usepackage{siunitx}

\setcitestyle{citesep={,}}

\usepackage{caption}
\usepackage{subcaption}
\usepackage[flushleft]{threeparttable}

\usepackage{setspace}

\usepackage[toc,page,header]{appendix}
\usepackage{minitoc}
\noptcrule 

\usepackage{mdframed}

\allowdisplaybreaks

\newtheorem{assumption}{Assumption}[section]

\newtheorem{definition}{Definition}[section]

\newtheorem{lemma}{Lemma}[section]

\newtheorem{proposition}{Proposition}

\newcounter{exbox}
\renewcommand{\theexbox}{\arabic{exbox}}

\newenvironment{proof}{\emph{Proof.} }{$\Box$}
\makeatletter
\newcommand{\tightmainparagraphs}{%
	\renewcommand\paragraph{\@startsection{paragraph}{4}{\z@}%
		{2.6ex \@plus .7ex \@minus .2ex}{-1em}{\normalfont\normalsize\bfseries}}%
}
\newcommand{\tightappendixparagraphs}{%
	\renewcommand\paragraph{\@startsection{paragraph}{4}{\z@}%
		{0pt \@plus .1ex}{-1em}{\normalfont\normalsize\bfseries}}%
}
\makeatother
\tightmainparagraphs
\AddToHook{cmd/appendix/before}{%
	\setcounter{equation}{0}%
	\renewcommand{\theequation}{\thesection.\arabic{equation}}%
	\setcounter{theorem}{0}%
	\setcounter{proposition}{0}
	\tightappendixparagraphs%
}

\title{\vspace*{-2cm}{\Huge{Capacity, Technology Portfolios, \\[0.3em]
			and the Paradox of Concentration}
	\thanks{
			We would like to thank Martino Banchio, Giacomo Calzolari, Estelle Cantillon, Thomas Chaney, Zoe Cullen, Áureo de Paula, Francesco Decarolis, Natalia Fabra, Christos Genakos, Joao Granja, Alessandro Iaria, Gaston Illanes, Rocco Macchiavello, Thierry Mayer, Nathan Miller, Francesco Nava, Volker Nocke, Marco Ottaviani, Fausto Panunzi, Mar Reguant, Geert Ridder, Jean-Marc Robin, Nicolas Schutz, Jesse Shapiro, Otto Toivanen, Mo Xiao, and numerous seminar and conference participants for helpful comments and discussions. We are extremely indebted to Jaime Castillo and Luis Guillermo V\'elez for introducing us to the Colombian electricity market and patiently addressing all our questions. Excellent research assistance was provided by Santiago Vel\'asquez Bonilla, Cristian Chica, Mathias Dachert, Brayan Perez, and Nicol\'as Torres. Michele Fioretti thanks the Sciences Po Advisory Board for financial support. All errors and omissions are ours.}
}}

\vspace{4ex}
\author{{Michele Fioretti}\thanks{Bocconi University,  IGIER, and CEPR. e-mail: \href{fioretti.m@unibocconi.it}{fioretti.m@unibocconi.it} }
	\and
	{Junnan He}\thanks{Sciences Po, Department of Economics. e-mail: \href{junnan.he@sciencespo.fr}{junnan.he@sciencespo.fr} } 
	\and
	{Jorge Tamayo}\thanks{Harvard University, Harvard Business School, Digital Reskilling Lab, and NBER. e-mail: \href{jtamayo@hbs.edu}{jtamayo@hbs.edu} }}

\date{June 8,  2026}

\begin{document}
	
	\maketitle
	\vspace{-1em}
	\begin{abstract} \singlespacing

		Does limiting the largest firm's capacity always lower prices? We model firms competing in supply schedules with multiple technologies, each with constant marginal cost up to capacity. In a tractable model, in which capacity and technological efficiency coexist as distinct sources of market power, we find that when the largest firm leads by efficiency, a small transfer of higher-cost capacity to the leader raises concentration yet lowers prices, contrary to standard antitrust intuition. Larger transfers raise prices, restoring standard intuition and tracing a non-monotone price-concentration relation. We prove existence and uniqueness of equilibrium, derive closed-form conditions for when transfers raise or lower prices, and extend the results to other oligopoly models. Evidence from Colombia's wholesale electricity market, where weather shocks shift hydropower capacity across technology-diversified firms, supports the pattern. Counterfactual transfers lower prices up to 30\% in the least concentrated markets. We draw implications for capacity caps, divestitures, and mergers.


		\medskip
		
		
	\end{abstract}
	
	\onehalfspacing
	
	\thispagestyle{empty}
	\newpage
	\pagenumbering{arabic}

	\newpage
	
	\section{Introduction}
	Regulators limit concentration in productive capacity—from airport runways to radio spectrum to, increasingly, AI compute—on the premise that it raises prices. In 2009, the UK Competition Commission forced BAA to divest Gatwick and Stansted, finding that its concentrated ownership of London runway capacity raised prices for airlines and passengers. US antitrust authorities likewise require divestitures of spectrum licenses in mobile carrier mergers, notably in T-Mobile/Sprint (2020). Colombia caps the share of installed electricity generation any one firm may own at 25\%. The 2026 Constellation--Calpine merger was approved only after a DOJ-ordered divestiture of 4.4~GW of gas generation. These interventions, and the emerging debate over concentration in cloud computing infrastructure \citep{korinek2025concentrating}, share a common premise: concentrating productive capacity in a few firms increases market power and thus prices.
	
	Yet there is little theoretical guidance for how capacity translates into market power, especially when, as in all these examples, firms operate multiple technologies with different costs and capacity constraints.\footnote{For instance, cloud computing shares the structural features we study: (i) a handful of firms such as Amazon, Microsoft, and Google own most large-scale data centers; (ii) each operates diversified portfolios of CPUs and GPUs with distinct marginal costs and capacities; (iii) demand for compute services is uncertain and (iv) pricing is a function of quantity of tokens.} In standard oligopoly models, capacity enters only at corners: either the constraint binds or it is irrelevant. These models say little about how the distribution of capacity across technologies and firms shapes competition when constraints are slack. Most settings compound the problem: data on technology-specific capacities and costs are rarely observed. Regulators therefore lack both a theory linking capacity to prices and the data to evaluate whether a proposed technology reallocation will raise or lower prices.
	
	This paper shows that, when firms hold more than one technology, the key distinction is not whether concentration is high or low, but \textit{why} the largest firm has market power. If market power comes from a firm's capacity (its size), concentrating capacity raises prices. If it comes from efficiency, transferring capacity to the efficient firm can lower prices even though this firm is the largest and concentration rises. When instead a firm operates a single technology, whether literally or because one of its technologies is effectively unlimited, efficiency has no role to play, and market power can only come from capacity.
	
	To establish this, we provide a formal link from capacity to market power in a supply-function equilibrium model \citep{klemperer1989supply}. Firms commit to a full supply schedule before demand is realized. A quantity offered at one price is therefore not a local decision: units offered at low prices remain inframarginal when demand is high and prices rise. With finite capacity, deploying a large share of capacity at low prices leaves less room to ramp up supply in high-demand states, so the inverse supply schedule steepens at the top. This link across price states is what makes an efficient firm constrained: the same low-cost technology that gives it incentives to undercut rivals also makes it exhaust capacity sooner along the schedule. We provide closed-form sufficient conditions for when capacity transfers between asymmetric, technology-diversified firms lower prices and when they raise them, and then test the mechanism empirically. The result is a paradox of concentration: prices can rise both when the dominant firm's capacity expands and when it contracts, even after controlling for the market's total available capacity.

	Evidence for this mechanism comes from Colombia's wholesale electricity market, where firms own multiple production technologies and weather-driven variation in hydropower capacity provides identification. The empirical results and counterfactual simulations show both regimes: additional thermal capacity can lower prices by letting the efficient leader compete more aggressively when hydropower is scarce, but can raise them when its hydropower‑capacity advantage is already large. Colombia caps any firm at 25\% of installed capacity (\citealp{alfaro2014deregulation}, CREG regulation); our results suggest that such caps can be counterproductive in the first case.\footnote{Ownership caps are common in deregulated electricity markets. For instance, in the UK, the 2008 EDF acquisition of British Energy required divestiture of the Eggborough coal plant to satisfy European Commission antitrust concerns.}

	Our theory shows that the standard intuition that capacity concentration raises prices holds when firms operate a single technology or when the expanding firm has enough capacity that its supply decisions are effectively unconstrained. The novel case arises when the largest firm also operates the low-cost technology. Because low costs make aggressive low-price bidding attractive, the largest firm bids a flat inverse supply function through most of its capacity, undercutting rivals to win market share. As quantity rises, the firm faces a trade-off: keep bidding low to win more share, or raise the bid and earn more on every inframarginal unit. The inverse supply function therefore steepens. At the capacity ceiling, the function is vertical: the firm cannot supply more regardless of price. At peak demand, the largest firm is pinned at capacity, so rivals supply the marginal units and set the clearing price.
	
	This is the sense in which the largest firm is ``constrained.'' The issue is not that it fails to earn the high market-clearing price on the units it already supplies; it does. The issue is that, once capacity binds, it cannot expand supply further to undercut rivals on additional units. Transferring some high-cost capacity from a rival to the largest firm holds total available capacity fixed but moves this constraint outward. Because transferred capacity operates at the same cost under either owner, these price effects do not come from scale economies or technological synergies, but from strategic changes in supply incentives. The largest firm now bids the transferred high-cost units at lower prices than the rival did, undercutting rivals on high-price marginal units, and can bid its low-cost technology even more aggressively at low prices. Rivals, facing a more aggressive competitor, expand output to defend market share, and prices fall despite rising concentration.

	Only when the transfer is large enough to weaken rivals substantially does the standard concentration effect dominate and prices rise, producing a U-shape. For policy, the relevant point is not the exact location of the minimum, but the existence of the left arm: some concentration-increasing, diversification-enhancing transfers are pro-competitive. We formalize this mechanism in a tractable supply-function equilibrium with multi-technology firms and a non-strategic fringe. 
	
		In the duopoly case, we prove existence, uniqueness, and closed-form comparative statics. Existence and uniqueness rest on two ingredients. Full support of the demand distribution ensures that each firm's supply schedule is a best response at every price, not just locally. Capacity constraints provide the boundary condition that selects a unique equilibrium: at the fringe entry price, any firm with idle capacity could profitably undercut, so at least one firm must exhaust its full capacity there. This boundary condition pins down the equilibrium schedules through the first-order conditions.\footnote{The key equilibrium force is strategic complementarity in supply schedules: in the first-order conditions, a steeper rival schedule raises a firm's optimal supply slope. A capacity change for one firm therefore propagates through equilibrium schedules.} We then extend the analysis beyond duopoly numerically and find the same non-monotonicity when capacity is reallocated to the leader.
	
	The closed-form comparative statics deliver two further ownership implications. When both firms hold identical technology portfolios, symmetric ownership is the most competitive structure: any reallocation away from symmetry raises prices, because the transferring firm cuts its low-cost supply at low prices while the receiving firm restricts supply to leverage its larger capacity. The comparison between specialized and diversified ownership, however, depends on demand: diversification creates within-technology rivalry that lowers prices at low demand, while specialized ownership of the low-cost technology serves high demand more cheaply, before high-cost technology is needed for the market to clear.
	
	The mechanism is not tied to a homogeneous-good supply-function environment. \citet{maggi1996strategic} studies a differentiated-product price-capacity model in strategic trade, where equilibrium moves between Bertrand and Cournot as capacity constraints become more important. A multi-technology extension of that pricing subgame delivers the same sign reversal: transfers to the larger efficient firm lower prices when they relax its constraint and raise prices when they mainly weaken rivals. Thus, the central insight extends beyond SFE to differentiated-product price competition, a standard setting in the trade literature.
	
	These predictions are not theoretical curiosities. To set the stage, we open the paper with reduced-form evidence from a large wholesale electricity market. Colombia is a natural setting to bring the theory to data: generation capacity is reported at the unit level, thermal plants and hydropower dams have distinct and observable marginal costs and capacities, firms are diversified, and submit daily supply schedules with only a forecast of hourly demand. Instead of relying on mergers, which are endogenous, weather-driven variation in dam inflows generates exogenous shifts in how capacity is distributed across firms. Controlling for total available capacity, prices are higher both when the dominant hydro firm faces drought (it becomes constrained, concentration falls) and when it benefits from abundance (the standard concentration effect takes over). The two forces trace out a U-shape, documented without imposing any model of firm behavior.
	
	Our analytical results are deliberately low-dimensional. They isolate the mechanism in closed form; the empirical setting requires solving the corresponding problem for all Colombian firms and their observed technology portfolios. We therefore extend the theoretical model to the Colombian setting, in which $N$ firms submit hourly supply schedules each day under uncertain demand, and recover technology-specific marginal costs from firms' first-order conditions. Counterfactual simulations reallocate thermal capacity to the market leader while holding effective costs fixed, isolating the effect of portfolio composition. Small transfers lower prices by up to 30\% in the least concentrated markets; large ones raise them, reproducing the U-shape.

	\paragraph{Related literature.} The prevailing view across several literatures is that concentration raises prices. Under Cournot, prices rise monotonically with concentration measured in market shares \citep[e.g.,][pp.\ 221--223]{tirole1988theory}; capacity-constrained Bertrand yields the same prediction, though the mechanism runs through capacity precommitment rather than direct quantity choice \citep{kreps1983quantity}. The corporate finance literature reinforces this view: mergers improve market outcomes primarily through efficiency gains, whether from more productive asset deployment \citep{maksimovic2001market}, quality convergence \citep{sheen2014real}, product-market synergies \citep{hoberg2010product}, or successful post-acquisition integration \citep{gokkaya2026when}. Absent such gains, as in our analysis, the presumption is that consolidation raises prices \citep{farrell1990horizontal,eckbo1983horizontal}.\footnote{The upward pricing pressure framework \citep{farrell2010antitrust,nocke2022concentration} formalizes this presumption as a merger screen for differentiated-product markets. \cite{greenfield2021upward} extend it to capacity-constrained Bertrand settings and find that constraints amplify upward pressure. In the Bertrand-Edgeworth tradition, \cite{chen2018horizontal} and \cite{compte2002capacity} show that mergers raise prices when they create slack capacity. \cite{friberg2015divestiture} and \cite{delaprez2024unveiling} study the competitive effects of divestitures; \cite{illanes2025large} provide a large-scale empirical evaluation of merger simulation predictions. Our results also connect to \cite{obrien2000competitive} and \cite{lopez2019overlapping} who show that ownership of financial stakes in rivals softens competition; we show that transfers of physical capacity shares can intensify it when the acquiring firm is constrained.} Empirically, however, the mapping from concentration to prices is less clean than this view suggests.\footnote{\citet{bushnell2008vertical} find that observed markup differences across three U.S. electricity markets reflect long-term contracts rather than concentration; we control for forward contracts.}
	
	Our asset-transfer experiment is related to work that treats mergers and remedies as reallocations of productive assets. \citet{perry1985oligopoly} make assets explicit by letting firms produce with labor and a fixed industry stock of capital; the technology has constant returns in output and capital jointly, so moving capital across firms shifts firm-level marginal costs. Recent work on structural remedies similarly studies divestitures that move assets to rivals or entrants \citep{nocke2026optimal,tesoriere2025merger}. These approaches are natural for merger policy, but they require assumptions about how transferred assets are used: whether the asset inherits the receiving firm's managerial productivity \citep{bloom2007measuring}, whether its own quality is preserved, and how heterogeneous capital goods are aggregated \citep{robinson1953production}. We take a complementary route. A technology is a capacity block with an owner-invariant marginal cost. This shuts down scale economies and integration synergies, and isolates the strategic effect of who owns the capacity.
	
	The question of how capacity shapes price competition has a long history, going back at least to \citet{edgeworth1925papers}, who first showed that capacity-constrained firms need not converge to the Bertrand outcome. A subsequent literature in the Bertrand-Edgeworth tradition \citep{levitan1972price,kruse1994bertrand} developed the point formally, but these models typically lack pure-strategy equilibria when firms differ in capacity, requiring  mixed-strategy approaches that make comparative statics on mergers and divestitures difficult \citep{froeb2003bertrand}.\footnote{See \cite{vives1993edgeworth} for a survey of Edgeworth's contributions to oligopoly theory, \cite{noel2008edgeworth} on dynamic Edgeworth cycles, and \cite{beckmann1965edgeworth,cheviakov2005beckmann} on characterizing mixed-strategy equilibria.} 
	
	The supply function equilibrium of \cite{klemperer1989supply} resolves this by expanding the strategy space from a single price to an entire schedule: firms choose output at every possible price, and demand uncertainty selects a unique equilibrium. This framework has been applied to electricity markets \citep{green1992competition,hortaccsu2008understanding}, strategic trade \citep{laussel1992strategic}, financial markets \citep{hortaccsu2018bid}, and government procurement \citep{delgado2004coalition}; see \cite{vives2011strategic} and \cite{holmberg2015supply} for surveys. With capacity constraints, \citet{holmberg2007supply} establishes uniqueness in the asymmetric-capacity, single-technology case, and \citet{genc2011supply} characterize refinements of the equilibrium set when uniqueness conditions fail. We analyze the framework in an asymmetric duopoly with two technologies, proving existence and uniqueness and deriving closed-form comparative statics on the effect of capacity transfers.\footnote{As argued by \cite{grossman1981nash}, firms can commit to supply schedules by specifying output at each price in contractual agreements \citep[see also][]{wilson1979auctions}.}
	
A related strand models electricity markets as multi-unit auctions, where firms submit price--quantity pairs for a known level of demand \citep[e.g.,][]{ausubel2014demand,fabra2006designing,crawford2007bidding}. Closest to us, \citet{fabra2024fossil} compare specialized and diversified ownership structures, showing that diversification fosters within-technology competition while specialization can preserve productive efficiency. Other work also shows that ownership of low-cost or renewable capacity matters when firms have market power \citep{genc2019should,bahn2021market}. Our question is different: rather than comparing ownership regimes, we ask whether reallocating a given amount of capacity from one firm to another has a monotone price effect. It does not. The sign depends on whether the receiving firm is capacity-constrained, which in turn depends on demand and on the firm's technology portfolio.\footnote{Under Cournot competition, giving a strategic firm a more diversified portfolio leads it to withhold output: \cite{acemoglu2017competition} show that firms with more renewables reduce thermal supply, and \cite{bushnell2003mixed} shows that diversified hydro-thermal firms shift hydropower away from peak periods. Our supply function equilibrium delivers the same logic when the leader has ample capacity, but reverses it when the leader is capacity-constrained.}  This has direct implications for antitrust and regulatory ownership caps: concentration alone is not a sufficient statistic for competitive harm.

\vspace{1em}
The paper proceeds in four stages. Sections~\ref{s:background} and~\ref{s:em} introduce the Colombian wholesale electricity market and document the empirical regime switch behind the U-shape. Section~\ref{s:framework} develops a static supply-function model that explains why capacity transfers lower prices in the efficiency regime and raise them in the capacity regime. Section~\ref{s:model} quantifies the mechanism in the Colombian market, using the four largest strategic firms and treating the estimated value of water as an effective cost held fixed across counterfactual reallocations. Section \ref{s:conclusion} concludes.

\section{The Colombian Wholesale Electricity Market}\label{s:background}

\begin{figure}[!ht]
	\centering
	\caption{Installed capacity and production volumes by technology}\label{fig:by_tech}
	\begin{subfigure}{.95\textwidth}\centering
		\caption{Total installed capacity by technology}\includegraphics[width=1\linewidth]{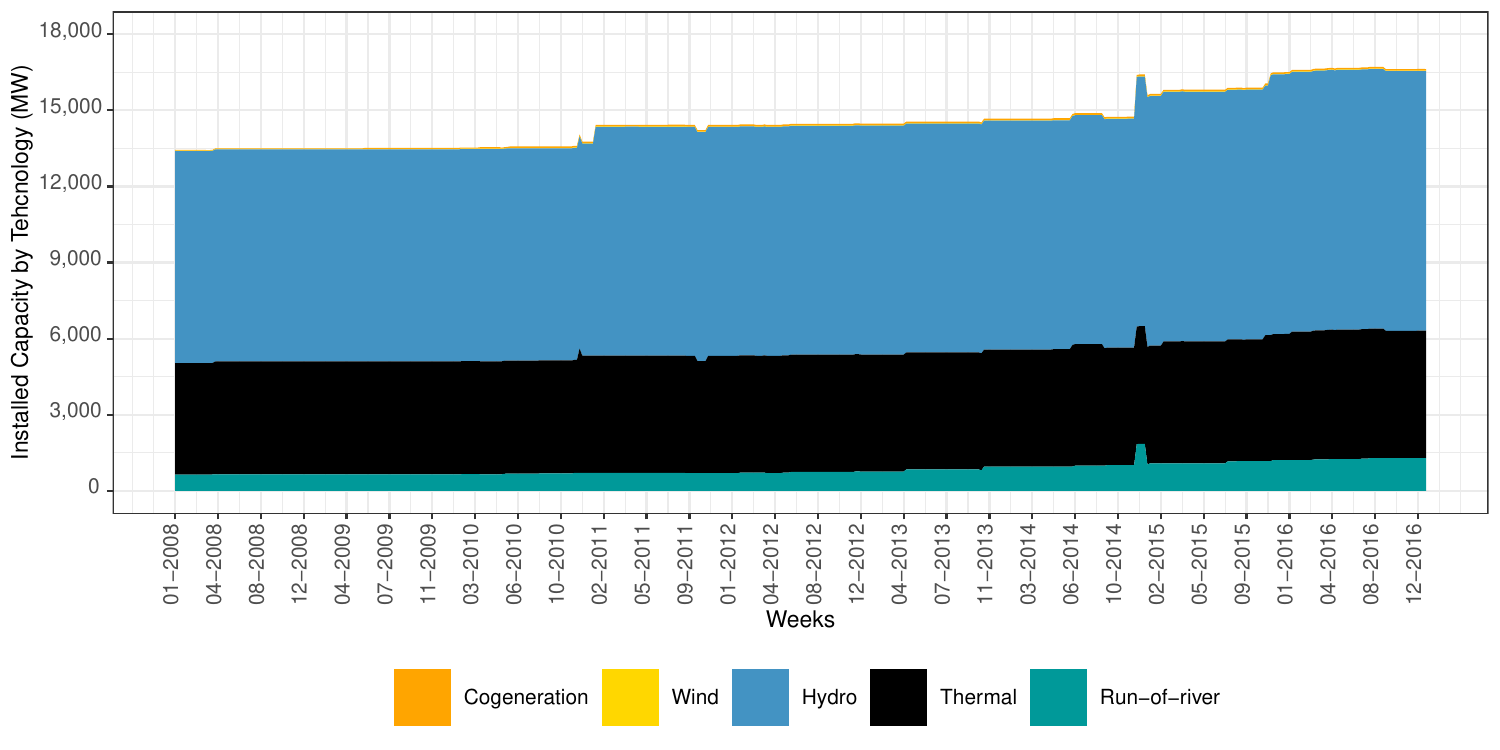} \hfill
		\label{fig:cap_by_tech}
	\end{subfigure}
	\begin{subfigure}{.94\textwidth}\centering
		\caption{Total weekly production by technology}
		\colorbox{white}{\includegraphics[width=1\linewidth]{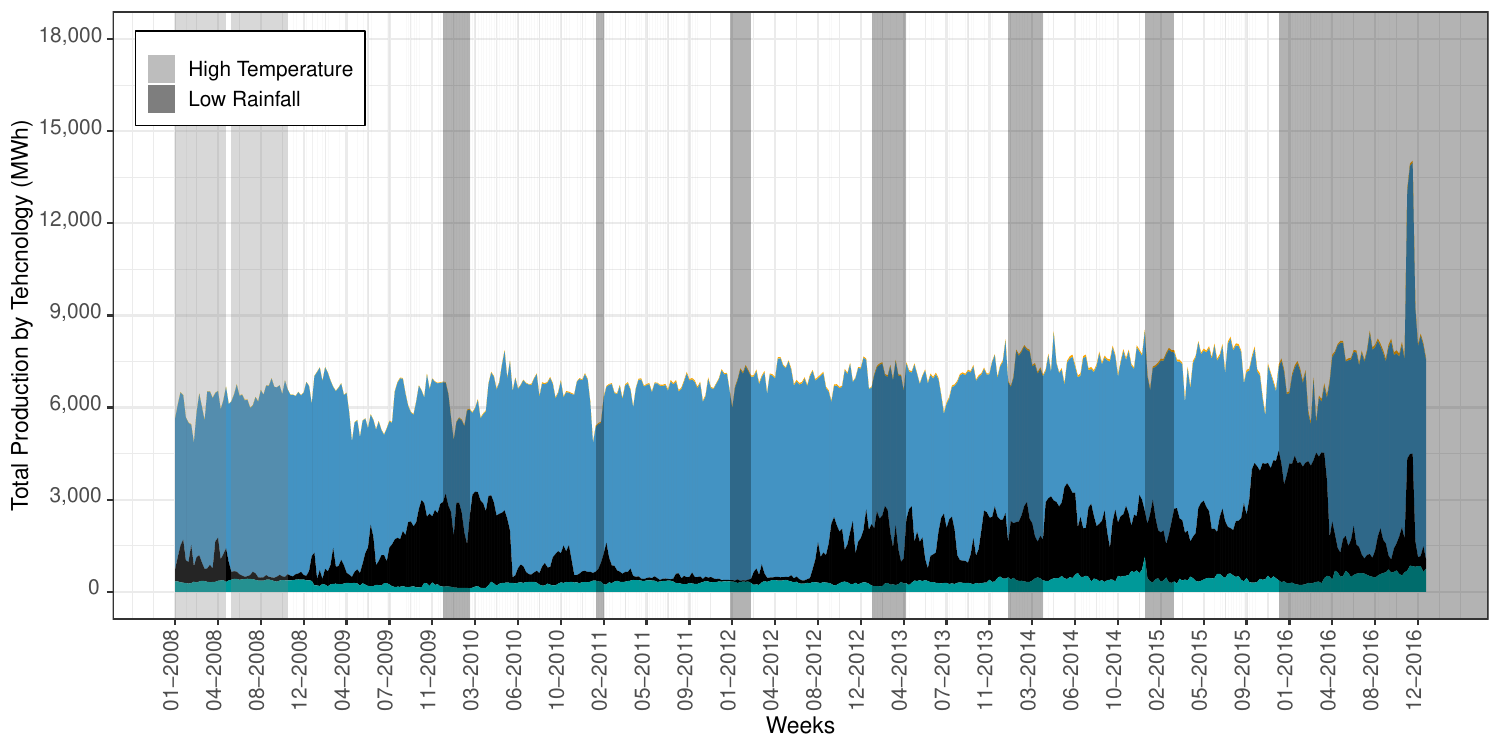}} \hfill
		\label{fig:prod_by_tech} 
	\end{subfigure}
	\begin{minipage}{1 \textwidth}
		\vspace*{-0.5em}
		{\footnotesize Note: Vertical bars in Panel~(b) mark periods where a dam experiences temperature (rainfall) more than one standard deviation above (below) its long-run average. \par}
	\end{minipage}
\end{figure}

During 2010--2015, Colombia’s electricity market comprised roughly 190 generation units operated by approximately 50 firms. Ownership is highly concentrated: six firms control more than 75\% of total installed capacity, while most firms operate a single, small-capacity unit. Firms hold diversified portfolios of large hydroelectric dams, coal and gas thermal plants, and run-of-river units. Panel~(a) of Figure~\ref{fig:by_tech} shows installed capacity by technology from 2008 to 2016: hydropower and thermal account for roughly 60\% and 30\% of total capacity respectively; solar, wind, and cogeneration are negligible. Hydropower dominates production, supplying on average 75\% of dispatched electricity. Dams owned by the same firm tend to cluster on adjacent rivers (Panel (a) of Appendix Figure~\ref{fig:omitted}).

\begin{figure}[!ht]
	\centering
	\caption{Market prices}\label{fig:prices}
		\includegraphics[width=1\linewidth]{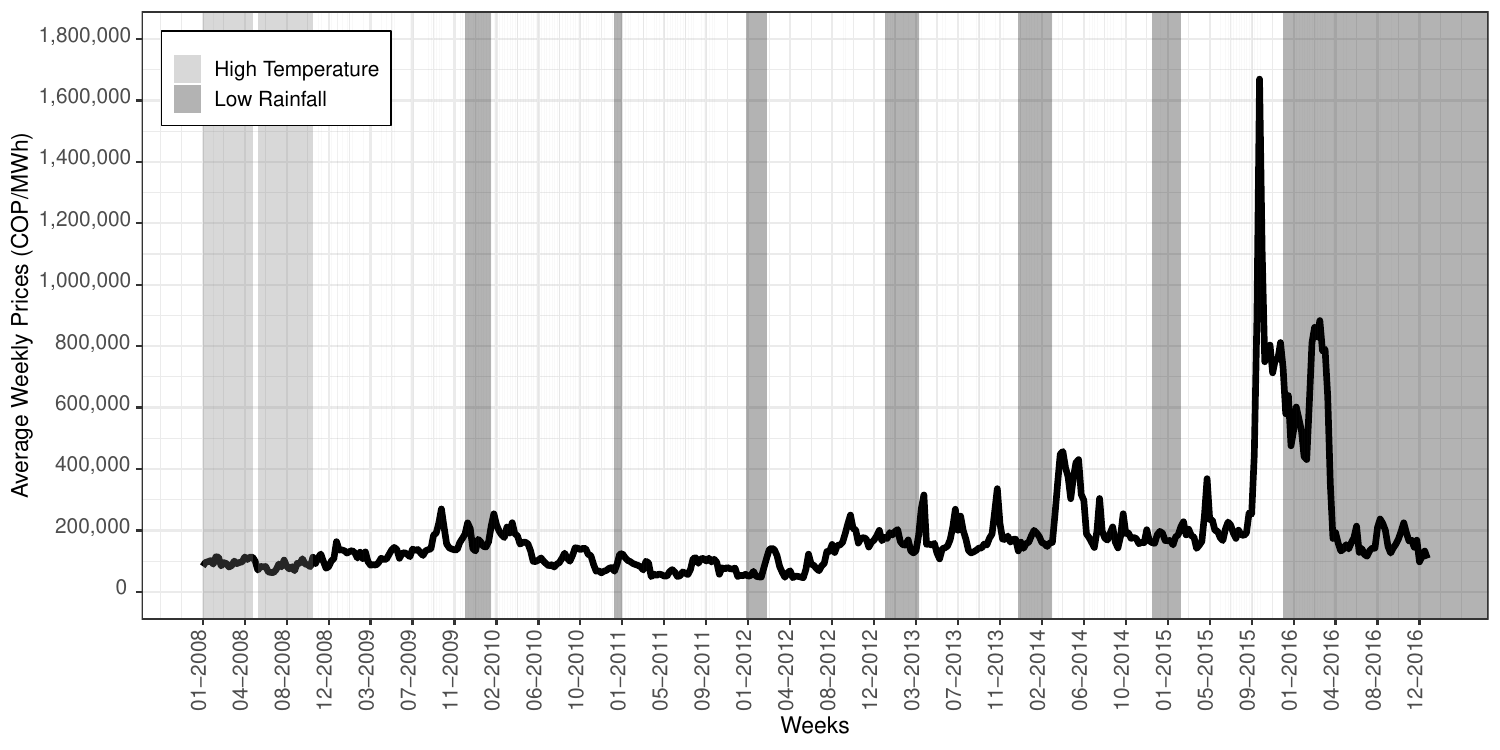} \hfill
	\begin{minipage}{1 \textwidth}
		\vspace*{-0.5em}
		{\footnotesize Note: Weekly average market prices. Vertical bars mark scarcity periods as in Figure~\ref{fig:by_tech}. 2,900 COP $\simeq$ 1 USD. \par}
	\end{minipage}
\end{figure}

Production varies sharply with hydrological conditions. Panel~(b) shows that hydropower output falls during dry spells (proxied by above-average temperatures or below-average rainfall at dams) while thermal generation rises to compensate. Swings are especially large during El~Ni\~no (2015--2016) and La~Ni\~na episodes. Wholesale prices reflect this substitution: Figure~\ref{fig:prices} shows prices more than doubling during droughts relative to normal conditions, with peak levels during the prolonged 2016 El~Ni\~no episode and the regular December--March dry season.

\subsection{Institutions}\label{s:institutions}

\noindent\textbf{Spot market.} The spot market is a multi-unit uniform-price auction operated daily by \textit{XM}. Each firm submits one price bid per unit per day and one quantity bid per unit per hour. Price bids state the minimum price at which a unit will produce; quantity bids state the maximum output at that price. \textit{XM} stacks bids in merit order to form aggregate supply, crosses it against projected hourly demand, and sets the market-clearing price at the bid of the marginal unit, which all dispatched units receive.\footnote{Thermal units may receive reimbursement for startup costs \citep{balat2022effects}. \cite{bernasconi2023relational} document collusion among thermal units until 2009, after which regulatory changes reduced coordination.}

\vspace*{1ex}
\noindent\textbf{Forward market.} Firms hedge spot price risk through bilateral forward contracts. We observe each unit's aggregate contract position per hourly market; forward contract prices are not publicly disclosed at the time of bidding.

\vspace*{1ex}
\noindent\textbf{Reliability mechanism.} Units with reliability charge obligations (\textit{cargo por confiabilidad}) must produce a committed quantity $\overline{q}_{ijt}$ whenever the spot price exceeds a monthly scarcity threshold $\overline{p}_t$. We treat both as given in the spot market \citep{cramton2007colombia,mcrae2024reliability}.

\subsection{Data}\label{s:data}

We use \textit{XM} data covering 2006--2017: unit-level quantity and price bids, forward contract positions, ownership, capacity, and daily water inflows and reservoir stocks for dams. Weather variables (rainfall and temperature) are inverse-distance-weighted averages of 303 meteorological stations within 120\,km of each unit, adjusted for orography using data from the \textit{Agust\'in Codazzi Geographic Institute}. Monthly NINO3.4 indices from the \textit{International Research Institute} at \textit{Columbia University} provide ENSO forecasts over a nine-month horizon (2004--2017), used to predict dam inflows. International daily prices of oil, gas, coal, liquid fuels, and ethanol complete the dataset.

\section{Prices and Concentration: A U-Shape}\label{s:em}

In models where firms differ only in size, such as Cournot, prices are monotonically increasing in market concentration: transferring capacity from a small firm to a large one raises concentration and raises prices. This monotone relationship underlies the widespread use of the Herfindahl-Hirschman Index (HHI) as a proxy for market power. Yet testing this prediction empirically is difficult, because market-share-based HHI depends on equilibrium prices, creating a simultaneity problem \citep{miller2022misuse}.

We provide causal evidence that the relationship is not monotone. Controlling for total available capacity, prices rise both when the dominant hydro firm faces drought (it becomes capacity-constrained and concentration falls) and when it benefits from abundance (concentration rises and the standard effect takes over). The two forces trace out a U-shape. This section documents the fact without imposing any model of firm behavior (Section~\ref{s:em_prices}), and shows that firms' supply adjustments are consistent with it (Section~\ref{s:empirical_strategy}).

\subsection{Empirical Evidence}\label{s:em_prices}

We leverage a capacity-based measure of concentration that does not depend on prices. Forecast inflows shift each firm's effective hydro capacity. Because dams owned by the same firm cluster on adjacent rivers while dams on distant rivers tend to belong to different firms, these shifts are approximately independent across firms conditional on season \citep[e.g.,][]{poveda2006annual}.

We classify firm $i$'s $l$-month-ahead forecast as \textit{adverse} (\textit{favorable}) if at least one of its dams faces a forecast more than one standard deviation below (above) its historical average for that horizon and month, and let
\[
\text{net adverse}_{i,t+l} \;\equiv\; \text{adverse}_{i,t+l} - \text{favorable}_{i,t+l},
\]
so that $\text{net adverse}_{i,t+l} \in \{-1,\,0,\,+1\}$. Our running variable weights each firm's forecast by the square of its capacity share $s_i$:
\begin{equation}\label{eq:delta}
	\Delta_{t+l} \;\equiv\; -\sum_{i} \text{net adverse}_{i,t+l} \cdot  s_{i,t-1}^2.
\end{equation}
The sum runs over firms with dams; each such firm's share $s_{i,t-1}$ is based on its full effective capacity, lagged hydro plus time-invariant thermal. Thus $\Delta_{t+l}$ has the sign of the forecast-induced change in capacity concentration: when $\Delta_{t+l}<0$, large firms receive adverse hydro forecasts, lose effective capacity relative to rivals, and concentration falls; when $\Delta_{t+l}>0$, large firms receive favorable forecasts, gain effective capacity, and concentration rises. 

To justify equation \eqref{eq:delta}, a first-order Taylor expansion shows that the forecast-induced change in the capacity-based HHI is proportional to $\Delta_{t+l}$.\footnote{The $s_i^2$ weighting in \eqref{eq:delta} follows from two facts: $\text{HHI}=\sum_i s_i^2$ is quadratic in the capacity shares $s_i=K_i/K$, and, as is standard in hydrology, inflows follow a location-scale family whose scale grows with dam size, so a standardized adverse forecast lowers effective capacity in proportion to its size. A first-order expansion then gives $\Delta\,\text{HHI}\approx \Delta_{t+l} + \text{HHI}\sum_i \text{net adverse}_{i,t+l}$, whose second term is mean-zero across weakly correlated dams as in our data. Using $\Delta\,\text{HHI}_{t+3}$ rather than $\Delta_{t+3}$ as the running variable in the regression discontinuity below yields consistent results: the estimated left and right slopes are $-4.05$ and $+2.46$ (bootstrap standard errors $0.15$ and $0.11$), with a statistically significant slope change. \label{foot:hhi}} Under the null, prices should increase as $\Delta_{t+l}$ moves from negative to positive values: negative values imply lower concentration, while positive values imply higher concentration.

\paragraph{Implementation.} Inflow forecasts are computed from an ARDL model estimated on a two-year rolling window, using past inflows, local temperature and rainfall, and ENSO probabilities as predictors; Appendix~\ref{apndx:forecast} provides details. We set $l=3$ months because, as the next section shows, this is the earliest horizon at which sibling thermal units begin adjusting price bids in response to adverse inflow forecasts, indicating that capacity constraints become empirically relevant at this horizon. The concentration measure $\Delta_{t+l}$ is defined in (\ref{eq:delta}), with capacity shares $s_i$ computed from thermal capacity (time-invariant) plus lagged hydro water stocks (time-varying).

\begin{figure}[!t]
    \caption{Market price and inflow-driven changes in concentration}
     \label{fig:empiricalU}
        \centering
        \includegraphics[width=.8\textwidth]{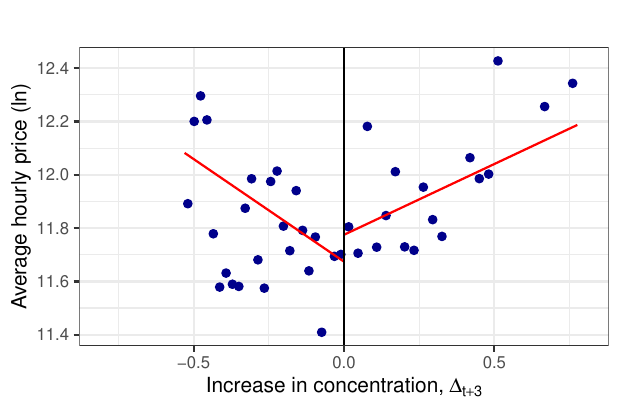}
\begin{minipage}{1 \textwidth}
{\footnotesize Notes: Binned scatter plot of weekly log market prices against $\Delta_{t+3}$, defined in (\ref{eq:delta}). Solid lines are local linear fits on each side of $\Delta_{t+3}=0$. Month-by-hour and year fixed effects are residualized from the dependent variable. Bootstrap 95\% CIs (300 repetitions): left slope $[-0.834,-0.649]$, right slope $[0.448,0.575]$, slope change $[1.148,1.367]$. \par}
\end{minipage}
\end{figure}

\paragraph{Results.} We first show this graphically. Figure~\ref{fig:empiricalU} plots log market prices, residualized of fixed effects, against $\Delta_{t+3}$ at the week-by-hour level. Weekly aggregation reduces serial correlation and within-week dispatch noise while preserving the forecast variation relevant for the design. The relationship is U-shaped: prices rise as $\Delta_{t+3}$ moves away from zero in either direction. The left slope is negative and significant (bootstrap 95\% CI: $[-0.834,-0.649]$), the right slope is positive ($[0.448,0.575]$), and the slope change is large and significant ($[1.148,1.367]$). 

The same U-shape holds at the daily frequency and at two- and three-month-ahead horizons (Appendix Figure~\ref{fig:empiricalU_robustness}), and when the $\pm 1$ standard-deviation indicator is replaced by the continuous standardized forecast (Appendix Figure~\ref{fig:empiricalU_continuous}). The left side is the key finding: when large firms lose effective capacity relative to rivals, concentration \textit{falls} and prices \textit{rise}, contradicting the monotone prediction.

\FloatBarrier

The binned scatter is descriptive: it residualizes fixed effects but not aggregate inflow conditions or lagged market controls. For inference, we estimate the corresponding pooled regression-kink specification:
\begin{equation}\label{eq:rd_piecewise}
	\begin{aligned}
		\ln(p_{th}) &= \gamma_1^-\cdot \Delta_{t+l}\mathds{1}[\Delta_{t+l}<0] + \gamma_1^+ \cdot \Delta_{t+l}\mathds{1}[\Delta_{t+l}\geq 0] \\
		&+ \alpha_0 \cdot \mathds{1}[\Delta_{t+l}>0] + \alpha_1 \cdot \text{HHI}_{t-1} + \alpha_2 \cdot \textstyle\sum_i \text{net adverse}_{i,t+l} \\
		&+ \mathbf{x}_{t-1,h}\,\boldsymbol{\alpha} + \mathbf{x}_{t-1,h}\,\boldsymbol{\alpha}^+\, \mathds{1}[\Delta_{t+l}>0] + \tau_{th} + \epsilon_{th},
	\end{aligned}
\end{equation}
where $\gamma_1^-$ and $\gamma_1^+$ are the slopes of log price in $\Delta_{t+l}$ below and above the cutoff; their difference $\gamma_1^+-\gamma_1^-$ is the kink of interest. A local U-shape corresponds to $\gamma_1^-<0<\gamma_1^+$, and therefore to a positive kink ($\gamma_1^+-\gamma_1^->0$). The indicator $\mathds{1}[\Delta_{t+l}>0]$ allows a level shift $\alpha_0$ at the cutoff, distinct from the change in slope. 

The running variable $\Delta_{t+l}=-\sum_i \text{net adverse}_{i,t+l}\cdot s_{i,t-1}^2$ is in effect an interaction between the forecasts and the squared capacity shares, so the second-row controls are its two main effects: the predetermined concentration level $\text{HHI}_{t-1}=\sum_i s_{i,t-1}^2$ and the aggregate forecast $\sum_i \text{net adverse}_{i,t+l}$. Without them the kink would conflate the share-weighted reallocation we are after with the direct price effects of higher concentration or of more adverse forecasts on average; with them, $\gamma_1^-$ and $\gamma_1^+$ identify the response to $\Delta_{t+l}$ alone. 

Finally, the vector $\mathbf{x}_{t-1,h}$ contains lagged log demand, log forward contracts, log price, and log total available capacity, and is interacted with $\mathds{1}[\Delta_{t+l}>0]$ to let its relationship with $p_{th}$ differ on each side of the cutoff. Controlling for total available capacity is what distinguishes the paradox in Figure \ref{fig:empiricalU} from a mechanical scarcity effect, which sets higher prices when water stocks are scarce. The terms $\tau_{th}$ are month-by-hour and year fixed effects, with day-of-week-by-hour fixed effects added in the daily specification. Standard errors are Newey--West to account for serial correlation.

Appendix Table~\ref{tab:empiricalU_piecewise} reports the estimates. Columns~(1) and~(3) use the binary net-adverse running variable in \eqref{eq:delta}. Columns~(2) and~(4) use the same capacity-share-squared weighting but keep the standardized forecast magnitude at each dam, rather than coarsening it to the $\{-1,0,1\}$ indicator; Appendix~\ref{apndx:additional_results} defines this continuous running variable. Columns~(3) and~(4) add the $\text{HHI}_{t-1}\times\text{forecast}_{t+3}$ control implied by the full Taylor expansion from a change in HHI (see footnote \ref{foot:hhi}). The slope change is stable across specifications: about $+0.11$ for the indicator measure and about $+0.03$ for the continuous measure. The two arms are asymmetric: the scarcity (left) slope is steep and precise, the abundance (right) slope flat and statistically indistinguishable from zero.\footnote{The flat right arm is not a matter of statistical power: more than half of the observations lie on the abundance side of the cutoff, with dispersion in $\Delta_{t+l}$ comparable to the scarcity side.} As we will see, this asymmetry is reflected in the underlying capacity mechanism of Section~\ref{s:framework} and replicated in the simulations of Section~\ref{s:simulations}.
	
The next section demonstrates that these results are due to the way firms employ their thermal and hydropower units ahead of adverse and favorable forecasts.

\subsection{Supply Responses}\label{s:empirical_strategy}
 
Why do prices rise on both sides of Figure~\ref{fig:empiricalU}? The answer should appear in firms' bid schedules. If the U-shape reflects capacity scarcity and abundance, diversified firms should substitute between hydro and thermal capacity in opposite directions across the two regimes. We document these responses with the following regression:
\begin{equation}\label{eq:as_inflow}
    y_{ij,th} = \beta_{\ell}^{low} \, \text{adverse}_{ij,t+\ell} + \beta_{\ell}^{high} \, \text{favorable}_{ij,t+\ell} + \textbf{x}_{ij,t-1,h}\mathbf{\alpha} + \mu_{ijt} + \tau_{t} + \tau_{h}+ \epsilon_{ij,th},
\end{equation}
estimated separately for each forecast horizon $\ell \in \{1,2,3,4\}$ months ahead, where $y_{ij,th}$ is the quantity or price bid of unit $j$ of firm $i$, aggregated over weeks $t$ per hour $h$. Weekly aggregation is the natural frequency: the ARDL forecast indicators are constant within each week, so daily observations add no treatment variation while introducing within-week dispatch noise.\footnote{Units bidding below the 5th percentile of their technology distribution are excluded to remove unobserved maintenance periods; results are qualitatively unchanged.} For dams, $\text{adverse}_{ij,t+\ell}$ and $\text{favorable}_{ij,t+\ell}$ are dam-level forecast indicators; for \textit{sibling} thermal units (thermal plants owned by the same diversified firm as at least one dam), they are defined at the firm level, aggregating across all owned dams. Controls $\textbf{x}_{ij,t-1,h}$ include lagged demand, water stocks, and forward positions; the slope of lagged water stock is allowed to vary across units (or across firms for sibling thermal regressions). Fixed effects are at the unit-by-month and firm-by-year (in $\mu_{ijt}$), hour ($\tau_h$), and week-by-year ($\tau_t$) levels. Standard errors are clustered by unit, month, and year.

The identification assumption is that a firm’s current supply schedule does not directly depend on the lagged weather variables used for forecasting, beyond what is captured by the lagged water stock. Due to their rural locations, weather conditions at dams are unlikely to affect market-level demand. Appendix Figure~\ref{fig:forecast_errors} confirms that units do not respond to forecast errors, indicating that our forecasts capture the information firms act on.\footnote{Appendix Figure~\ref{fig:comp_responses} shows that units do not respond to \textit{competitors’} inflow forecasts.}

\begin{figure}[!t]
    \caption{Responses to inflow forecasts}
     \label{fig:responses}
    \centering
    \textbf{Top panel:} Responses of dams to own inflow forecasts\par\smallskip
    \begin{subfigure}{0.48\textwidth}
        \centering
        \includegraphics[width=\textwidth]{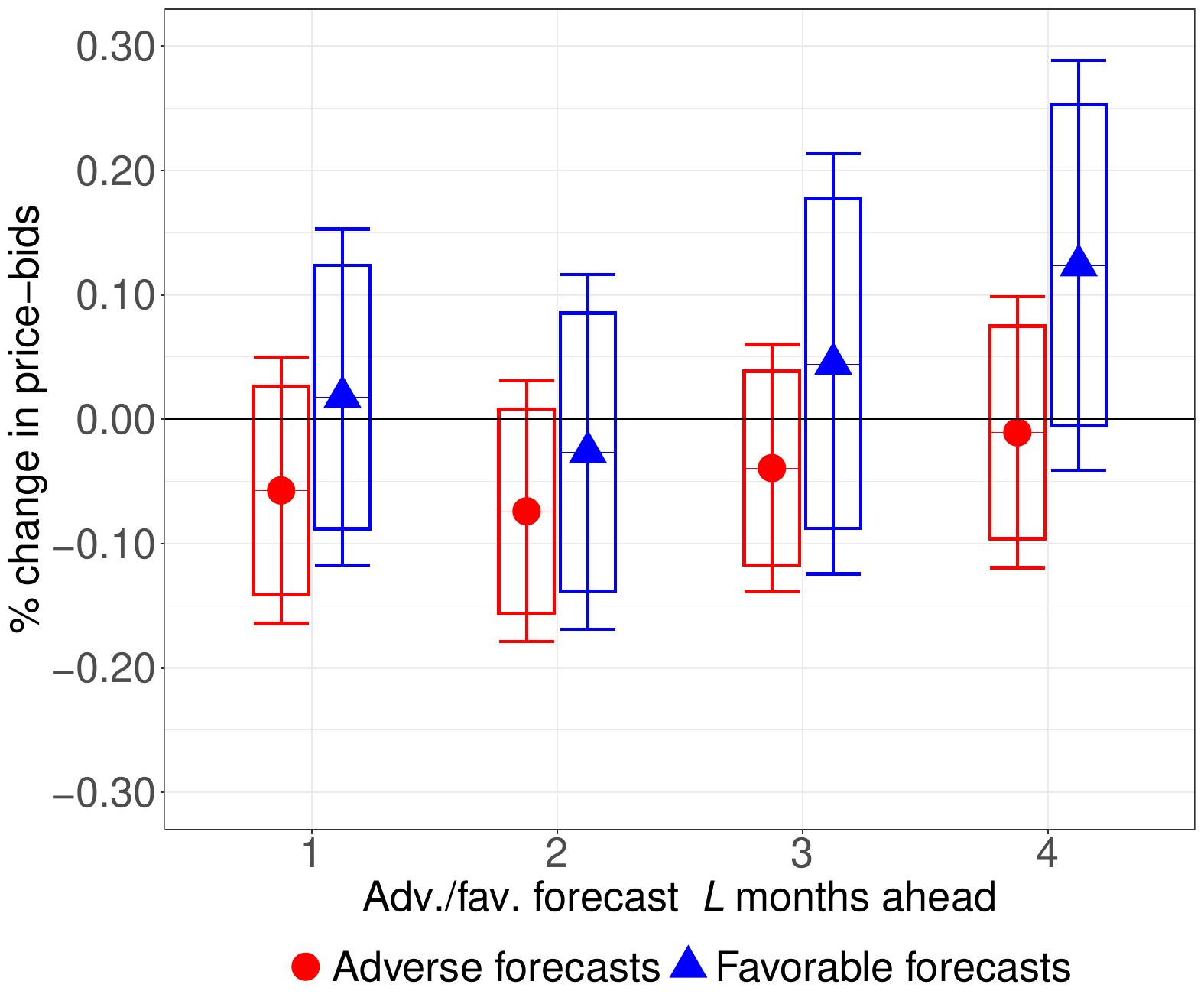}
        \captionsetup{width=0.9\linewidth, justification=centering}
        \caption{Price bids}
    \end{subfigure}\hfill
    \begin{subfigure}{0.48\textwidth}
        \centering
        \includegraphics[width=\textwidth]{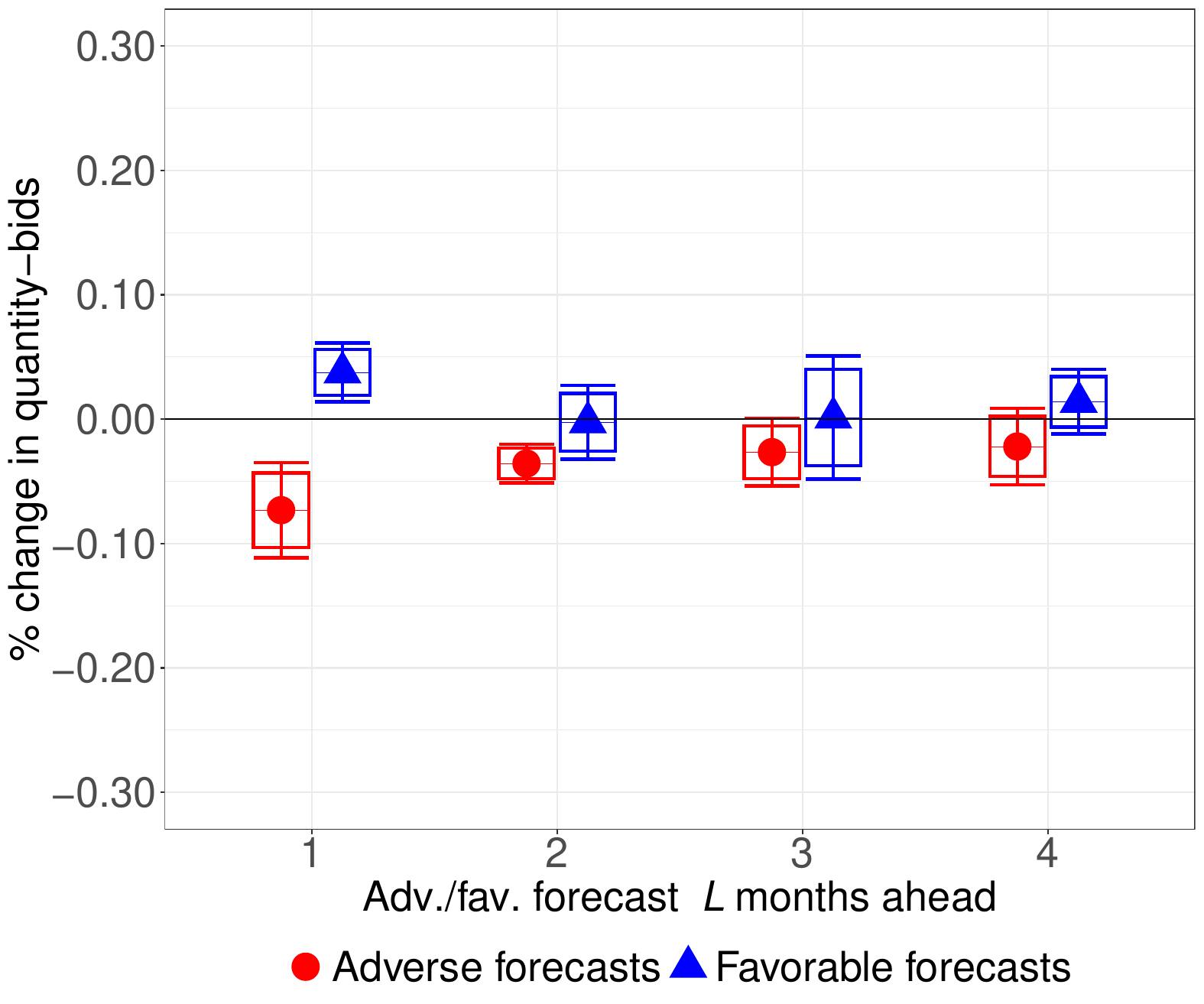}
        \captionsetup{width=0.9\linewidth, justification=centering}
        \caption{Quantity bids}
    \end{subfigure}
    \par\medskip\textbf{Bottom panel:} Responses of thermal units to the forecasts of their sibling dams\par\smallskip
    \begin{subfigure}{0.48\textwidth}
        \centering
        \includegraphics[width=\textwidth]{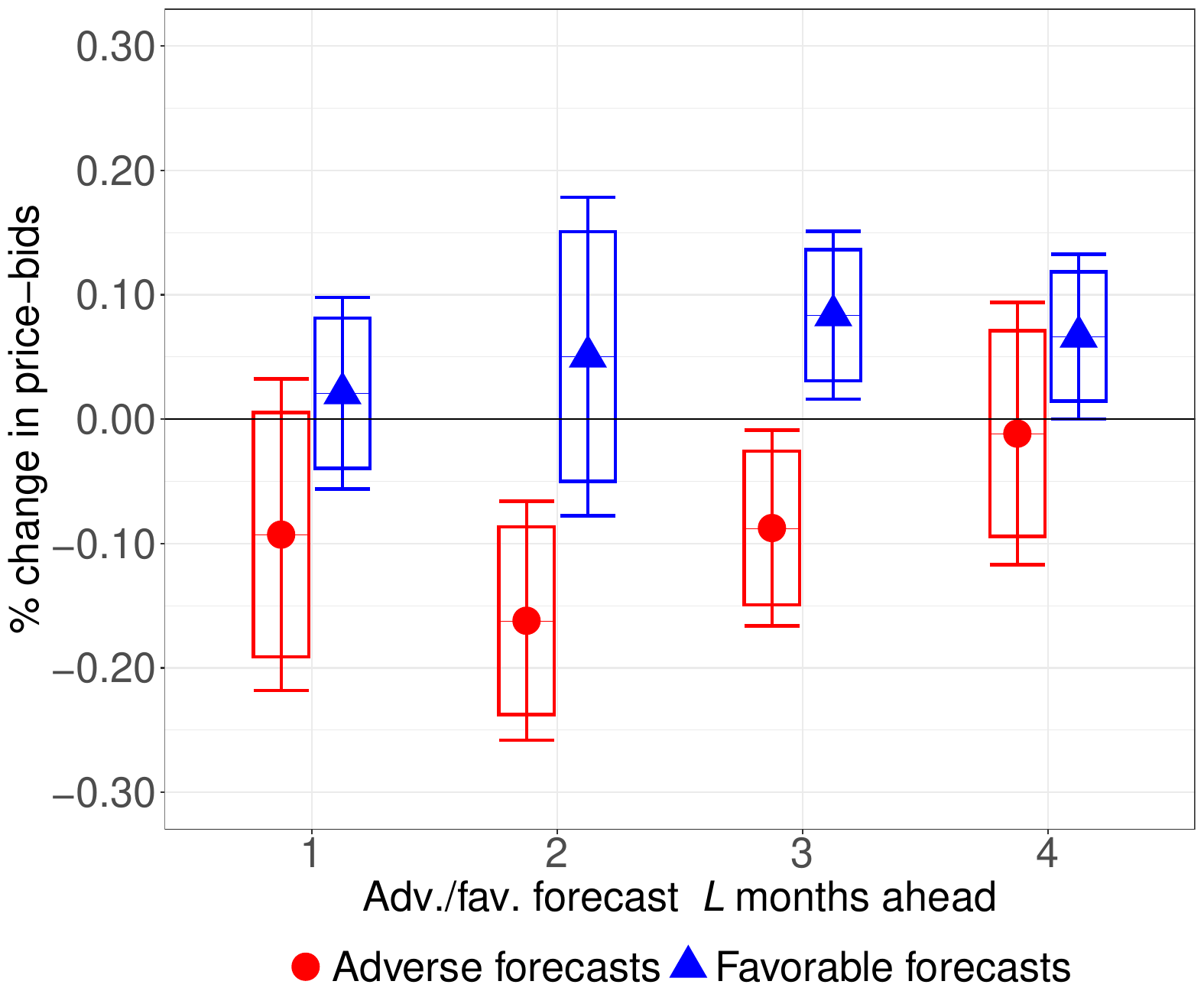}
        \captionsetup{width=0.9\linewidth, justification=centering}
        \caption{Price bids}
    \end{subfigure}\hfill
    \begin{subfigure}{0.48\textwidth}
        \centering
        \includegraphics[width=\textwidth]{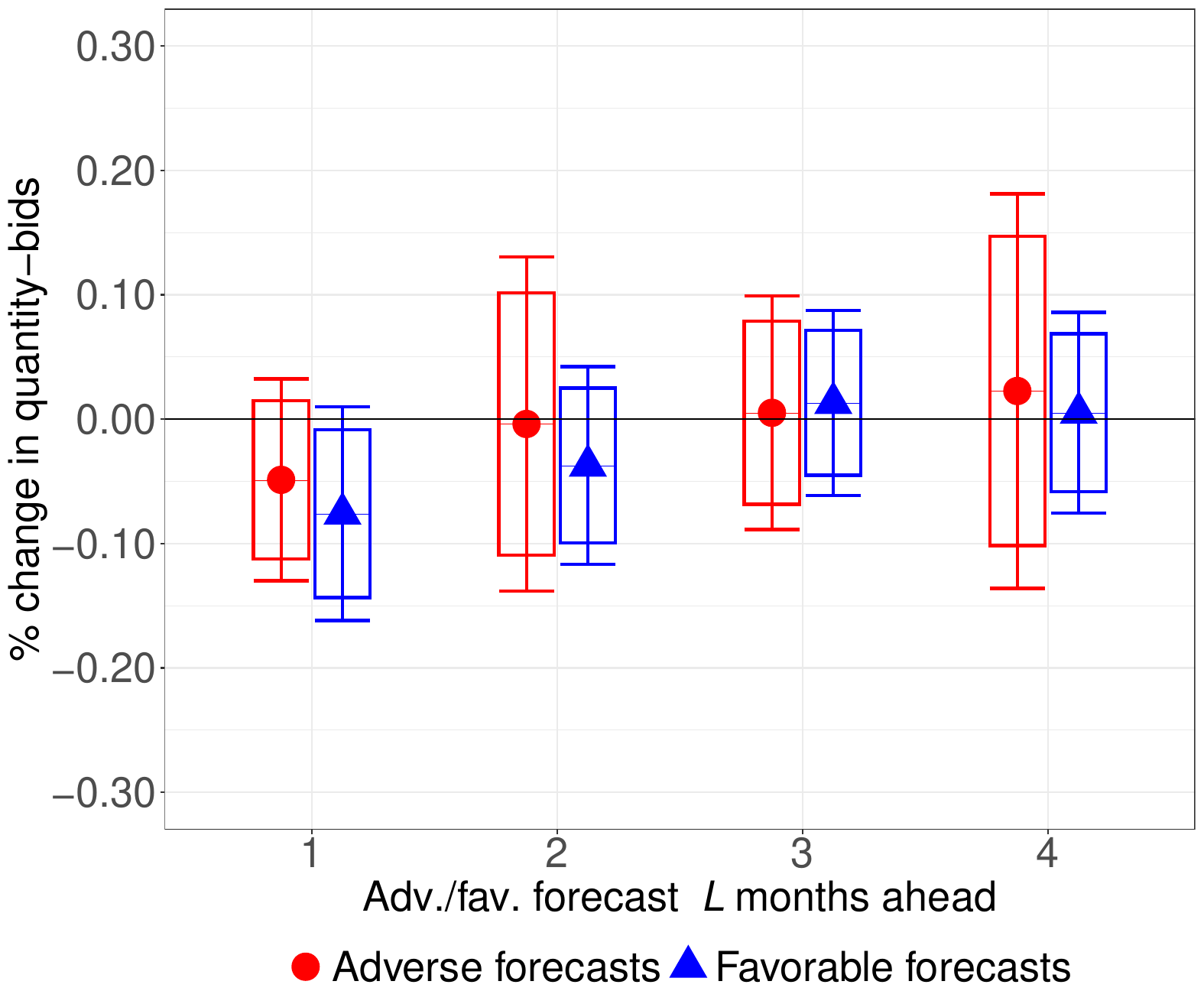}
        \captionsetup{width=0.9\linewidth, justification=centering}
        \caption{Quantity bids}
    \end{subfigure}
    \begin{minipage}{1 \textwidth}
        {\footnotesize Notes: Estimates of $\{\beta^{low}_\ell\}$ (red) and $\{\beta^{high}_\ell\}$ (blue) from (\ref{eq:as_inflow}) at one to four months ahead. Error bars (boxes) report 95\% (90\%) confidence intervals. FEs: unit-by-month, firm-by-year, hour, week-by-year. SEs clustered by unit, month, and year.\par}
    \end{minipage}
\end{figure}

\paragraph{Results.} Figure~\ref{fig:responses} reports the estimates. The two technologies move on different margins: dams adjust quantity bids and sibling thermal units' daily price bids, reflecting hydro's low operating cost and thermal's ramping costs. Under adverse hydro forecasts, dam quantity bids fall and sibling thermal price bids fall: the firm moves thermal capacity down the merit order to replace missing hydro. Under favorable forecasts, dam quantity bids rise and sibling thermal price bids rise: the firm lets cheap hydro serve demand and holds expensive thermal capacity for higher-price states. The thermal price-bid response is significant through $\ell=3$.

These two regimes put upward pressure on prices for different reasons. Under scarcity, the firm replaces cheap hydro with more expensive thermal capacity; lower thermal bids mitigate the shock but do not undo the loss of low-cost supply. Under abundance, the firm has ample cheap hydro and bids sibling thermal units at higher prices, so the standard concentration force dominates.

This bid evidence motivates the model in Section~\ref{s:framework}: capacity reallocations have opposite price effects depending on whether the leader's market power comes from efficiency under scarcity or from capacity under abundance.

\section{Explaining the U-Shape}\label{s:framework}

To explain the U-shape documented in Section~\ref{s:em}, we strip the problem to its essentials: a static competition in supply functions with two diversified firms and constant marginal costs. This minimal setting admits closed-form solutions and isolates the mechanism. When the dominant firm's advantage stems from the \textit{efficiency} of its low-cost technology, a small capacity transfer from competitors \textit{lowers} prices through strategic complementarities. When its advantage stems from sheer \textit{capacity}, the same transfer \textit{raises} them, producing the U-shape purely from oligopolistic competition.

\vspace{1ex}
\noindent\textbf{Setup.} We consider a static market with a single period, no forward contracts or reliability charges, and no continuation value. Each firm $i$ chooses a non-decreasing supply schedule $S_i(p)$ before observing demand $D(\epsilon)$. The market clears at
\begin{equation}
    \label{eq:profit_conceptual1}
   p^*(\epsilon)=\min\left\{p\geq 0 \middle| \sum_{j=1}^N S_j(p)\geq D(\epsilon)\right\}.
\end{equation}

Following \cite{klemperer1989supply}, when firm $i$'s supply $S_i(p)$ clears the residual demand $D^R_i(p,\epsilon)$ at price $p$, its profit from the spot market as in \eqref{eq:profit} equals
\begin{equation}\label{eq:profits.mkt.clearing}
    \pi_i(p) = p\cdot D^R_i(p,\epsilon) - C_i ( D^R_i(p,\epsilon)),
\end{equation}
where $D^R_i(p,\epsilon)=\max\{0, D(\varepsilon)-\sum_{j\neq i} S_j(p)\}$, and $C_i(q)$ is the cost of producing $q$. As $\epsilon$ varies, the profit-maximizing quantity-price pairs trace out an ex-post optimal supply schedule, i.e., the same schedule that an expected-profit-maximizing firm would bid before demand realizes.

\begin{definition}{\textbf{Supply Function Equilibrium (SFE).} }\label{apndx:def:SFE}
An SFE is an $N$-tuple $(S_i)_{i=1}^N$ such that, given $S_j$ for all $j\neq i$, firm $i$'s supply $S_i$ maximizes its ex post profit (\ref{eq:profits.mkt.clearing}) at every $D(\epsilon)$.
\end{definition}

A feature of SFE over Cournot or Bertrand is that each firm maximizes profit for \textit{every} demand realization, not just in expectation. Ex-post optimality implies ex-ante optimality, with no need to specify beliefs about $\epsilon$, simplifying our analysis.

Whenever supply functions are differentiable, the FOC of (\ref{eq:profits.mkt.clearing}) yields
\begin{equation}\label{eq:foc_main}
(p - C_i') \cdot S_{-i}'(p) = S_i(p) \qquad \forall\, p,
\end{equation}
where $S_{-i}'(p) \equiv \sum_{j \neq i} S_j'(p)$. This has a useful reformulation:

\begin{proposition}\textbf{Markup.} \label{prop:markup} Firm $i$'s markup satisfies
\begin{equation}\label{eq:market.power}
    \frac{p - C_i^\prime(S_i(p))}{p} = \frac{s_i(p)}\eta\,\,
    {\left(1 - \frac{S^\prime_{i}(p)}{ {D^R_i}^\prime(p)}\right)},
\end{equation}
where $s_i(p)$ is market share and $\eta$ is the price elasticity of demand.
\end{proposition}
\begin{proof}
Rewrite (\ref{eq:foc_main}) using $D^R_i = D(\epsilon) - S_{-i}(p)$ and $\eta \equiv \frac{p}{\sum_l S_l(p)} \frac{d}{dp}\sum_l S_l(p)$, which holds by market clearing.
\end{proof}

The factor $(1 - S'_i/{D^R_i}')$ distinguishes the SFE markup from the standard Lerner index. In differentiated Bertrand, supply is horizontal ($S'_i = 0$), the factor equals one, and the markup reduces to the textbook inverse-elasticity rule. In Cournot, supply is vertical ($S'_i \to \infty$) and the formula is degenerate: the firm commits to a fixed quantity for all prices, so the supply-slope channel shuts down entirely. In an SFE, supply can often have finite positive slope, and the factor lies strictly between zero and one: the more aggressively the firm bids (larger $S'_i$), the more its markup is compressed.

The markup depends on competitors' supply slopes, not just their market shares. Totally differentiating (\ref{eq:foc_main}) with respect to $p$ gives
\begin{equation}\label{eq:strat_comp}
S_i'(p) = S_{-i}'(p) + (p - C_i') \cdot S_{-i}''(p).
\end{equation}
With constant marginal costs, the equation decomposes into two channels. First, a more responsive rival supply ($S_{-i}'$ larger) directly raises firm $i$'s own slope: when rivals supply more at each price, so does firm $i$. Second, a more convex rival supply ($S_{-i}'' > 0$) also raises $S_i'$, as rivals become less responsive at higher prices, making firm $i$ bid more aggressively. Together, these \textit{strategic complementarities in supply slopes} create an amplification mechanism central to the U-shape. As we show next, a capacity transfer that leads one firm to expand supply triggers a matching expansion by the rival, and a transfer that leads one firm to restrict supply triggers a matching restriction. The amplification works in both directions, producing the nonmonotone price response.

\subsection{Analytical Characterization of the U-Shape}\label{s:duopoly}

In this section, we solve analytically for the equilibrium of a duopoly with heterogeneous technologies. The market features three constant-marginal-cost technologies: $l$ for low-cost technology, $h$ for high-cost technology, and $f$ for fringe-cost technology, and their costs satisfy $0 \leq c^l < c^h < c^f < \infty$. A tech-portfolio $K_i = (K_i^l, K_i^h, K_i^f) \in \mathbb{R}_+^3$ summarizes firm $i$'s capacity in each. For $\tau\in\{l,h,f\}$, $S_i^\tau$ denotes $i$'s supply function in technology $\tau$. Total production costs $C_i = \sum_\tau c^\tau S_i^\tau(p)$ when $ S_i^\tau(p) \in [0, K_i^\tau]$ for all $\tau$, and $C_i=\infty$ otherwise. 

Two strategic firms compete. Firm~1 holds low- and high-cost capacity, $K_1 = (K_1^l,  \delta, 0)$, and Firm~2 holds only high-cost capacity, $K_2 = (0, K_2^h - \delta, 0)$, with $K_1^l > K_2^h > 0$: the largest firm is also the most efficient, as in Colombia where the dominant firms combine large, cheap dams with smaller, expensive thermal units. The parameter $\delta \geq 0$ captures a transfer of high-cost capacity from Firm~2 to Firm~1. If $ K_1^h=0$, both firms specialize in a single technology when $\delta = 0$; Firm~1 is diversified when $\delta > 0$. Thus, the transfer raises concentration and diversifies Firm~1 while holding total industry capacity fixed. 

There are infinitely many fringe firms and they only supply with fringe technology. These firms are small and competitive, and they are assumed to dispatch all capacity at price $c^f$. Hence all demands are cleared at this price. Strategic firms are dispatched before the fringe.\footnote{Equivalently, strategic firms can undercut fringe firms by an infinitesimal amount.}

We solve and characterize the equilibrium in Appendix \ref{apndx:theory}, here we summarize the main features of the analysis. We also propose numerical examples.

\begin{figure}[ht!]
    \caption{Abundance: equilibrium before and after capacity transfer}
    \label{fig:equilibrium_abundance}\label{fig:equilibrium}\label{fig:cases}
    \centering
    \begin{subfigure}{0.48\textwidth}
        \centering
        \includegraphics[width=\textwidth]{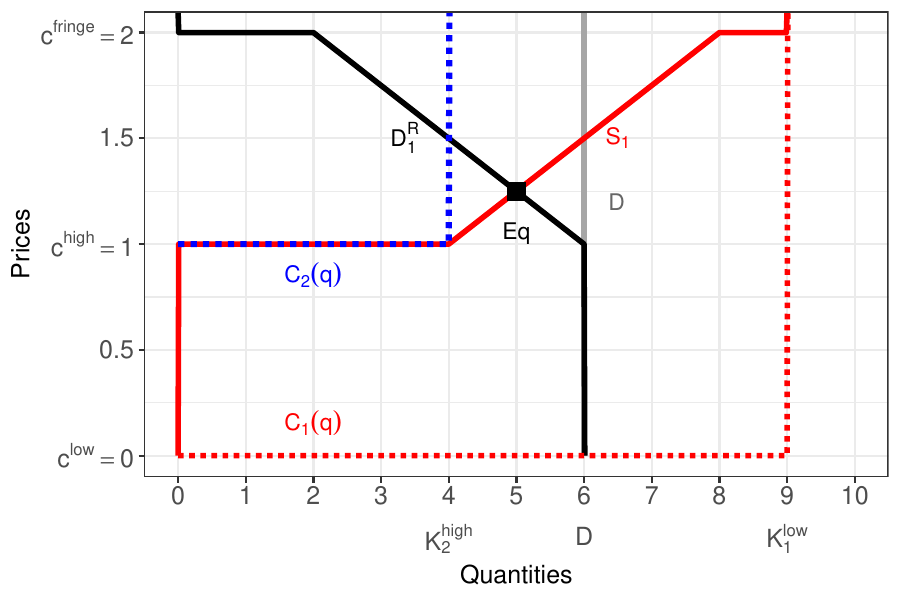}
        \captionsetup{width=0.9\linewidth, justification=centering}
        \caption{Baseline: $K_1 = (9, 0, 0)$, $K_2 = (0, 4, 0)$}
    \end{subfigure}\hfill
    \begin{subfigure}{0.48\textwidth}
        \centering
        \includegraphics[width=\textwidth]{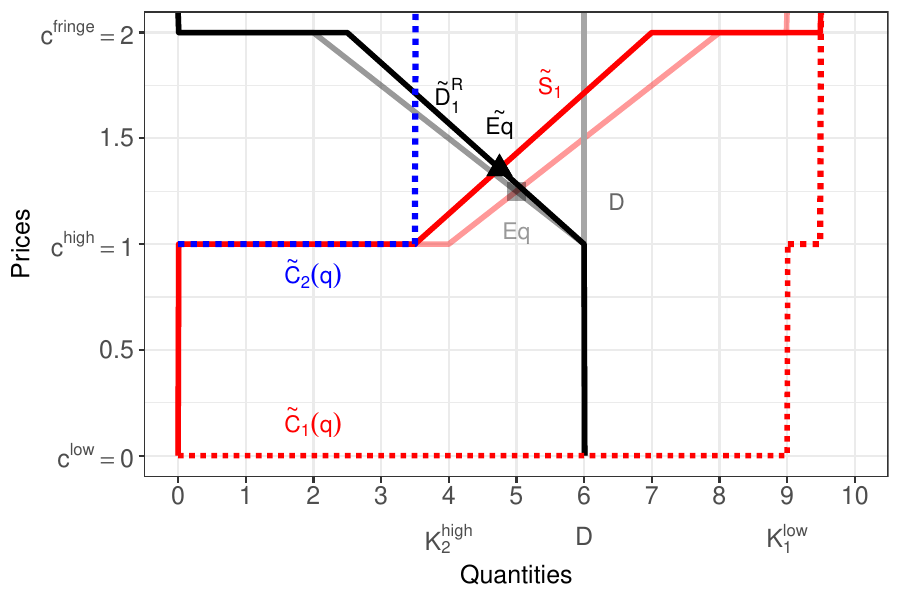}
        \captionsetup{width=0.9\linewidth, justification=centering}
        \caption{After transfer: $\tilde{K}_1 = (9, 0.5, 0)$, $\tilde{K}_2 = (0, 3.5, 0)$}
    \end{subfigure}
    \begin{minipage}{1\textwidth}
        {\footnotesize Notes: Equilibrium from Firm~1's perspective under abundance ($K_1^l = 9$). Red solid line: $S_1(p)$. Red step function: $C_1(q)$. Blue dashed step function: $C_2(q)$. Black solid line: $D_1^R(p)$ at $D = 6$. Panel~(a): baseline with $Eq$ (square). Panel~(b): post-transfer with $\tilde{Eq}$ (triangle); faded lines and square reproduce the baseline for comparison. Gray vertical line: $D = 6$. Parameters: $c^l = 0$, $c^h = 1$, $c^f = 2$.}
    \end{minipage}
\end{figure}

\begin{figure}[ht!]
    \caption{Scarcity: equilibrium before and after capacity transfer}
    \label{fig:equilibrium_scarcity}\label{fig:transfer}
    \centering
    \begin{subfigure}{0.48\textwidth}
        \centering
        \includegraphics[width=\textwidth]{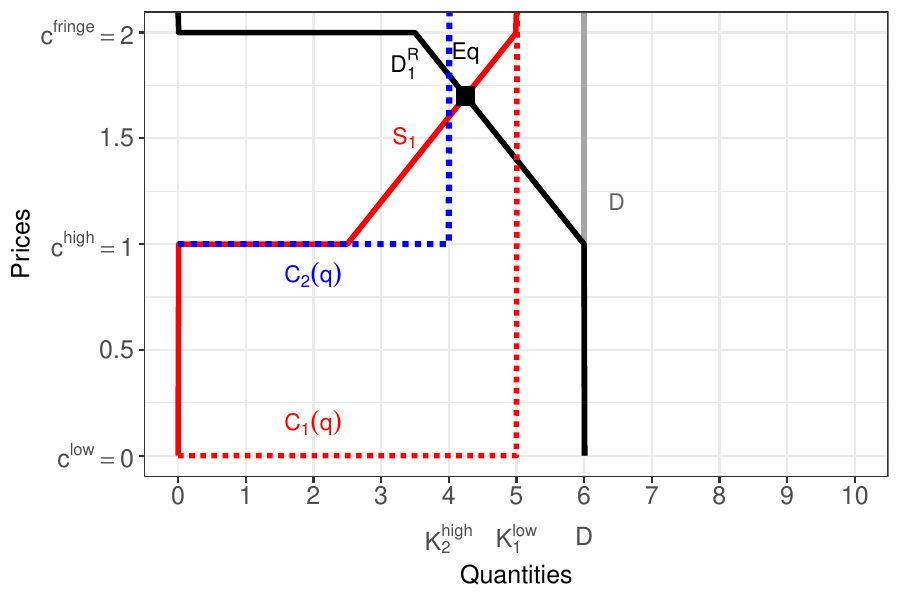}
        \captionsetup{width=0.9\linewidth, justification=centering}
        \caption{Baseline: $K_1 = (5, 0, 0)$, $K_2 = (0, 4, 0)$}
    \end{subfigure}\hfill
    \begin{subfigure}{0.48\textwidth}
        \centering
        \includegraphics[width=\textwidth]{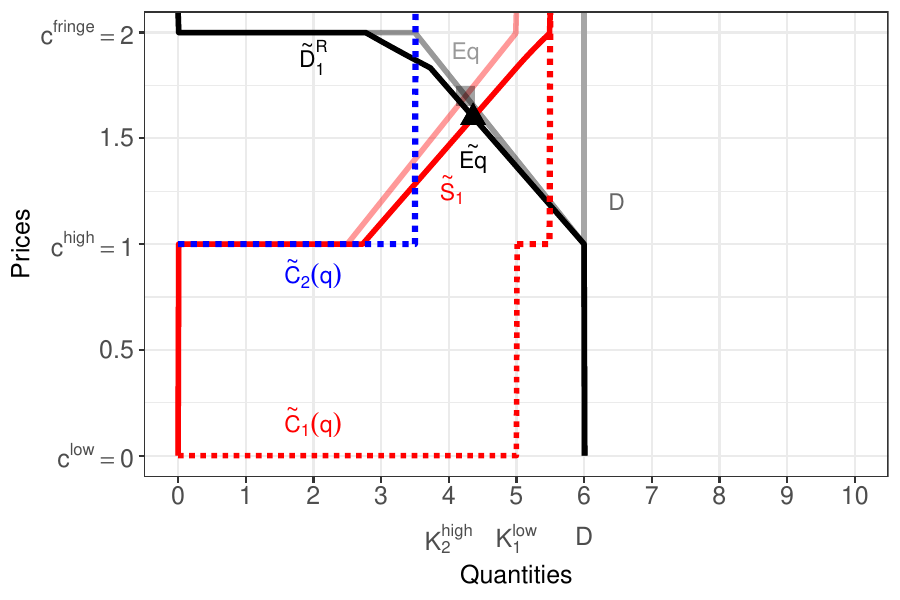}
        \captionsetup{width=0.9\linewidth, justification=centering}
        \caption{After transfer: $\tilde{K}_1 = (5, 0.5, 0)$, $\tilde{K}_2 = (0, 3.5, 0)$}
    \end{subfigure}
    \begin{minipage}{1\textwidth}
        {\footnotesize Notes: Equilibrium from Firm~1's perspective under scarcity ($K_1^l = 5$). Red solid line: $S_1(p)$. Red step function: $C_1(q)$. Blue dashed step function: $C_2(q)$. Black solid line: $D_1^R(p)$ at $D = 6$. Panel~(a): baseline with $Eq$ (square). Panel~(b): post-transfer with $\tilde{Eq}$ (triangle); faded lines and square reproduce the baseline for comparison. Gray vertical line: $D = 6$. Parameters: $c^l = 0$, $c^h = 1$, $c^f = 2$.}
    \end{minipage}
\end{figure}

\paragraph{Equilibrium and comparative statics.} In equilibrium, Firm~2 does not produce for $p < c^h$ and Firm~1 best-responds with zero output. For $p \in [c^h, c^f)$, the FOC (\ref{eq:foc_main}) yields a system of two ODEs whose unique non-negative, non-decreasing solution, when Firm~1 has large enough low-cost capacity, satisfies
\begin{equation}\label{eq:system1_main}
	S_1(p) = \kappa_1 \cdot (p - c^l), \qquad S_2(p) = \kappa_1 \cdot (p - c^h).
\end{equation}
Geometrically (Panels~(a) of Figures~\ref{fig:cases} and~\ref{fig:transfer}), neither firm produces below $c^h$: Firm~2 has no profitable supply there, and Firm~1, with no rival to undercut, has no incentive to. At the entry price $c^h$, Firm~1 enters with a head-start of $\kappa_1(c^h - c^l)$ units, the slope times its cost-advantage interval; Firm~2 enters at zero. Above $c^h$ both schedules rise in parallel.

The common slope $\kappa_1$ emerges from strategic complementarity in (\ref{eq:foc_main}): a steeper rival schedule serves more demand at high prices, pushing the focal firm to match aggressiveness or lose market share. To pin its value, note that a firm with idle capacity at $c^f$ could profitably deviate by steepening just below; at least one firm must therefore exhaust its capacity as $p \to c^f$.

Which firm exhausts first? Suppose Firm~2 exhausts first. Then $S_2(c^f) = K_2^h$, which pins down the common slope: $\kappa_1 = K_2^h/(c^f - c^h)$. Firm~1 has been supplying with the same slope since the price $p = c^l$, so under this hypothesis its output as $p\to c^f$ would be $\kappa_1 \cdot (c^f - c^l) = \Gamma \cdot K_2^h$. We call this product Firm~2's \textit{effective capacity},
\[
	K_2^{\mathrm{eff}} \equiv \Gamma \cdot K_2^h, \qquad
	\Gamma \equiv \frac{c^f - c^l}{c^f - c^h}.
\]

If $K_1^l > K_2^{\mathrm{eff}}$, this output is feasible, Firm~2 indeed binds first, and (\ref{eq:system1_main}) is the equilibrium. We call this the case of \textit{abundance} of Firm~1's low-cost capacity (see numerical Example~\ref{ex:abundance} below). If $K_2^{\mathrm{eff}} > K_1^l$, the implied output exceeds $K_1^l$: Firm~1 must bind first instead, and (\ref{eq:system1_main}) is no longer the equilibrium. We call this case \textit{scarcity} (Example~\ref{ex:scarcity}).

Under scarcity, Firm~1 must exhaust its low-cost capacity at a switching price $\hat{p} < c^f$ and transition to high-cost technology. The ODE solutions on $[\hat{p}, c^f)$ satisfy
\begin{equation}\label{eq:system2_main}
	S_1(p) = \kappa_2(p - c^h) + \frac{\kappa_3}{p - c^h}, \qquad S_2(p) = \kappa_2(p - c^h) - \frac{\kappa_3}{p - c^h}.
\end{equation}
The two schedules are no longer parallel: Firm~1 expands output less aggressively with price (conserving its remaining high-cost capacity), while Firm~2, with more slack, expands more aggressively. The coefficients $\kappa_2$, $\kappa_3$, and $\hat{p}$ are pinned down from continuity and differentiability at the switching price (Appendix~\ref{apndx:proof.2}).

The next proposition establishes existence and uniqueness of an equilibrium for all $\delta \geq 0$, and performs comparative statics with respect to $\delta$.

\begin{proposition}\label{prop:comp} \textbf{Existence, uniqueness, and comparative statics.}
Let $K_1=(K_1^l,K_1^h+\delta,0)$, $K_2=(0,K_2^h-\delta,0)$, $K_j=(0,0,K_j^f)$ for fringe $j \geq 3$, with $0 \leq c^l < c^h < c^f < \infty$, $K_2^h \geq \delta$, and $K_2^{\mathrm{eff}}\equiv \frac{c^f-c^l}{c^f-c^h}K_2^h$.
\begin{enumerate}
    \item There exists a unique SFE for every $\delta \geq 0$.
    \item Let $S^\delta(p) = \sum_i S_i^\delta(p)$ denote equilibrium market supply. For $\delta_1 > \delta_2 > 0$:
\begin{itemize}
    \item[a.] $S^{\delta_1} < S^{\delta_2}$ when $K_1^l > K_2^{\mathrm{eff}}$ \quad (abundance: prices rise);
    \item[b.] $S^{\delta_1} > S^{\delta_2}$ when $K_1^l < K_2^{\mathrm{eff}}$ and $K_1^h+\delta_1$ small enough \quad (scarcity: prices fall).
\end{itemize}
\end{enumerate}
\end{proposition}
\begin{proof}
See Appendix~\ref{apndx:proof.2}.
\end{proof}

\textit{Part~1} extends \cite{klemperer1989supply} to asymmetric step-function marginal costs. Similar to their uniqueness results, we use full support of $D(\epsilon)$ to rule out supply pairs that are only local best responses. The decreasing residual demand for strategic firms at $p=c^f$ fixes a boundary condition for the ODE system that uniquely pins down the equilibrium pair. The step-function cost structure then creates multiple equilibrium regimes connected by continuity at the switching price $\hat{p}$, and the closed-form characterization across regimes is what enables the comparative statics on capacity transfers, on which we turn next.\footnote{\cite{holmberg2007supply} proves uniqueness for asymmetric capacities with identical constant marginal costs; our model nests this as the single-technology case. \cite{anderson2013existence} proves existence with heterogeneous convex costs but without capacity constraints, so the equilibrium ODE has no terminal boundary condition and uniqueness is not obtained.}

\textit{Part 2} characterizes the comparative statics. Under abundance ($K_1^l > K_2^{\mathrm{eff}}$, \textit{Part 2.a}), Firm~2 exhausts first at $c^f$. A transfer shrinks the binding firm, lowering $\kappa_1$. By strategic complementarity, both firms bid less aggressively, and prices rise. Thus, our model replicates the standard effect of concentration.

Under scarcity (\textit{Part~2.b}), prices drop after the reallocation.\footnote{In Proposition \ref{prop:comp} (\textit{Part 2.b}), the bound on $K_1^h + \delta_1$ does not admit a closed-form expression because it depends on $\tilde{S}_1(c^f)$, which is solved numerically from the ODE system. The condition requires $K_1^h < \underline{S}_1(c^f) - K_1^l$, where $\underline{S}_1(c^f)$ is the output Firm~1 would need to produce at $c^f$ if Firm~2 were the capacity-constrained firm. Firm~1's high-cost capacity must be small enough that it cannot sustain that output level, forcing Firm~1 (not Firm~2) to be the firm that exhausts capacity at $c^f$. This is the regime where transfers lower prices.}  The intuition is a commitment problem of efficiency: the efficient firm has too little capacity relative to its cost advantage. Because Firm~1 produces from $c^l$ while Firm~2 produces only from $c^h$, Firm~1 commits more capacity at every price and depletes it faster; with $K_1^l$ barely above $K_2^h$, it runs out at high prices while its high-cost rival still holds idle capacity. The same cost advantage that lets Firm~1 undercut at the bottom of the schedule thus leaves it constrained at the top. A small transfer of high-cost capacity moves units from where they sit idle to where they are needed, freeing Firm~1 to bid aggressively across the whole price range; by strategic complementarity, Firm~2 then expands to defend its share, both firms produce more, and prices fall. 

\paragraph{Numerical examples.} We illustrate with $c^l = 0$, $c^h = 1$, $c^f = 2$, $K_1^h=0$, $K_2^h = 4$, $D = 6$, and set $\delta =0$ before the transfer and $\delta=0.5$ after it. Here, $\Gamma = 2$, so $K_2^{\mathrm{eff}}=8$ and abundance requires $K_1^l > K_2^{\mathrm{eff}}$. Example \ref{ex:abundance} sets $K_1^l = 9$ (abundance); Example \ref{ex:scarcity} sets $K_1^l = 5$ (scarcity). After-transfer variables get a superscript $\tilde{}$.\footnote{Figures \ref{fig:cases} and \ref{fig:transfer} plot the equilibria in Example  \ref{ex:abundance}  and \ref{ex:scarcity}, respectively, from the perspective of Firm~1. Appendix Figure~\ref{apndx:fig:cases_competitors} shows Firm~2's equilibrium supply schedules for both examples.}

\begin{mdframed}[linewidth=0pt, backgroundcolor=gray!5, innertopmargin=6pt, innerbottommargin=6pt]
\refstepcounter{exbox}\label{ex:abundance}
\noindent\textbf{Example \theexbox: Abundance (\textit{Part 2.a}).} Figure~\ref{fig:equilibrium_abundance} shows the equilibrium before and after transferring $\delta = 0.5$ units from Firm~2 to Firm~1. In the figure, dashed step functions show marginal costs and capacities; solid lines show supply and residual demand. Red denotes Firm~1, blue Firm~2, and black Firm~1's residual demand. The equilibrium is drawn from Firm~1's viewpoint.

\textit{Baseline} (Panel~a). Before the transfer, $\delta = 0$. Firm~1 has far more capacity ($K_1^l = 9$) than Firm~2 ($K_2^h = 4$). Both firms share slope $\kappa_1$, but Firm~2 produces over a narrower price range ($[c^h, c^f) = [1, 2)$ versus $[c^l, c^f) = [0, 2)$ for Firm~1). With less room to accumulate output, Firm~2 hits its capacity ceiling first: $S_2(c^f) = \kappa_1(c^f - c^h) = K_2^h = 4$, giving $\kappa_1 = 4$. As a result:
\[
S_1(p) = 4\,(p - c^l), \qquad S_2(p) = 4\,(p - c^h), \qquad \text{for } p \in [c^h, c^f).
\]
Firm~1 has idle capacity at $c^f$: $S_1(c^f) = 8 < 9 = K_1^l$. Hence, it holds one unit in reserve for demand realizations above $D = 8$.

\textit{After transfer} (Panel~b). We  now set $\delta = 0.5$. The transfer shrinks the binding firm: $\tilde{K}_2^h = 3.5$. Since $\kappa_1$ is pinned by Firm~2's exhaustion condition, reducing $\tilde{K}_2^h$ directly lowers $\kappa_1$ to $3.5$:
\[
\tilde{S}_1(p) = 3.5\,(p - c^l), \qquad \tilde{S}_2(p) = 3.5\,(p - c^h), \qquad \text{for } p \in [c^h, c^f).
\]
Both firms bid less aggressively at every price. Firm~1 does not deploy the transferred capacity at the market clearing price; it simply holds a larger reserve for extreme demand states. The price increase comes entirely from weakening Firm~2: a smaller rival means less competitive pressure. This is the standard concentration effect.
\end{mdframed}

\begin{mdframed}[linewidth=0pt, backgroundcolor=gray!5, innertopmargin=6pt, innerbottommargin=6pt]
\refstepcounter{exbox}\label{ex:scarcity}
\noindent\textbf{Example \theexbox: Scarcity (\textit{Part 2.b}).} Figure~\ref{fig:equilibrium_scarcity} shows the equilibrium before and after transferring $\delta = 0.5$.

\textit{Baseline} (Panel~a). With $K_1^l = 5 < K_2^{\mathrm{eff}} = 8$, Firm~1 cannot sustain the required output at $c^f$ and binds first: $S_1(c^f) = \kappa_1(c^f - c^l) = K_1^l = 5$, giving $\kappa_1 = 2.5$:
\[
S_1(p) = 2.5\,(p - c^l), \qquad S_2(p) = 2.5\,(p - c^h), \qquad \text{for } p \in [c^h, c^f).
\]
Firm~2 has idle capacity: $S_2(c^f) = \kappa_1(c^f - c^h) = 2.5 < 4 = K_2^h$. Compared to abundance (Example \ref{ex:abundance}), both firms bid less aggressively ($\kappa_1 = 2.5$ versus $4$) because less total capacity is available in the market.

\textit{After transfer} (Panel~b). After the transfer, Firm 1 is diversified with $\tilde{K}_1 = (5, 0.5, 0)$, while Firm~2 has $\tilde{K}_2 = (0, 3.5, 0)$. Yet, Firm~1 remains the binding firm: it still exhausts all capacity at $c^f$. The $\delta = 0.5$ units of high-cost capacity serve as a buffer at high prices, freeing Firm~1 to bid more aggressively throughout the price range. With this reserve, Firm~1 optimally depletes its low-cost capacity at the switching price $\hat{p}$, where $S_1(\hat{p}) = \kappa_1(\hat{p} - c^l) = K_1^l$ from (\ref{eq:system1_main}). For $p \in [\hat{p}, c^f)$, supply follows the second regime (\ref{eq:system2_main}) with coefficients pinned by continuity at $\hat{p}$ and the boundary condition $S_1(c^f) = K_1^l + \delta = 5.5$. Solving yields $\kappa_1 = \frac{30}{11} \approx 2.73$, $\hat{p} = \frac{11}{6}$, $\kappa_2 = \frac{48}{11}$, and $\kappa_3 = \frac{25}{22}$. The slope $\kappa_1$ rises from $2.5$ to $2.73$: Firm~1 supplies more at every price in $[c^h, \hat{p})$.

Because Firm~1 expands supply, strategic complementarity (\ref{eq:strat_comp}) implies Firm~2 also produces more. Both firms expand at every price, and the market-clearing price falls despite higher concentration.\footnote{Appendix~\ref{apndx:verification} verifies these are equilibrium strategies, one can check that each firm's supply function traces out its ex-post optimal quantity at every demand realization.}
\end{mdframed}

In both examples, Firm~1 uses only its low-cost technology at the market clearing price, so its average cost is unchanged by the transfer. The price changes are entirely strategic, reflecting market power rather than returns to scale or synergies. Scarcity is the efficiency regime: Firm~1's market power comes from its low cost, but its low-cost capacity is small relative to Firm~2's effective capacity. Abundance is the capacity regime: Firm~1 is no longer constrained, so transfers mainly reduce the capacity of its rival.

\begin{figure}[htbp]
    \centering
     \caption{Numerical illustration of the U-shape mechanism with transfer share $\kappa$    \label{fig:u_shape_n2n3}}
    \includegraphics[width=0.72\textwidth]{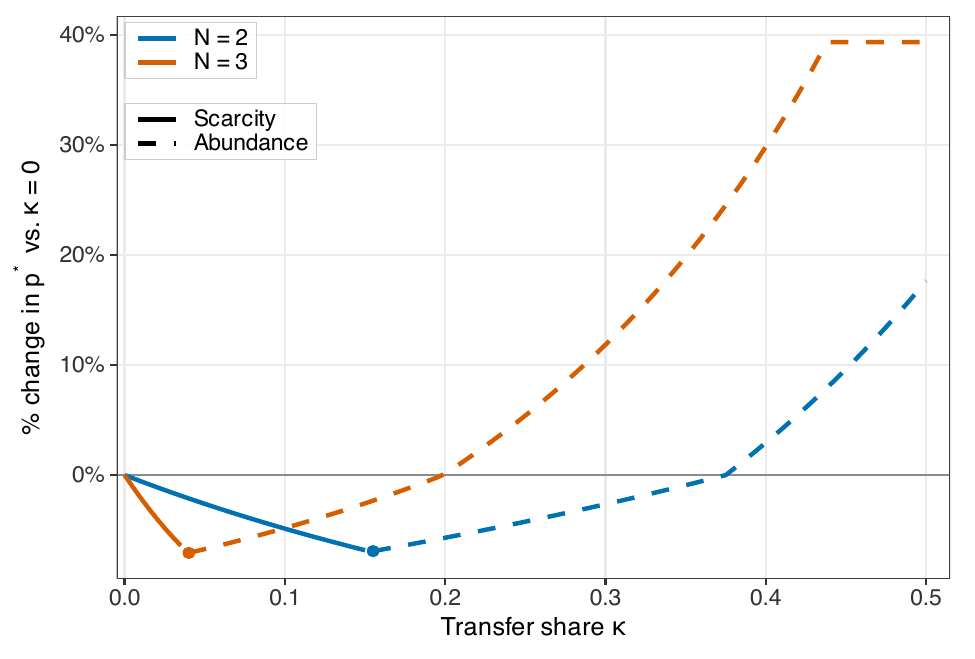}
 \begin{minipage}{1\textwidth}
 	{\footnotesize Notes: Percentage change in the market-clearing price relative to $\kappa=0$. For $N=2$, the calibration is Example~\ref{ex:scarcity}: $(c^l,c^h,c^f)=(0,1,2)$, $K_1^l=5$, $K_2^h=4$, and $D=6$. For $N=3$, Firm~1 has $K_1^l=5$, the two strategic rivals have $K_i^h=4$, $(c^l,c^h,c^f)=(0.72,1,2)$, and $D=8$. Solid (dashed) lines indicate the scarcity (abundance) case.}
 \end{minipage}
\end{figure}

For large transfers, the scarcity mechanism breaks down. As $\delta$ grows, Firm~1 accumulates enough high-cost capacity that it is no longer constrained at $c^f$. Beyond this point, Firm~2 becomes the binding firm, and we are back to the abundance case: further transfers weaken Firm~2, both firms restrict supply, and prices rise. In the limit where Firm~1 acquires all of Firm~2's capacity, it prices at $c^f$, just undercutting the fringe. Small transfers lower prices under scarcity (solid blue line in Figure~\ref{fig:u_shape_n2n3}); large ones raise them (dashed blue line). The result is a U-shape. This non-monotonicity is not confined to the two-strategic-firm case: the same pattern appears numerically with three strategic firms (orange line). Appendix~\ref{apndx:u_shape_n2n3} describes the numerical solution. Extending the characterization to four or more strategic firms requires considering more cases: the equilibrium can have several switching prices, one per firm as it exhausts a technology, and which firm exhausts at each depends on the cost and capacity configuration, so each ordering induces a distinct ODE system. We leave this for future work.

\subsection{Capacity and Efficiency Beyond Supply Functions}\label{s:maggi_extension}
Our insight is not specific to supply functions. We illustrate this by introducing technology-specific capacity transfers into an asymmetric version of the price subgame in \citet{maggi1996strategic}, an influential differentiated-product oligopoly model with capacity constraints. Maggi's full model has capacity choice followed by price competition; we take capacities as given and study the price subgame. Capacity is soft: a firm produces at low marginal cost up to capacity and at a higher marginal cost beyond it. For any pair of active marginal costs, differentiated-Bertrand prices are increasing in both firms' active costs. A transfer therefore affects prices by changing which cost tier is active. In Appendix~\ref{apndx:maggi_extension}, the efficient recipient initially produces beyond its low-cost capacity, so its active cost is high. A small transfer lets the recipient replace production above its original low-cost capacity with the transferred block, whose marginal cost is below the recipient's overflow cost, while the rival's active marginal cost is unchanged. Prices fall. A large transfer instead removes enough of the rival's low-cost capacity, moving the rival's production to its higher marginal-cost tier. Prices rise above the baseline. Our result therefore survives in a setting with soft rather than hard capacity and price rather than schedule competition, though the channel differs: it runs through active marginal costs rather than strategic complementarity in supply slopes.\footnote{Appendix~\ref{apndx:maggi_extension} also discusses when this second-stage pattern can be embedded in a full capacity-price game. Because capacity is soft in this setting, capacity shares describe low-cost production ranges rather than hard production limits; this makes concentration changes less directly comparable to those in our SFE model.}

\subsection{Mapping Theory to Data}\label{s:theory_data}
The theoretical mechanism outlined in this section aligns well with the empirical bid responses in Figure~\ref{fig:responses}. Under abundance, Figure~\ref{fig:equilibrium_abundance} shows that the transferred high-cost capacity is placed at $c^f$: Firm~1 holds it in reserve at the top of the schedule, whereas before the transfer Firm~2 had been supplying this capacity at lower prices along the slope $\kappa_1$. Panel~(c) of Figure~\ref{fig:responses} shows the empirical counterpart: thermal generators at firms facing favorable hydro forecasts bid systematically higher than those at firms in normal conditions.

Under scarcity, Figure~\ref{fig:equilibrium_scarcity} shows the mirror image: the constrained leader uses the transferred capacity at lower prices, where Firm~2 had left it idle at $c^f$. Panel~(c) of Figure~\ref{fig:responses} confirms: thermal generators at firms facing adverse hydro forecasts bid at lower prices.

The static model thus reproduces the direction of the firm-level bid responses observed in the data under both regimes. As the theory also explains that the effect of capacity transfers is not monotone in the capacity of the receiving firm, these responses are the micro-level foundations behind the market-level U-shape of Figure~\ref{fig:empiricalU}. The counterfactual simulations in Section~\ref{s:simulations} reconstruct this link by applying the theoretical thermal-transfer experiment to the Colombian market, showing price declines up to 30\% in the least concentrated markets but price increases up to 40\% in the most concentrated ones.

\subsection{The Role of Diversification}\label{s:symmetric}\label{s:discussion}

The U-shape in Section~\ref{s:duopoly} required technology heterogeneity and demand uncertainty: the transfer lowered prices because it relaxed the constrained firm's binding constraint on its most efficient technology, providing a buffer at high prices that freed it to bid aggressively throughout the price range. Without multiple technologies, this mechanism disappears. If all firms share the same technology, the smallest firm is always the most constrained at the highest price, and any transfer simply tightens its binding constraint. But heterogeneity is necessary, not sufficient. If both firms already hold both technologies in equal amounts, a transfer cannot create new options. It simply makes one firm larger and the other smaller, dragging supply slopes down. Prices rise.\footnote{Under Cournot or Bertrand, firms cannot adjust how aggressively they bid across different price levels, so the mechanism has no channel to operate through.}

\begin{proposition}\label{prop:symmetry}
   \textbf{Symmetric firms.} Let $K_1 = (K^l, K^h + \delta, 0)$ and $K_2 = (K^l, K^h - \delta, 0)$ with $\delta \geq 0$ and $K^h \geq \delta$. There exists a unique SFE for every $\delta$. Equilibrium market supply $S^\delta$ is monotonically decreasing in $\delta$.
\end{proposition}
\begin{proof}
See Appendix~\ref{apndx:proof:sym}.
\end{proof}

When firms hold identical portfolios ($\delta = 0$), both bind simultaneously at $c^f$ and supply is at its most aggressive. Any transfer of high-cost capacity ($\delta > 0$) tightens the constraint on the firm that loses it while creating idle capacity for the firm that gains it. Since the supply slope is determined by the tighter constraint, the net effect is always to lower supply and raise prices.

\begin{figure}[ht!]
    \caption{Equilibrium before and after reallocation between symmetric firms}
    \label{fig:realloc_eq}
    \centering
    \begin{subfigure}{0.48\textwidth}
        \centering
        \includegraphics[width=\textwidth]{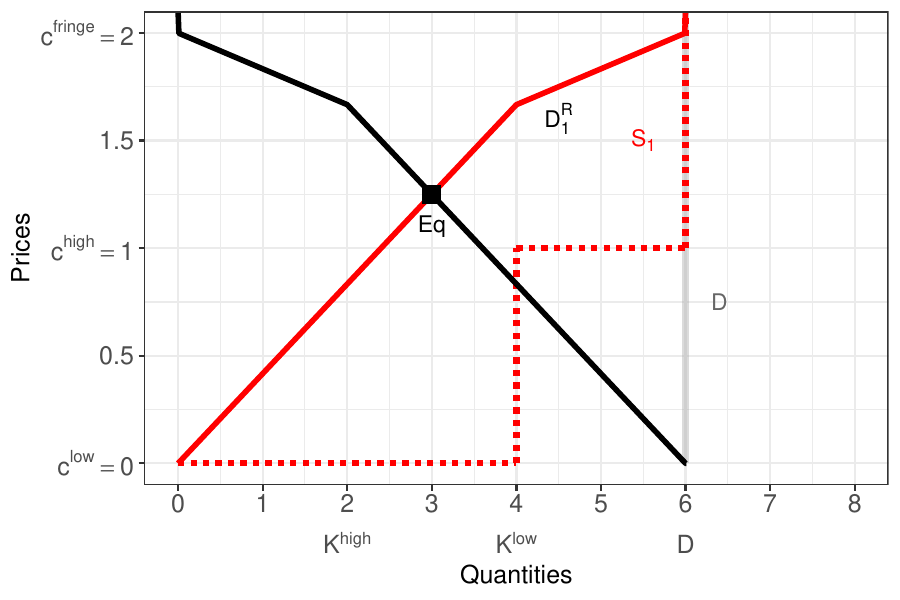}
        \captionsetup{width=0.9\linewidth, justification=centering}
        \caption{Symmetric baseline: $K_1 = K_2 = (4, 2, 0)$}
    \end{subfigure}\hfill
    \begin{subfigure}{0.48\textwidth}
        \centering
        \includegraphics[width=\textwidth]{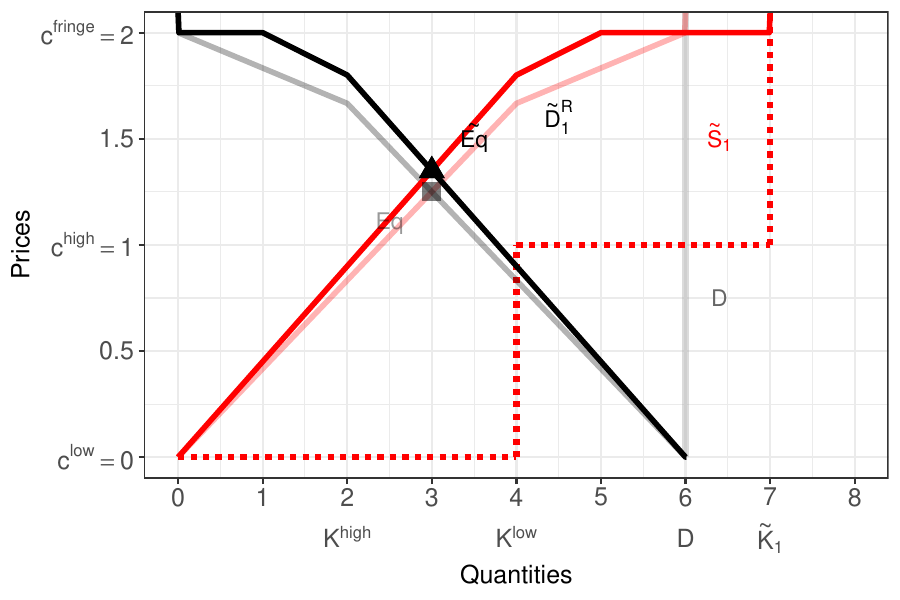}
        \captionsetup{width=0.9\linewidth, justification=centering}
        \caption{After transfer ($\delta = 1$): $\tilde{K}_1 = (4, 3, 0)$, $\tilde{K}_2 = (4, 1, 0)$}
    \end{subfigure}
    \begin{minipage}{1\textwidth}
        {\footnotesize Notes: Equilibrium from Firm~1's perspective. Red solid line: Firm~1's supply $S_1(p)$; red dashed step function: Firm~1's marginal cost $C_1(q)$. Black solid line: residual demand $D_1^R(p)$ at $D = 6$. Panel~(a): symmetric baseline with $Eq$ (square). Panel~(b): post-transfer ($\delta = 1$) with $\tilde{Eq}$ (triangle); faded lines and square reproduce the baseline for comparison. Gray vertical line: $D = 6$. Parameters: $c^l = 0$, $c^h = 1$, $c^f = 2$.}
    \end{minipage}
\end{figure}

\begin{mdframed}[linewidth=0pt, backgroundcolor=gray!5, innertopmargin=6pt, innerbottommargin=6pt]
\refstepcounter{exbox}\label{ex:symmetry_realloc}
\noindent\textbf{Example \theexbox: Reallocation between symmetric firms.} Figure~\ref{fig:realloc_eq} illustrates with $K^l = 4$, $K^h = 2$, and $D = 6$. Since both firms have the same cost structure, we only present Firm~1's marginal cost with a dotted red line. Panel~(a) shows the symmetric baseline where both firms submit identical schedules and exhaust all capacity at $c^f$. Panel~(b) shows the equilibrium after transferring $\delta = 1$: $\tilde{K}_1 = (4, 3, 0)$, $\tilde{K}_2 = (4, 1, 0)$. Firm~2 binds with less capacity, lowering $\kappa_1$. Both firms bid less aggressively, and the market-clearing price rises from $p^*(D=6) = 1.25$ to $\tilde{p}^*(D=6) = 1.35$.
\end{mdframed}

Proposition~\ref{prop:symmetry} shows that when both firms hold both technologies, the most competitive outcome is an equal split. But is equal diversification always best for consumers?

\begin{figure}[ht!]
    \caption{Specialization vs.\ diversification: total supply}
    \label{fig:specialization}
    \centering
    \includegraphics[width=0.6\textwidth]{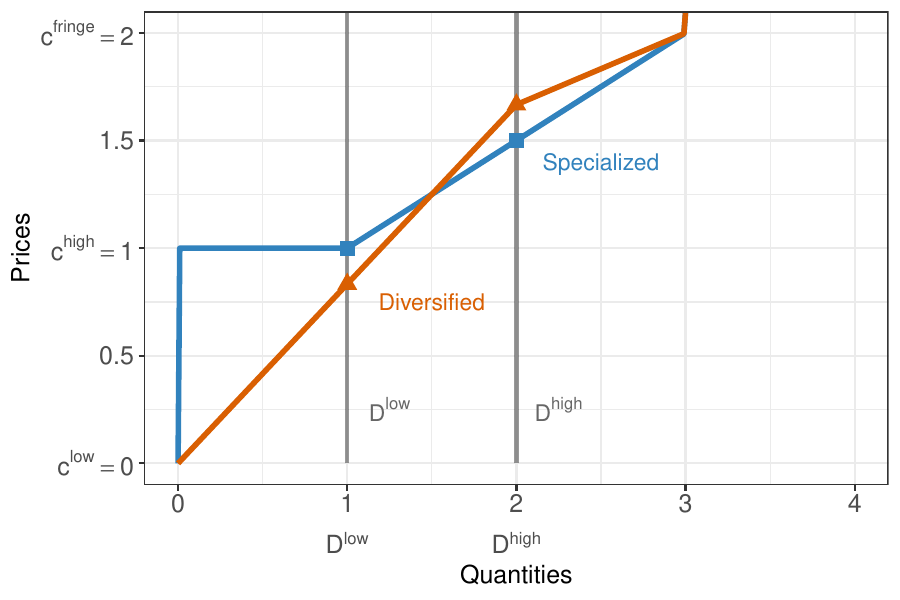}
    \begin{minipage}{1\textwidth}
        {\footnotesize Notes: Total equilibrium supply $S(p) = S_1(p) + S_2(p)$ in two markets with identical total capacity (3 units) and $c^l = 0$, $c^h = 1$, $c^f = 2$. Blue: Market~1 (specialized), $K_1 = (2, 0, 0)$, $K_2 = (0, 1, 0)$. Orange: Market~2 (diversified), $K_1 = K_2 = (1, \tfrac{1}{2}, 0)$. Vertical lines mark two demand realizations. At $D^{low} = 1$, Market~2 clears at a lower price (squares and triangles). At $D^{high} = 2$, Market~1 clears at a lower price.}
    \end{minipage}
\end{figure}

\begin{mdframed}[linewidth=0pt, backgroundcolor=gray!5, innertopmargin=6pt, innerbottommargin=6pt]
\refstepcounter{exbox}\label{ex:specialization}
\noindent\textbf{Example \theexbox: Specialization vs.\ diversification.} Consider two markets with $c^l = 0$, $c^h = 1$, $c^f = 2$, and identical total capacity (3 units). 

\textit{Market~1} (specialized): $K_1 = (2, 0, 0)$, $K_2 = (0, 1, 0)$. Both firms exhaust capacity at $c^f$. Total supply is $S(p) = 2p - 1$ for $p \in [1, 2)$.

\textit{Market~2} (diversified): $K_1 = K_2 = (1, \tfrac{1}{2}, 0)$. Both firms bid identically. Total supply is $S(p) = \tfrac{6}{5}p$ for $p \in [0, \tfrac{5}{3})$ and $S(p) = 3(p-1)$ for $p \in [\tfrac{5}{3}, 2)$.

The comparison depends on demand. Figure~\ref{fig:specialization} plots total supply for both markets: at $D = 1$, Market~2 clears at $p = \tfrac{5}{6}$ while Market~1 clears at $p = 1$: diversified firms compete from $c^l$, lowering prices; at $D = 2$, Market~1 clears at $p = \tfrac{3}{2}$ while Market~2 clears at $p = \tfrac{5}{3}$. The specialized market's concentrated low-cost capacity is more effective at high demand.
\end{mdframed}

Example~\ref{ex:specialization} illustrates a different tradeoff: competition within versus across technologies.\footnote{Closely related, \citet{fabra2024fossil} compare diversified and specialized ownership with the same aggregate capacity in each technology. In their multi-unit auction, diversification weakly lowers prices for any demand, while specialization can preserve productive efficiency when thermal capacity dominates. Example~\ref{ex:specialization} shows a different SFE margin: continuous supply schedules can make the price ranking switch with demand, as concentrated low-cost capacity can clear at a lower price at high demand. The ownership comparison is ancillary here; our main focus is the non-monotone price effect of reallocating capacity between firms.} In the specialized market, Firm~1 (low-cost) competes against Firm~2 (high-cost), but only for $p \geq c^h$. Below $c^h$, Firm~1 faces no rival. In the diversified market, both firms hold low-cost capacity and compete against each other from $c^l$, creating rivalry at every price. At low demand, this within-technology competition drives prices below what the specialized market can achieve. At high demand, however, the specialized market's concentrated low-cost capacity serves more demand cheaply before expensive technology is needed. Whether diversification helps depends on the distribution of demand.

The low-demand logic in Example~\ref{ex:specialization} extends beyond duopoly. If a single firm holds all low-cost capacity, it has no within-technology rival on $[c^l,c^h)$, so its equilibrium schedule is flat below $c^h$ and jumps at $c^h$, the smallest price at which any rival enters. Thus there exists a positive threshold $\bar D_0$ (e.g., $\bar D_0=1$ in Figure~\ref{fig:specialization}), equal to the size of this jump, such that every demand realization $D<\bar D_0$ clears at $c^h$. By contrast, when two or more firms hold low-cost capacity, the discontinuity disappears: competition within the low-cost technology drives equilibrium supply strictly above zero on $(c^l,c^h)$, so sufficiently low demand clears below $c^h$. In general, ranking equilibrium supplies with many technologies and arbitrary capacity distributions across firms requires solving the corresponding $N$-firm SFE, which we leave for future research.

These examples share a common lesson: the relevant question for welfare is not whether the largest firm grows, but whether the most constrained firm gains access to capacity it lacks. Reallocating capacity reshapes the competitive landscape: sometimes tightening rivalry (as in Example~\ref{ex:scarcity}), sometimes loosening it (as in Examples~\ref{ex:abundance} and~\ref{ex:symmetry_realloc}), and sometimes shifting the competitive advantage across demand states (as in the comparison of Examples~\ref{ex:symmetry_realloc} and~\ref{ex:specialization}, where the symmetric-diversified baseline dominates at low demand but the specialized configuration dominates at high demand).

\paragraph{Ownership vs. production reallocation.} The capacity transfers we have studied operate on prices through firms' capacity portfolios: who owns each unit determines how it is bid into the market. A second mechanism, long-term bilateral contracting, operates instead through firms' spot-price exposure \citep{bushnell2008vertical}: a firm that has pre-sold output earns a fixed price on those volumes regardless of how the spot price moves, and so has less incentive to push the spot price up. Contracted firms therefore mark up less, and equilibrium prices are lower at every level of market concentration. This is a genuine pro-competitive effect, and we do not argue against it.

What bilateral contracts and reliability charges do not do is change which firm owns which capacity. They reallocate production across firms without reallocating ownership. Under a bilateral contract, the seller still owns and dispatches its units, but a counterparty (often a retailer who resells the output downstream) collects the spot revenue at a fixed contract price. Colombia's reliability charge \citep{cramton2007colombia,mcrae2024reliability} works on the same margin from a different angle: it forces a firm to dispatch a committed volume at a fixed scarcity price during high-demand periods. In both cases the physical capacity remains with its original owner.

The U-shape arises from how a constrained efficient firm allocates its scarce low-cost capacity across price states, an allocation problem that depends on which firm owns each unit. Reallocating production without reallocating ownership leaves this allocation problem unchanged. Bilateral contracts and reliability charges therefore shift the \textit{level} of the price-concentration curve down; capacity transfers change its \textit{shape}. All three can lower prices, but the U-shape, and in particular the left arm where rising concentration is pro-competitive, cannot be produced without reallocating ownership.

\section{Quantification}\label{s:model}
We now quantify the mechanism in the Colombian market, where four diversified strategic firms interact and the duopoly closed form no longer applies. We estimate firms' marginal costs from observed supply-function first-order conditions and use them to compute counterfactual ex-post optimal supply schedules after thermal-capacity reallocations.

Each firm $i$ operates a set $\mathcal{D}_i$ of dams and a set $\mathcal{T}_i$ of thermal units with marginal costs $c^{\mathsf{D}}$ and $c^{\mathsf{T}}$; firm $i$ is diversified if both sets are non-empty. Let $\mathcal{J}_i \equiv \mathcal{D}_i \cup \mathcal{T}_i$ denote firm $i$'s set of generation units. Hydro capacity on day $t$ is $K^{\mathsf{D}}_{it}=w_{it}=\sum_{j\in\mathcal{D}_i}w_{ijt}$, where $w_{ijt}$ is the water stock at dam $j$ of firm $i$ at time $t$; thermal capacity $K^{\mathsf{T}}_i=\sum_{j\in\mathcal{T}_i}K^{\mathsf{T}}_{ij}$ is time-invariant. The system operator clears hourly markets by crossing supply schedules against demand:
\begin{equation} \label{eq:market.price}
D(\epsilon_{ht}) = \sum_{i=1}^N S_{iht}(p_{ht}(\epsilon_{ht})), \;\;\; \mathrm{for\ all}\; h = \{0,...,23\} \, \ \mathrm{and} \ \, t,
\end{equation}
where $\epsilon_{ht}$ is a random variable with full support, meaning that $D_{ht}(\epsilon)$  is the realized market demand in hour $h$ of day $t$ after firms submitted their bids.\footnote{Firm-level demand forecasts are not available, but the residual variation in realized demand puts a lower bound on the uncertainty firms face at the bidding horizon. After demeaning by month $\times$ day-of-week $\times$ hour, the residual in hourly demand has a standard deviation of roughly 500 MW on a 7,000 MW mean. The upper tail is material: the top 5\% (1\%) of demand states exceed the conditional mean by 800 MW (1,000 MW).}

Hourly profits combine spot profits, forward contracts, and a reliability charge:
\begin{equation}\label{eq:profit}
\pi_{iht} = \underbrace{D^R_{iht}(p_{ht},\epsilon_{ht}) \cdot p_{ht} - C_{iht}}_{\text{Spot market from }\eqref{eq:profits.mkt.clearing}} + \underbrace{(PC_{iht}-p_{ht}) \cdot  QC_{iht}}_{\text{Forward market}} + \underbrace{\mathds{1} _{[p_{ht} > \overline{p}_t]} (\overline{p}_t-p_{ht}) \cdot \overline{q}_{ijt}}_{\text{Reliability charge}},
\end{equation}
Here $QC_{iht}$ is firm $i$'s contracted quantity and $PC_{iht}$ the corresponding unobserved contract price. The reliability charge requires production $\overline{q}_{ijt}$ whenever the spot price exceeds the scarcity threshold $\overline{p}_t$.\footnote{Scarcity prices, $\overline{p}_{ijt}$, are updated monthly as a heat rate times a gas/fuel index, plus other variable costs. Firm energy obligations, $\overline{q}_{ijt}$, are set through periodic auctions \citep{cramton2007colombia,mcrae2024reliability}. We treat both variables as given in the spot market.}

Unlike the static framework in Section \ref{s:framework}, production updates hydropower capacity across adjacent markets. Water stocks evolve as \citep[e.g.,][]{lloyd1963probability}:
\begin{equation}\label{eq:water.balance}
w_{it+1} = w_{it} - \underbrace{\sum_{h=0}^{23} S_{iht}^{\mathsf{D}}(p_{ht}(\epsilon_{ht}))}_{\text{Water used in production}}   + \underbrace{\sum_{j\in \mathcal{D}_i} \xi_{ijt}}_{\text{Water inflows}},
\end{equation}
where $\xi_{ijt}$ denotes inflows at dam $j$. We model inflows at the firm level: dams of the same firm cluster on adjacent rivers (Panel (a) of Appendix Figure~\ref{fig:omitted}), so inflows are correlated within firms but nearly independent across firms. The cross-firm inflow correlation, conditioning on seasons and lagged inflows, is $< 0.2$. 

Firm $i$'s program is to choose the daily vector of hourly schedules $S_{iht}(p)=\sum_{j\in\mathcal{J}_i} q_{ijht} \mathds{1}_{\{p\geq b_{ijt}\}}$, for $h=0,\ldots,23$, before demand is realized:
\begin{align} \label{eq:inter.opt}
V(\textbf{w}_{t}) 	= \max_{\{q_{ijht},b_{ijt}\}_{j\in\mathcal{J}_i,\,h=0,\ldots,23}} \mathbb{E}_{\mathbf{\epsilon}}\left[ \sum_{h=0}^{23} \pi_{iht}
+ \beta \int_\mathbb{W} V(\textbf{u}) \, f \left( \textbf{u}\big|\mathbf{\Omega}_{t} \right)\, \textnormal{d} \textbf{u}\right],
\end{align}
where $\textbf{w}_t\in\mathbb{W}\equiv\{\mathcal{W}_i\}^N_i$ is the vector of water stocks, and $f(\cdot|\mathbf{\Omega}_{t})$ is the transition matrix governing the probability of next-period stocks given current stocks, hydro supply, and exogenous inflow factors $\mathbf{Z}_t$ (e.g., El Niño probabilities), so that $\mathbf{\Omega}_{t}=\{\mathbf{w}_{t},\mathbf{S^{\mathsf{D}}}_t,\mathbf{Z}_t\}$, where $\mathbf{S^{\mathsf{D}}}_t$ denotes the vector of hydro supplies at $t$.

\subsection{From Dynamic SFE to Estimating Equations}\label{s:identification}

The counterfactual object of interest is the supply-function equilibrium induced by a new capacity vector. In that equilibrium, a firm's schedule changes every rival's residual demand, hydro output changes future water stocks, and the value of those stocks feeds back into future schedules. A full counterfactual would therefore require a fixed point in firms' supply schedules and value functions. The dimensionality is large: in an SFE the control is a function rather than a scalar, whereas standard dynamic-game estimators \citep{rust1987optimal,hotz1993conditional,bajari2007estimating} are designed for finite-dimensional actions.\footnote{The four strategic firms operate roughly seven units each, or about $4\times 7\times 2\approx 56$ continuous controls per market-hour (quantity and price bids). Even a binary grid over these controls yields $2^{56}\approx 7\times 10^{16}$ candidate bid profiles, roughly the number of seconds since the Big Bang; a ten-level grid yields $10^{56}$, roughly the number of atoms in a million Earths.} 

The issue is not only the number of controls. A unit's quantity at price $p$ is linked to the quantities the same firm plans to offer at prices above and below $p$, and to the quantities offered by its other units, through the firm's supply schedule. SFE estimation is therefore about a function, not an isolated action. Even in static applications, the literature typically imposes parametric restrictions on schedule shape, such as linear slopes, or on cost structure, while taking competitors' supplies as given \citep{green1992competition,hortaccsu2008understanding}.

We therefore take a local route. We use the observed equilibrium first-order conditions to estimate marginal costs and the effective cost of water, and then compute counterfactual ex-post best responses after capacity reallocations, holding rivals' observed schedules and the estimated effective-cost function fixed. The exercise is analogous to an upward-pricing-pressure calculation \citep{farrell2010antitrust,nocke2022concentration,greenfield2021upward}: it asks how prices change when physical capacity is reallocated, taking the rest of the strategic environment as given. This isolates the portfolio-composition channel in the static model while allowing the Colombian application to include storage, contracts, reliability obligations, and the observed bid format.

The bridge from Section~\ref{s:framework} is the schedule optimality condition. In the static model, equation~\eqref{eq:foc_main} is the pointwise condition that a marginal perturbation of firm $i$'s supply schedule has zero payoff at each residual-demand realization. Here the same logic applies to a richer payoff. Forward contracts and reliability obligations enter marginal revenue, while the intertemporal value of water enters hydro's effective marginal cost: hydro production has direct cost $c^{\mathsf{D}}$ plus the opportunity cost of reducing future water stocks. We refer to the estimated mapping from water stocks to this opportunity-cost component as the effective-cost function. Conditional on these effective costs, firms' daily bid schedules satisfy the same ex-post optimality logic as in \citet{klemperer1989supply}. Appendix~\ref{apndx:bellman_expost} gives sufficient conditions for this daily ex-post Bellman interpretation.

To take this schedule optimality condition to the observed bids, we smooth supply schedules using kernel methods, as is standard in this literature \citep{wolak2003measuring}. Each firm's smoothed supply function is $S_{iht}(p) = \sum_{j\in\mathcal{J}_i} q_{ijht}\, \mathcal{K}\!\left(\frac{p - b_{ijt}}{bw}\right)$, where $\mathcal{K}$ is the standard normal CDF and $bw$ is the bandwidth (Appendix~\ref{apndx:smooth}). Quantity bids and price bids are two finite-dimensional perturbations of the same schedule. At an interior optimum, both FOCs hold; we use the quantity projection below because it is better identified in the data, but the price bids appearing in it are themselves optimal.

To gain intuition, consider the quantity-first-order condition from \eqref{eq:inter.opt}:

\begin{align}\label{eq:FOC.q}
\frac{\partial V(\mathbf{w}_t)}{\partial q_{ijht}} = 0 \; : \quad & \underbrace{\left(p_{ht}\frac{\partial D^R_{iht}}{\partial p_{ht}} + D^R_{iht}\right)\frac{\partial p_{ht}}{\partial q_{ijht}} - \left(QC_{iht} + \mathds{1}_{[p_{ht}>\overline{p}]}\overline{q}_{ijt}\right)\frac{\partial p_{ht}}{\partial q_{ijht}}}_{\text{Marginal revenue}} \nonumber\\[6pt]
& - \underbrace{\sum_{\tau\in\{\mathsf{D},\mathsf{T}\}} \left(\frac{\partial S^\tau_{iht}}{\partial q_{ijht}} + \frac{\partial S^\tau_{iht}}{\partial p_{ht}}\frac{\partial p_{ht}}{\partial q_{ijht}}\right) c^\tau}_{\text{Marginal cost}} \nonumber\\[6pt]
& + \underbrace{\left(\frac{\partial S^{\mathsf{D}}_{iht}}{\partial q_{ijht}} + \frac{\partial S^{\mathsf{D}}_{iht}}{\partial p_{ht}}\frac{\partial p_{ht}}{\partial q_{ijht}}\right) \int_\mathbb{W} \beta\, V(\mathbf{u})\frac{\partial f(\mathbf{u}|\mathbf{\Omega}_t)}{\partial S^{\mathsf{D}}_{iht}}\,\textnormal{d}\mathbf{u}}_{\text{Marginal value of holding water}} \nonumber\\[6pt]
& + \underbrace{\frac{\partial p_{ht}}{\partial q_{ijht}} \sum_{l\neq i}^N \frac{\partial S^{\mathsf{D}}_{lht}}{\partial p_{ht}} \int_\mathbb{W} \beta\, V(\mathbf{u})\frac{\partial f(\mathbf{u}|\mathbf{\Omega}_t)}{\partial S^{\mathsf{D}}_{lht}}\,\textnormal{d}\mathbf{u}}_{\text{Marginal value from competitor $l$'s holding water}} \; = \; 0,
\end{align}
where we write $\tau$ for $\tau_{ij}$ ($\mathsf{D}$ for dams, $\mathsf{T}$ for thermal). The FOC equates marginal revenue (spot revenue net of forward contracts and reliability payments) to marginal cost and the shadow value of water. As $\left|\frac{\partial p_{ht}}{\partial q_{ijht}}\right|$ increases, the firm cuts supply to raise the marginal revenue on inframarginal units; this markup mechanism applies uniformly across technologies. The (unobserved) forward contract price $PC_{iht}$ drops out because it enters profits only through $p_{ht}$, not through $\frac{\partial p_{ht}}{\partial q_{ijht}}$.

The connection to the U-shape in Figure~\ref{fig:empiricalU} runs through the third row. Under abundance (the right orthant of Figure~\ref{fig:empiricalU}, where the large firm holds ample water), the derivative $\frac{\partial f(\cdot|\mathbf{\Omega}_t)}{\partial S^{\mathsf{D}}_{iht}} \simeq 0$ and the opportunity cost of water vanishes. The firm responds mainly to residual demand (top panels of Figure~\ref{fig:cases}), and prices rise with concentration through the markup alone. Under scarcity (the negative orthant, where the large firm faces drought), $\frac{\partial f(\cdot|\mathbf{\Omega}_t)}{\partial S^{\mathsf{D}}_{iht}} < 0$ and water becomes valuable. Thermal units cannot directly deplete the water stock ($\frac{\partial S^{\mathsf{D}}_{iht}}{\partial q_{ijht}} = 0$ for $j \notin \mathcal{D}_i$), so the third row is non-negative for them: diversified firms expand thermal output to preserve hydro capacity, as in Figure~\ref{fig:responses}. This can lower prices relative to a firm without thermal capacity. The last line captures the effect on competitors' water stocks through $\frac{\partial p_{ht}}{\partial q_{ijht}}$; since it does not differentiate hydro from thermal units, we do not emphasize it further.

\subsection{Estimation and Counterfactuals}\label{s:simulations}

\paragraph{Estimation.} The first-order condition in \eqref{eq:FOC.q} provides moment conditions for technology-specific marginal costs and for the effective-cost component associated with water storage. El Ni\~no probabilities and lagged inflows shift the shadow value of water through the transition density in the third row of \eqref{eq:FOC.q}, separately from contemporaneous production costs; within-firm variation in residual demand identifies $c^{\mathsf{D}}$ and $c^{\mathsf{T}}$.\footnote{In particular, the transition matrix and all the supply slopes, whose functional form is in Appendix~\ref{apndx:smooth}, can be computed from the data.}

$V(\cdot)$ is a function on the state space, so we represent it parsimoniously. We approximate $V$ with a five-knot spline (details below); we assume marginal costs are constant within each technology, $c^{\mathsf{D}}$ and $c^{\mathsf{T}}$; and we assume each firm's value function depends only on its own water stock. This last restriction removes the final term in (\ref{eq:FOC.q}); the dimensionality gain is substantial (a tensor-product basis in all four firms' stocks would require $5^4 = 625$ parameters).\footnote{An alternative would be an oblivious-equilibrium-style state reduction \citep{weintraub2008markov}, with $V$ depending on firm $i$'s own stock $w_{it}$ and an aggregate rival state $w_{-it}$; this would require the joint transition of $(w_{it},w_{-it})$ and interaction terms in $V(\cdot)$. We instead let firm-specific transition matrices and time fixed effects proxy for that flexibility, modeling $f_i(\cdot)$ as the ARDL forecast of Section~\ref{s:em} with Pearson Type~IV residuals, widely used in hydrology for its asymmetric tails (Appendix~\ref{apndx:forecast}). Appendix Figure~\ref{fig:comp_responses} supports dropping the rival state: firm $i$'s bids do not respond to rivals' inflow forecasts up to fixed effects. Given the model fit below, we keep this one-state representation.}

We rearrange (\ref{eq:FOC.q}) so that marginal revenue is on the left-hand side and the cost and value-function terms are on the right. Approximating $\beta \cdot V(w)$ with a five-knot spline $\sum_{r=1}^5 \gamma_r B_r(w)$ absorbs the discount factor and yields an estimating equation\label{eq:2sls} with seven parameters: $c^{\mathsf{D}}$, $c^{\mathsf{T}}$, and $\{\gamma_r\}_{r=1}^5$. All other terms (the supply derivatives, the transition-matrix integrals, and marginal revenue) are constructed from observed bids and the ARDL forecasting model. Fixed effects absorb persistent differences across firms, units, and technology-specific time trends.

We instrument for the supply derivatives using temperature at dam locations and global gas price fluctuations interacted with monthly indicators, and for thermal start-up costs using competitors' lagged thermal utilization rate interacted with lagged gas prices.\footnote{Although domestic fuel prices are regulated, global prices pass through via the government's benchmark-indexed pricing formulas \citep{abdallah2019reforming}: the elasticity of Colombian electricity prices to global gas prices is 0.8 (S.E.\ 0.02, $R^2=53\%$).}

A final specification choice concerns which moment conditions we use. We estimate the quantity FOC rather than the price FOC because the two, while both valid, differ sharply in identifying variation. Both arise from projecting the same functional optimality condition, $\frac{\delta V}{\delta S_i(p)} = 0$, onto different test functions: $\frac{\partial V}{\partial q_j} = \int \frac{\delta V}{\delta S_i(p)} \cdot \mathcal{K}\left(\frac{p-b_j}{bw}\right) dp = 0$ weights by the kernel CDF, while $\frac{\partial V}{\partial b_j} = -\int \frac{\delta V}{\delta S_i(p)} \cdot \frac{q_j}{bw}\cdot\mathcal{K}'\left(\frac{p-b_j}{bw}\right) dp = 0$ weights by the kernel PDF. The CDF assigns non-degenerate weight to every dispatched unit; the PDF is near-zero for units whose bids are far from the clearing price.\footnote{Geometrically, with units ordered by bid, raising $q_j$ shifts the supply curve rightward at every price above $b_j$ (a global perturbation), whereas lowering $b_j$ is equivalent to raising $q_j$ and lowering $q_{j+1}$ by the same amount, a differenced perturbation that cancels above $b_{j+1}$. The price FOC thus retains only this local effect.} In our data, 57.5\% of unit-hours have bids far enough from the clearing price that the PDF weight is effectively zero, so the price FOC has far less identifying power. In addition, Colombian market rules require hourly quantity bids but only daily price bids, providing richer within-day variation for the quantity margin.

\paragraph{Results.}\label{s:results}
We estimate the model using daily data from 2010 to 2015 via two-stage least squares. Table~\ref{tab:estimates} reports results across four specifications: Columns (1)--(2) with week fixed effects, Columns (3)--(4) with daily fixed effects, and Columns (2) and (4) adding month-by-technology fixed effects to account for seasonal patterns that differentially affect hydro and thermal units. All columns include firm- and unit-fixed effects.

Focusing on Columns (2) and (4), thermal marginal costs are approximately 140,000 COP per MWh, roughly equal to the average market price (Figure~\ref{fig:prices}), confirming that thermal units are dispatched only during scarcity (Figure~\ref{fig:by_tech}, Panel b). Hydropower has substantially lower operating costs, making it the inframarginal technology. Our thermal estimates, ranging from \$45.57 to \$70.44 per MWh, are consistent with engineering benchmarks of \$20--\$40 for coal and \$40--\$80 for gas \citep[e.g.,][]{blumsack2023basic}.\footnote{As an illustration, \cite{reguant2014complementary} estimates thermal production costs in Spain at \euro 30--\euro 36 per MWh in 2007, when oil averaged \$72/barrel versus \$84.70 during our sample period.} The $\gamma_r$ coefficients lack a direct economic interpretation due to the spline approximation. Appendix Table~\ref{tab:estimates_p4} shows that using four knots instead of five yields similar results; Appendix Table~\ref{tab:estimates_normal} confirms robustness using a normal transition matrix instead of the Pearson Type IV distribution. 

\input{Tables/structural_model/res2_p5_rob}

\paragraph{Simulation algorithm.} To simulate prices, we compute the focal firm's daily ex-post best-response schedule conditional on rivals' observed schedules and the estimated effective-cost function $\{\hat{\gamma}_r\}$. At each residual-demand grid point, the firm chooses output to maximize current profits plus continuation value at the price implied by residual demand; solving this problem across grid points traces out the simulated supply schedule.

We implement this problem as a mixed-integer linear program. Residual demand is discretized into $Z$ steps, each unit's supply into $G$ steps, and the value function into $M$ steps; we set $G=M=Z=10$ using the parameters from Column~(4) of Table~\ref{tab:estimates}. Each supply step has length equal to the unit's capacity divided by $G$; the firm can produce any fraction of a step, but must fill lower steps before activating higher ones. To reduce computation, we aggregate daily observations to week-by-hour cells and solve \eqref{eq:inter.opt} for each of the four largest firms in each of the roughly 7,500 week-by-hour cells from 2010 to 2015.\footnote{The constraints are: (i)~hourly market clearing; (ii)~monotonicity of residual demand; (iii)~monotonicity of each unit's supply schedule; (iv)~hourly capacity bounds per unit; (v)~a water balance linking total hydro output across hours to the water stock, and bounded by the minimum of hydro capacity and available water. Deficits are not a concern: a dam exhausts its water stock on only four days in six years, and the 1st percentile of the water stock is over ten times the median hydro-quantity bid.}

\begin{figure}[!t]
	\centering
	\caption{Model fit: simulated and observed average weekly prices}\label{fig:model_fit}
		\includegraphics[width=.95\linewidth]{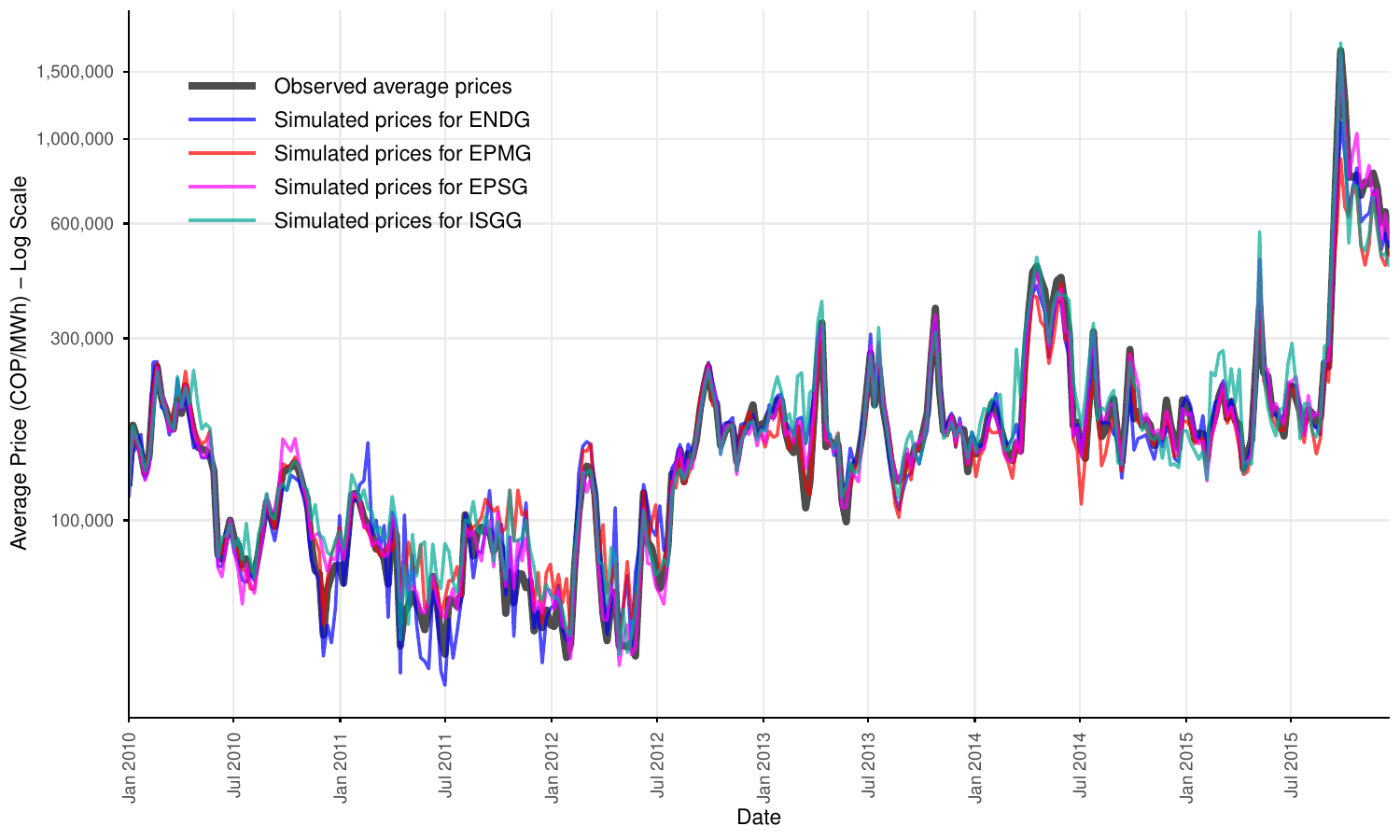} \hfill
	\begin{minipage}{1 \textwidth}
		{\footnotesize Note: Thick gray line: observed average market price. Colored thin lines: simulated prices from solving each firm's profit maximization problem (Section~\ref{s:simulations}) for the four largest firms (ENDG, EPMG, EPSG, ISGG). Ten discretization steps ($G=M=Z=10$). 2,900 COP $\simeq$ 1 US\$.  \par}
	\end{minipage}
\end{figure}

\paragraph{Model fit.}
Figure~\ref{fig:model_fit} compares observed average weekly prices (thick gray line) with simulated prices for each of the four largest firms. The model captures price volatility well across all firms. Appendix Table~\ref{tab:model_fit_hp} shows that simulated hourly prices are on average within 14\% of observed prices. Appendix Table~\ref{tab:model_fit_reg} further validates the model by regressing observed outcomes on their simulated counterparts: the simulations track prices, quantities, and water usage closely. Overall, despite relying on only seven parameters, the model captures the main movements in Colombia's wholesale electricity market over six years.

\paragraph{Counterfactual exercises.}\label{s:counter}
The counterfactual exercises relax a firm's thermal capacity constraint by reallocating capacity from its competitors. When transferring $\kappa\%$ of capacity from unit $k$, we leave its supply unchanged if the unit retains sufficient unused capacity; otherwise, we proportionally reduce its quantity bid.\footnote{We define a unit's capacity as its maximum observed quantity bid, although results are qualitatively similar when using technical capacity. Units that do not submit bids in a market are excluded from the simulations, as their absence may be due to technical constraints, which would otherwise inflate total capacity in the counterfactual relative to the data.}

\begin{figure}[!t]
	\caption{Counterfactual capacity transfers}
	\label{fig:counter_price}
	\centering
	\begin{minipage}{0.48\textwidth}
		\centering
		\includegraphics[width=\textwidth]{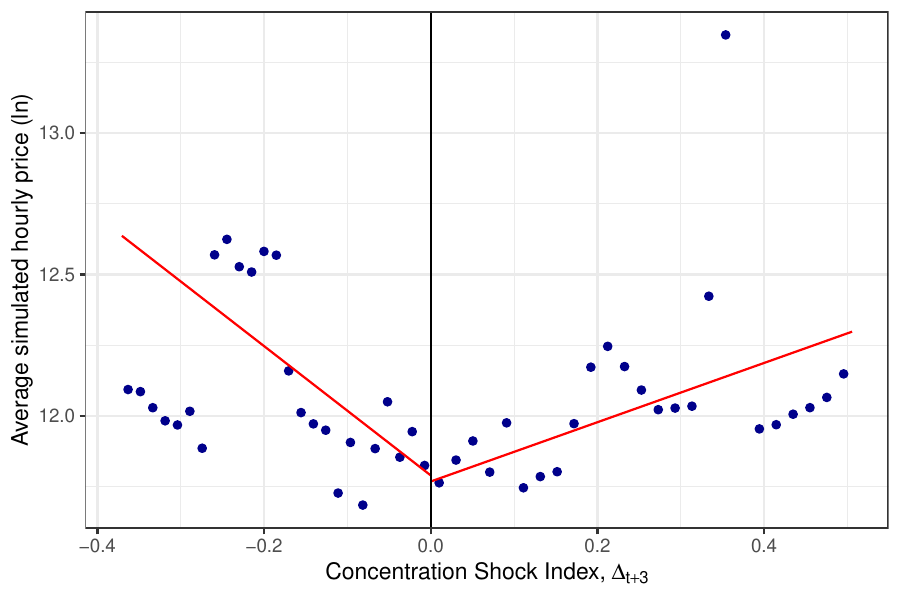}
		\subcaption{Simulated U-shape (cf.\ Figure~\ref{fig:empiricalU})}
	\end{minipage}\hfill
	\begin{minipage}{0.48\textwidth}
		\centering
		\includegraphics[width=\textwidth]{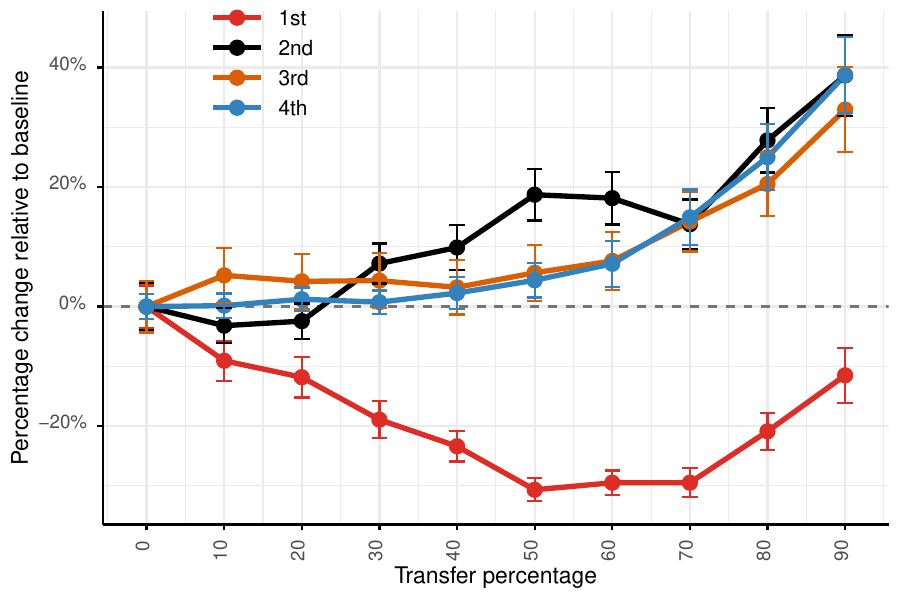}
		\subcaption{Price changes by HHI quartile}
	\end{minipage}
	\begin{minipage}{1 \textwidth}
		{\footnotesize Notes: Panel~(a): regression-discontinuity plot of log simulated prices on the Concentration Shock Index $\Delta_{t+3}$ (25 bins, cutoff at zero), replicating the reduced-form U-shape of Figure~\ref{fig:empiricalU} with model-generated data. Panel~(b): average percentage price change relative to the no-transfer baseline ($\kappa=0$) as a function of the transfer share $\kappa$ (0--90\% of competitors' thermal capacity transferred to the focal firm), pooling all four largest firms as the focal firm. Markets are grouped into quartiles of the counterfactual HHI, recomputed at each transfer level. \par}
	\end{minipage}
\end{figure}

Figure~\ref{fig:counter_price} summarizes the counterfactual results. Panel~(a) replicates the reduced-form analysis of Figure~\ref{fig:empiricalU} using simulated rather than observed prices: binned averages of log simulated prices are plotted against the same concentration shock index $\Delta_{t+3}$ as in Figure~\ref{fig:empiricalU}, with a cutoff at zero. The U-shape reappears in the simulated data, showing that the quantitative model reproduces the central empirical pattern.

Panel~(b) decomposes the price effects by market concentration. The $x$-axis varies the transfer share $\kappa$ from 0 to 90\% of competitors' thermal capacity, pooling the simulations of all four largest firms as the focal firm. Markets are grouped into quartiles of the counterfactual HHI, recomputed at each transfer level; because even the lowest-concentration quartile sees rising concentration as $\kappa$ grows, all four lines eventually slope upward.

In already concentrated markets (3rd and 4th HHI quartiles), transfers are unambiguously anticompetitive. Prices increase steadily with $\kappa$, reaching nearly 40\% above the no-transfer baseline for the most concentrated quartile. The initial increases are modest because the receiving firm holds ample water in these markets, so the added thermal capacity has little effect on its bidding strategy and competitors may retain enough unused capacity that their supply schedules barely adjust. Indeed, even relocating half of competitors' thermal capacity raises prices only modestly; the sharp increases come only at the largest transfers. This muted response is the structural counterpart of the flat, statistically insignificant right (abundance) slope in Appendix Table~\ref{tab:empiricalU_piecewise}, discussed in Section \ref{s:em_prices}, where greater concentration likewise barely moves prices near the baseline.

In less concentrated markets, the pattern is markedly different. For the lowest HHI quartile, transfers reduce prices by as much as 30\% as the constraint-relief effect from the added capacity dominates. Prices then rise as $\kappa$ increases further, tracing a U-shape that mirrors the reduced-form evidence in Figure~\ref{fig:empiricalU}. The second quartile exhibits a similar U-shaped pattern, though the price gains from transfers are exhausted sooner. Since the percentage changes in Panel~(b) are computed relative to each group's own baseline price at $\kappa=0$, a larger percentage increase for the second quartile does not imply higher price levels than the fourth quartile.

One possible concern is that intertemporal errors in simulated water stocks could drive the U-shape. To address this without further complicating the simulation algorithm, we remove the intertemporal feedback of the water stock: in each period the focal firm is endowed with the water stock observed in the data and solves the same optimization problem described above, without carrying simulated stock errors into the next period. Appendix Figure~\ref{fig:counter_price_myopic} replicates the analysis under this specification and confirms that the results are robust.

\subsection{Discussion}
Taken together, the simulations put magnitudes on the mechanism behind the paradox of concentration: prices can rise whether the dominant firm grows or shrinks even after controlling for total available capacity. We draw out its implications for diagnosis, policy, and the scope of the result.

\paragraph{Diagnosing market power.} In a single-technology industry, market power comes from capacity: a firm marks up the residual demand that rivals cannot serve. With multiple technologies, efficiency is another source, as a low-cost technology lets its owner undercut rivals and capture demand, though finite capacity limits how far that pressure reaches. Our estimates show both regimes. Where the leader's cheap hydro capacity is scarce, relieving its constraint lets it compete more aggressively and lowers prices by up to 30\%. Where that capacity is ample, extra units no longer sharpen its efficiency advantage: the leader's constraint is slack, and the added capacity mainly crowds out rivals, raising prices by nearly 40\% through the standard concentration channel. What a transfer does therefore turns on which firm becomes constrained, not on concentration alone. 

\paragraph{Capacity caps.} A cap lowers prices when the firm's dominance rests on sheer capacity, but raises them when it tightens the already-binding constraint of a firm whose cheap capacity is scarce (Proposition~\ref{prop:comp}). The same logic that makes added capacity pro-competitive on the left arm of the U-shape makes a cap counterproductive there.

\paragraph{Divestitures.} Divestiture design should be organized around the technology it shifts, not only around whether the buyer can operate the asset, because ownership sets pricing incentives under capacity constraints. A divestiture that leaves a constrained efficient firm less diversified can backfire (Proposition~\ref{prop:symmetry}); one that gives a rival a technology it lacks can instead create within-technology rivalry, though whether diversification beats specialization depends on demand (Example~\ref{ex:specialization}). What should discipline the design is therefore not concentration but the firms' cost and capacity primitives, which pin down the regime a transfer triggers (Proposition~\ref{prop:comp}). Because the response is U-shaped, such design need not be exact, since some diversification-enhancing transfers lower prices even as they raise concentration.

\paragraph{Beyond supply functions.} The mechanism is not tied to SFE. Supply functions are natural for our application because electricity firms bid schedules, and because the model lets the supply response be determined in equilibrium rather than imposed as Bertrand or Cournot conduct, which assume either a horizontal or a vertical supply in equilibrium. But we show that the same capacity-versus-efficiency logic also appears in a multi-technology extension of \citet{maggi1996strategic}'s differentiated-product price subgame, where low-cost capacity is a soft constraint: firms can produce beyond it at a higher marginal cost (Section~\ref{s:maggi_extension}). There too, a transfer lowers prices when it relaxes the efficient firm's constraint and raises prices when it mainly weakens rivals.

These findings turn capacity regulation into a technology-allocation problem. Share limits remain useful when capacity is the operative source of market power, but they are incomplete when firms differ in both costs and constraints. The next step is to develop tractable screens that combine ownership shares with technology-specific costs and capacities, so that remedies identify not only how much capacity a firm controls, but which constraint the allocation relaxes or tightens.

\section{Conclusion}\label{s:conclusion}

The analysis in this paper is motivated by an empirical paradox: exploiting exogenous, weather-driven variation in hydro availability, we find that prices can fall as concentration rises. We propose a model in which two sources of market power, technological efficiency and productive capacity, coexist, and explain the paradox in closed form: the largest firm in a market can be capacity-constrained if it has the most efficient technologies because of incentives to undercut competitors. In this case, reallocating competitors' capacity results in lower market prices as it relaxes the leader's constraint. Similar reallocations raise prices when that firm already dominates on capacity. We estimate the model on Colombian wholesale data and quantify both regimes. The mechanism is not special to supply-function competition, as we show that it carries over to more standard differentiated-price oligopoly. In addition, it speaks directly to merger remedies and capacity regulation: concentration is not a sufficient statistic for competitive harm when firms hold multiple technologies, so the design of divestitures and capacity caps should turn on which firm a reallocation constrains.

\begin{spacing}{0.8}
	\bibliographystyle{ecca}
	\bibliography{sample.bib}
\end{spacing}

\FloatBarrier
\clearpage
\newpage
\appendix
\setlength{\parskip}{0pt}
\renewcommand*{\thefootnote}{\fnsymbol{footnote}}
\setcounter{footnote}{0}

\newpage\clearpage
\pagenumbering{arabic}
\setcounter{page}{1}
\renewcommand{\thepage}{OA-\arabic{page}}
\section*{\centering \Huge Online Appendix}
\addcontentsline{toc}{section}{Online Appendix}

\setlength{\abovedisplayskip}{0.55\abovedisplayskip}
\setlength{\belowdisplayskip}{0.55\belowdisplayskip}
\setlength{\abovedisplayshortskip}{0.55\abovedisplayshortskip}
\setlength{\belowdisplayshortskip}{0.55\belowdisplayshortskip}

\renewcommand*{\thefootnote}{\arabic{footnote}}
\setcounter{footnote}{0}
\setcounter{table}{0}
\renewcommand{\thetable}{A\arabic{table}}
\setcounter{figure}{0}
\renewcommand{\thefigure}{A\arabic{figure}}
\setcounter{equation}{0}
\renewcommand{\theequation}{A\arabic{equation}}
\section{Theoretical Appendix}\label{apndx:theory}

\subsection{Preliminaries and Lemmas} \label{apndx:primitives}

\vspace*{1ex}
\noindent\textbf{Notation.} For notation brevity, the indices $i,-i\in\{1, 2\}$ are reserved for the strategic firms and not for fringe firms. In the general setting of Section \ref{s:framework} with many non-strategic fringes, the market clearing condition, the residual demands, and the profits have the following characterization.

\vspace{1ex}
\noindent\textbf{Market clearing.}  With many non-strategic firms supplying at $c^f$,
 the market always clears at some price $p \leq c^f$. 
 Hence, for any realization $D$ of $D(\epsilon)$, given the supply schedules $S_i(p)$, it follows that the market clears at 
\begin{equation}\label{apndx:clearing}
p^* = \min\Big\{c^f, \min\big\{p\big| \sum_i^2 S_i(p) \geq D \big\}  \Big\}.
\end{equation}

\vspace{1ex}
\noindent\textbf{Residual demand.} 
Since the strategic firms have priority in production over the fringe firms at price $p=c^f$, 
Firm $i$'s residual demand function is derived as
\begin{equation}\label{apndx:DR}
    D_i^R(p,\epsilon) = 
\begin{cases} 
\max\{0, D(\varepsilon)- S_{-i}(p)\} & \text{ for } p\in [0,c^f],\\
0, & \text{ for } p >c^f.
\end{cases}
\end{equation}   

\vspace{1ex}
\noindent\textbf{Profits and best responses.} Given the marginal costs in Section \ref{s:duopoly}, the profit (\ref{eq:profits.mkt.clearing}) of firm $i$ given the opponent's strategy, when the market clears at price $p$ is 
\begin{equation} \label{apndx:eq:profit_asym}
    \pi_i(p) = D_i^R(p,\epsilon)  \cdot p -  \sum_{\tau \in \mathcal{T}} c^\tau S_i^\tau(p).
\end{equation} 
It follows from Definition (\ref{apndx:def:SFE}) that an SFE is then a pair of supply functions $(S_1,S_2)$, so that for $i\in \{1,2\}$  $S_i(p)=\sum_\tau S_i^\tau(p)$, the market clearing $p^*(\epsilon)$ in (\ref{apndx:clearing}) 
maximizes $i$'s ex post profit (\ref{apndx:eq:profit_asym}) at every  $D(\epsilon)$.

\vspace{1ex}
The propositions in Appendices \ref{apndx:proof.2} and \ref{apndx:proof:sym} rely on the following lemmas.
\begin{lemma}\label{increasing}
   Suppose $(S_1,S_{2})$ is an SFE. Then for $i\in\{1,2\},$ $S_i(p) < \sum_\tau K_i^\tau$ for every $p< c^f$. Moreover, $p<c^f$ and $S_i(p)>0$ implies $S_i(p')>S_i(p)$ whenever $p'>p.$
\end{lemma}
\begin{proof}
To prove the first part, suppose for contradiction that at some $p<c^f$, firm $i$ exhausts all its capacity, $S_i(p) = \sum_\tau K_i^\tau$. By right-continuity of $S_i$, there is $\underline p=\min\{p|S_i(p) = \sum_\tau K_i^\tau\}.$ In this case, one can check that the best response to $S_i$ by $-i$ satisfies $S_{-i}(p)=S_{-i}(\underline p)$ for all $p\in (\underline p,c^f)$.

Now that $S_{i}(p) =S_{i}(\underline p)$ on $p \in(\underline p,c^f)$ for both $i\in \{1,2\}$,  this contradicts ex post optimality. In the event that $D=S_1(\underline p)+S_2(\underline p)$ realizes so that $p^*=\underline p$ in (\ref{apndx:clearing}) clears the market, Firm $i$ can reduce production by $\epsilon$ to $\min\{0,S_i(p)-\epsilon\}$, so that $D>S_i(p)-\epsilon +S_2(p)$ for all $p\in(\underline p,c^f)$, causing the clearing price to jump up to $c^f$, a profitably deviation: exhausting capacity at $p<c^f$ is not ex post optimal.

To see the second part, suppose for contradiction, there exist two prices $p < p'<c^f$   such that $0<S_i(p)= S_i( p')< K_i$. Suppose $\underline p=\min \{r|S_i(r)=S_i(p')>0\}.$ 
One can similarly check that the best response to $S_i$ by $-i$ satisfies
$S_{-i}(r)=S_{-i}(\underline p)$ for all $r\in[\underline p,p']$.  Similarly, consider the demand $D$ such that $\underline p$ clears the market, Firm $i$ can slightly reduce production to cause the clearing price to jump up to at least $p'$, a profitable deviation. This completes the proof.
\end{proof}

 \begin{lemma}\label{bertrand}
 In SFE, there exists $i\in\{1,2\}$ such that $\lim_{p \to c^{f-}} S_i(p) =\sum_\tau K_i^\tau$. 
\end{lemma}

\begin{proof}
Since fringe firms enter and there will be no residual demand for $p>c^f$, any remaining capacity of $i$ will be produced at $c^f$, so $S_i(c^f) =\sum_\tau K_i^\tau$ for $i=1,2.$ Suppose for contradiction $\lim_{p\to c^f-} S_i(p)<\sum_\tau K_i^\tau$ for both $i=1,2,$ then both supply schedules jump discretely at $c^f$. In the event that market clears at $c^f$ but with realized demand $D<\sum_{i,\tau} K_i^\tau$, firm $i$ can deviate by supplying all its capacity at $c^f-\epsilon $. This increases $i$'s ex post profit by capturing additional demand with a unit price arbitrarily close to $c^f.$
\end{proof}

\begin{lemma}\label{changepoint} In SFE, the production cost for firm $i\in\{1,2\}$ at different market-clearing prices satisfies 
\[ C_i =\begin{cases}
   c^l S_i^l(p) & \text{ iff } S_i^l(p)< K_i^l ;\\
   c^l K_i^l +c^h S_i^h(p) & \text{ iff } S_i^l(p)= K_i^l . 
\end{cases}\]
\end{lemma}

\begin{proof}
   Since $C_i$ is the minimal cost function for producing a given quantity, and because $c^l<c^h$,  Firm $i$ will necessarily first produce with $\tau=l$ and only start $\tau=h$ production when $K_i^l$ is exhausted. 
\end{proof}

\begin{lemma}\label{smooth}
In SFE, for $i\in\{1,2\}$
\begin{enumerate}
\item  both $S_i$'s are continuous on $(c^h,c^f)$;
    \item if $S_i(p)\neq K_i^l$ for both $i\in\{1,2\}$, then both $S_i$'s are continuous at $p$;
    \item if $S_i(p)>0$ and $S_i(p)\neq K_i^l$ for both $i\in\{1,2\}$, then both $S_i$'s are continuously differentiable at $p$.
\end{enumerate}
\end{lemma}

\begin{proof}
We begin with \textit{Part 2}. first.
Suppose there exists $\hat p\in(\underline p, c^h)$ at which at least one of $S_i$ for  $i=1,2$ is discontinuous. Denote by $\hat q_i$ the left limit $\hat q_i=\lim_{p\to \hat p-}S_i(p)$ for $i=1,2.$ When the realized demand equals the sum of limits $\hat D=\hat q_1+\hat q_2$, the market clears at $\hat p$ by (\ref{apndx:clearing}).

Case 1. Suppose for contradiction, let $S_i(\hat p)<K_{i}^l$ for both $i\in\{1,2\}$. Then $C'_{i}(\hat q_{i})=c^l<\hat p$ by Lemma \ref{changepoint}. Consider $\delta>0$ small enough so that we have 
\[
\begin{cases}
    \delta<\max_i\{S_i(\hat p)-\hat q_i \}; \\
    C'_{i}(\hat q_{i}+\delta)=c^l<\hat p \ \ \text{ holds for both }\ i\in\{1,2\}.
\end{cases}
\]
In the event the realized demand is $\hat D+\delta$, the market still clears at $\hat p$ due to the first inequality above, with $i$ supplying $\hat q_i+\delta_i$ and $-i$ supplying $\hat q_{-i}+\delta_{-i}$ where $\delta_i+\delta_{-i}=\delta$ for some non-negative $\delta_i $ and $\delta_{-i}$. WLOG, let $\delta_i<\delta$. At the realization of demand $\hat D+\delta$, $i$ can locally increase supply to $S_i+\delta$ to under cut $-i$ and capture the entire demand of $\delta$ at the clearing price $\hat p$ with the same marginal cost $c^l$, a profitable deviation. This proves the first part of the Lemma.

Case 2. Suppose $S_i(\hat p)>K_{i}^l\geq 0$ for one of $i\in\{1,2\}$. Optimality of $S_i$ implies $\hat p>c^h$. Continuity follows from an analogous argument as above. Since if there is a discontinuity at $\hat p$, one of the firms will benefit from locally increasing supply by $\delta$ to undercut the opponent, at the marginal profit of at least $\hat p-c^h$.
 
\textit{Part 1} follows from an analogous argument as above since any price $\hat p\in(c^h,c^f)$ are above the marginal costs. Hence, if there is a discontinuity on this interval, one of the firms will benefit from a local $\delta$ increase in supply to undercut its opponent.

\textit{Part 3}. When $S_i(p)>0$, then $p>c^l$. By Lemma \ref{changepoint}, it holds that and $S_i(p)\neq K_i^l$ for $i\in\{1,2\}$,  for any small enough interval $U \ni p$, we have  $C'_i(S_i(p))=c^\tau_i\geq c_i^l$ is constant on $U$. So for all $p'$ on $U$ 
\begin{equation}
C_i(S_i(p)) - C_i\left(S_i(p) + S_{-i}(p)-S_{-i}(p')\right)=-c_i^\tau \left( S_{-i}(p)-S_{-i}(p') \right). \label{eq:marginalcost}
\end{equation} 
since $S_{-i}$ is continuous at $p$ from \textit{Part 2}.

Now consider the demand $D(\epsilon)=S_1(p)+S_2(p)$ such that $p$ clears the market, firm $i$ maximizes by producing $S_i(p)$. Therefore, any $p'\in U$  satisfies
\[ S_i(p)\cdot p-C_i(S_i(p)) \geq [D(\epsilon)-S_{-i}(p')]\cdot p' - C_i\left(D(\epsilon)-S_{-i}(p')\right).  \]
Substitute in $D(\epsilon)$ and apply (\ref{eq:marginalcost}), we have  
\[\frac{   S_i(p)  }{p'-c_i^\tau} (p-p') \geq     S_{-i}(p)-S_{-i}(p') .\]
Similarly, consider market clearing at $p'$ gives 
\[  \frac{   S_i(p')  }{p-c_i^\tau} (p'-p)  \geq     S_{-i}(p')-S_{-i}(p) . \]
Therefore, for any $p,p'\in U$ we have
\[  \frac{S_i(p)  }{p'-c_i^\tau}   \geq    \frac{ S_{-i}(p)-S_{-i}(p') }{p-p'} \geq  \frac{   S_i(p')  }{p-c_i^\tau}. \]
Since this holds for all $p,p'\in U$, continuity of $S_i$ 
from Part 2 implies
$S_{-i}$ is continuously differentiable. The exception is when $S_i(p)=K_i^l$, then $c_i^\tau$ changes discontinuously to another constant as $S_i(p')\to K_i^l$. 
\end{proof}

\begin{lemma} \label{monotone} 
    Using $\kappa_2,\kappa_3$ solved in $\kappa_1$ as in (\ref{coefficients}), whenever $\kappa_1 \geq \frac{K_1^l}{c^f-c^l} $ we have 
    \begin{enumerate}
        \item $\kappa_2(c^f-c^h)-\kappa_3/(c^f-c^h) \geq \kappa_1(c^f-c^h)$ where the equality holds iff $\kappa_1 = \frac{K_1^l}{c^f-c^l} $;
        \item $\kappa_2(c^f-c^h)+\kappa_3/(c^f-c^h) \geq \kappa_1(c^f-c^l)$ where the equality holds iff $\kappa_1 = \frac{K_1^l}{c^f-c^l} $;
        \item $\kappa_2(c^f-c^h)-\kappa_3/(c^f-c^h)$  and $\kappa_2(c^f-c^h)+\kappa_3/(c^f-c^h)$, as functions in $\kappa_1$, are continuous and strictly increasing to $\infty$.
    \end{enumerate}
\end{lemma}

\begin{proof} \vspace{1ex}
   \noindent \textit{Item 1}: Using the above notations, we have 
      \begin{align*}
       \kappa_2(c^f-c^h)-\kappa_3/(c^f-c^h) - \kappa_1(c^f-c^h)= \frac{\kappa_1}{2}   \frac{\kappa_1}{\alpha  - \kappa_1}   (c^f-c^h)  - \frac{(c^h-c^l)^2}{2} \frac{ \alpha -\kappa_1}{c^f-c^h},
   \end{align*}
   which is an increasing function in $\kappa_1$. Its minimum is at $\kappa_1 = \frac{K_1^l}{c^f-c^l} $, for which  we have
   \begin{align*} 
     \frac{\kappa_1}{2}   \frac{\kappa_1}{\alpha  - \kappa_1}   (c^f-c^h)  - \frac{(c^h-c^l)^2}{2} \frac{ \alpha -\kappa_1}{c^f-c^h} 
   =  \frac{\kappa_1}{2}   (c^h-c^l)  - \frac{(c^h-c^l) }{2} \frac{K_1^l}{c^f-c^l} = 0.
   \end{align*}

    \vspace{1ex}
   \noindent \textit{Item 3}: The first part of Item 3 is straightforward. To see the second part, we have
   \begin{align*}
        \kappa_2(c^f-c^h)+\kappa_3/(c^f-c^h)   
       =  \frac{\kappa_1}{2} \left(2 + \frac{\kappa_1}{\alpha  - \kappa_1}  \right) (c^f-c^h) + \frac{(c^h-c^l)^2}{2} \frac{ \alpha -\kappa_1}{c^f-c^h},
   \end{align*}
   Differentiate with respect to $\kappa_1$ gives
   \[  c^f-c^h  + \frac{2\kappa_1 (c^f-c^h)}{2(\alpha - \kappa_1)} + \frac{ \kappa_1^2 (c^f-c^h)}{2(\alpha - \kappa_1)^2} - \frac{(c^h-c^l)^2}{2(c^f-c^h)}, \]
   which is increasing in $\kappa_1$. Substitute in $\kappa_1 = \frac{K_1^l}{c^f-c^l} $ to obtain its lower bound as
      \begin{align*}
        \left(c^f-c^h  + \frac{2\kappa_1 (c^f-c^h)}{2(\alpha - \kappa_1)} \right) 
       + \frac{ \left(\frac{K_1^l}{c^f-c^l} \right)^2 (c^f-c^h)}{2(\alpha - \kappa_1)^2} - \frac{(c^h-c^l)^2}{2(c^f-c^h)} 
       =   c^f-c^h  + \frac{2\kappa_1 (c^f-c^h)}{2(\alpha - \kappa_1)}.
   \end{align*}
   To finish the proof for Item 3, it is easy to see as $\kappa_1 \to \alpha$, both expressions in Item 3 go to $\infty$.

   \vspace{1ex}
   \noindent \textit{Item 2}: we have that 
      \begin{align*}
       \kappa_2(c^f-c^h)+\frac{\kappa_3}{c^f-c^h} - \kappa_1(c^f-c^l)
       = \frac{\kappa_1}{2}   \frac{\kappa_1}{\alpha  - \kappa_1}   (c^f-c^h)  + \frac{(c^h-c^l)^2}{2} \frac{ \alpha -\kappa_1}{c^f-c^h} - \kappa_1(c^h-c^l).
   \end{align*}
   Its derivative is increasing in $\kappa_1$ and hence
      \begin{align*}
       \frac{2\kappa_1 (c^f-c^h)}{2(\alpha - \kappa_1)} + \frac{ \kappa_1^2 (c^f-c^h)}{2(\alpha - \kappa_1)^2} - \frac{(c^h-c^l)^2}{2(c^f-c^h)} - (c^h-c^l) 
       \geq \frac{ \kappa_1 (c^f-c^h)}{ (\alpha - \kappa_1)}- (c^h-c^l)= 0.
   \end{align*}
   Therefore $\kappa_2(c^f-c^h)+\kappa_3/(c^f-c^h) - \kappa_1(c^f-c^l)$ is increasing in $\kappa_1$, and its minimum is at $\kappa_1 = \frac{K_1^l}{c^f-c^l} $ which is
      \begin{align*}
       \frac{\kappa_1}{2}   \frac{\kappa_1}{\alpha  - \kappa_1}   (c^f-c^h)  + \frac{(c^h-c^l)^2}{2} \frac{ \alpha -\kappa_1}{c^f-c^h} - \kappa_1(c^h-c^l) =0.
   \end{align*}
   \end{proof}

\subsection{Proof of Proposition \ref{prop:comp}}\label{apndx:proof.2}
First, we prove part \textit{1}. in Section \ref{apndx:proof.2.1} and then \textit{2.a} and \textit{2.b} in Sections \ref{s:reallocation.abundance} and \ref{s:reallocation.scarcity}, respectively. The proofs rely on the Lemmas and definitions listed in Appendix \ref{apndx:primitives}.

\subsubsection{Proof of Proposition \ref{prop:comp}.1 \textit{(Existence and Uniqueness)}}\label{apndx:proof.2.1}

\begin{proof} 
Firm $i\in \{1,2\}$'s profits are as follows,
\begin{equation}\label{apndx:profits}
    \max_p \pi_i(p) := p\cdot D^R_i(p,\epsilon) - C_i (D^R_i(p,\epsilon)).
\end{equation} 
This maximization can be solved by the following FOCs almost everywhere due to differentiability (Lemma \ref{smooth}),
\begin{equation}\label{eq:focs}
p \frac{\partial D_i^R(p,\epsilon) }{\partial p}  + D_i^R(p,\epsilon)  -  C_i' \cdot \frac{\partial D_i^R(p,\epsilon)}{\partial p}   =0. \end{equation} 
Since $C^\prime_i = c^\tau_i$ for the appropriate $\tau\in\{l,h\}$ and $S_i(p)=D_i^R(p)$ when market clears at $p$, firm $i$'s best response to $-i$ satisfies
\begin{equation}\label{eq:focs-simple}
S_i(p) = (p - c^\tau_i) S_{-i}^\prime(p)
\end{equation}
for both $i\in\{1,2\}$.
To better determine the appropriate $c_i^\tau$, we analyze the supply functions by partitioning the domain into four intervals: 1. $p< c^h$; 2. $p \in [c^h, c^f)$ and $S_1^l(p) < K_1^l$; 3. $p \in [c^h, c^f)$ and $S_1^l(p) = K_1^l$;\footnote{The difference between intervals 2. and 3. is whether Firm 1 has exhausted its low-cost capacity.} and 4. $p\geq c^f$.
By Lemma \ref{smooth}, $S_i(p)$ must be smooth in these intervals, and thus we proceed to characterize $S_i(p)$ in each interval separately.

\begin{enumerate}
    \item \textit{Interval: } $p< c^h$. Firm~2 does not produce using the $c^h$ technology for prices $p < c^h$, hence $S_2(p) = 0$ and $S_1$ best response with $S_1(p) = 0$ as can be solved from (\ref{eq:focs-simple}) with $S_2'=0$. The equilibrium supply functions in this interval are:
    \begin{equation}\label{eq:system0}
        \begin{cases}
            S_1(p) = 0, &\textnormal{if } p< c^h,  \\
            S_2(p) = 0, &\textnormal{if } p< c^h. 
        \end{cases}
    \end{equation}
    \item \textit{Interval:} $p\geq c^h$ and $S_1^l(p)<K_1^l$. From the FOCs (\ref{eq:focs}), Firm 1 solves:
    \begin{equation*}
        S_1(p) = S_1^l(p) = S_2^\prime(p) \cdot (p-c^l),
    \end{equation*}
    as Firm 1 will spend only its low-cost capacity before moving to the high-cost one (Lemma \ref{changepoint}). Firm~2's supply solves:
    \begin{equation*}
        S_2(p) = {S_1^l}^\prime(p) \cdot (p- c^h).
    \end{equation*}
    Solving these two partial differential equations gives 
    \begin{equation*}
        \begin{cases}
            S_1^l(p) &= \tilde{\kappa} \frac{ (p - c^l) \log(p - c^h) + (c^l - p) \log(p - c^l) - c^l + c^h}{(c^l - c^h)^2} + \kappa_1 (p - c^l),  \\
            S_2(p) &= \tilde{\kappa} \frac{ (p - c^h) \left(\ln(p-c^h)-\ln(p-c^l)\right) +c^h-c^l}{(c^l - c^h)^2} + \kappa_1 (p - c^h), 
        \end{cases}
    \end{equation*}
    where $\kappa_1$ and $\tilde{\kappa}$ are the two undetermined coefficients. Since the supply functions are non-negative and non-decreasing, we have $\tilde{\kappa}=0$. Therefore, the supply schedules in this interval are,  
    \begin{equation}\label{eq:system1}
        \begin{cases}
            S_1^l(p) = \kappa_1 (p-c^l) \\
            S_2(p) = \kappa_1 (p-c^h) 
        \end{cases} \textnormal{when } p\geq c^h \, \& \, S_1^l(p)<K_1^l,  
    \end{equation}
    where both $S_1^l(p)$ and $S_2(p)$ are non-negative and non-decreasing for an undetermined coefficient $\kappa_1$ to be solved for later.
    \item \textit{Interval:} $p\geq c^h$ and $S_1^l(p)=K_1^l$. Firm 1 exhausted its low-cost technology, so that $S_1^l(p) = K_1^l$ in this interval. It follows from (\ref{eq:focs-simple}) that
    \begin{equation*}
        S_1(p) = S_1^l(p) + S_1^h(p)=  K_1^l + S_1^h(p) = S_2^\prime(p) \cdot (p-c^h).
    \end{equation*}
    At the same time, Firm~2's supply solves
    \begin{equation*}
    S_2(p) = {S_1^h}^\prime(p)  \cdot (p-c^h).
    \end{equation*}
    Solving this system of differential equations obtains the following solutions with undetermined coefficients $\kappa_2$ and $\kappa_3$ (we solve for $\kappa_2$ and $\kappa_3$ later):
    \begin{equation}\label{eq:system2}
        \begin{cases}
            S_1(p) =  \kappa_2 \left(p-c^h  \right) + \kappa_3  \frac{1}{p-c^h} \\
            S_2(p) = \kappa_2 \left(p-c^h \right) - \kappa_3 \frac{1}{p-c^h} 
        \end{cases} \textnormal{when } p\geq c^h \, \& \, S_1^l(p)=K_1^l.
    \end{equation}
    \item \textit{Interval}: $p\geq c^f$. To prevent the entry of fringes, it is optimal to exhaust capacity at $p=c^f$,
    \begin{equation}\label{eq:system3}
        \begin{cases}
            S_1(p) =  \sum_\tau K_i^\tau, &\textnormal{if } p\geq c^f,  \\
            S_2(p) = \sum_\tau K_i^\tau, &\textnormal{if } p\geq c^f.
        \end{cases}
    \end{equation}
\end{enumerate}

\vspace{1ex}
To solve for $\{\kappa_1,\kappa_2,\kappa_3\}$. By continuity of Lemma \ref{smooth}.1, the boundary condition at $\hat p$, defined implicitly by $S_1(\hat p)=K_1^l$, together with (\ref{eq:system1}) and (\ref{eq:system2}) yield the following system of equations at the price $\hat{p}$
\begin{equation*}
\begin{cases}
K_1^l &= \kappa_1(\hat{p}-c^l), \\
\kappa_1(\hat{p}-c^l) &= \kappa_2(\hat{p}-c^h)+\kappa_3\frac{1}{\hat{p}-c^h}, \\
\kappa_1(\hat{p}-c^h) &= \kappa_2(\hat{p}-c^h)-\kappa_3\frac{1}{\hat{p}-c^h}.
\end{cases} 
\end{equation*}
Subtract the third line from the second line and rearrange to obtain
\begin{equation}
    \hat{p} = \frac{2\kappa_3}{\kappa_1(c^h-c^l)}+c^h.
    \label{switchingP}
\end{equation} Then, substitute this equation in the first and third line to obtain
\begin{equation}
\begin{cases}
\kappa_3 &= \frac{(c^h-c^l)^2}{2} \left( \alpha -\kappa_1\right), \\
\kappa_2 &= \frac{\kappa_1}{2} \left(1 + \frac{\alpha}{\alpha  - \kappa_1}  \right), \\
\hat{p} - c^h &= \frac{(c^h-c^l)}{\kappa_1} \left( \alpha -\kappa_1\right),
\end{cases} \label{coefficients}  
\end{equation}
where we defined $\alpha\equiv \frac{K_1^l}{c^h-c^l}$. 

The solution in (\ref{eq:system1}) implies $\kappa_1>0$, or else the supply function will not be strictly increasing, violating Lemma \ref{increasing}. On the other hand, if $\kappa_1>\alpha$, substitute in the price $p=c^h$ into (\ref{eq:system1}) to see that
\[ S_1^l(c^h) = \kappa_1 (c^h-c^l)>K_1^l.\]
This exceeds the capacity, implying when $\kappa_1>\alpha$, Firm 1 will exhaust its low-cost capacity at some $p<c^h$. This contradicts Lemma \ref{increasing}, since Firm 1 will not use its $K^h_1$ for $p < c^h$, and hence its supply function cannot be strictly increasing.

Consider the values of the (left) limits $\lim_{p\to c^{f-}} S_1(p)$ and $\lim_{p\to c^{f-}} S_2(p)$ as functions of $\kappa_1\in (0,\alpha)$. Substituting $\kappa_3$ into (\ref{switchingP}) under $\kappa_1< \frac{K_1^l}{c^f-c^l}<\alpha$, yields 
$\hat{p} = \frac{     K_1^l }{\kappa_1 } +c^l>c^f$. 
With this value of $\kappa_1$, Firm 1 does not exhaust its low-cost capacity when $p\to c^f$ from below; both left limits $\lim_{p\to c^{f-}} S_1(p)$ and $\lim_{p\to c^{f-}} S_2(p)$ are defined by the solution in (\ref{eq:system1}) and are increasing in $\kappa_1$ on the interval $(0,\frac{K_1^l}{c^f-c^l})$.

On the other hand, when $\kappa_1\in(\frac{K_1^l}{c^f-c^l},\alpha)$, we have $\hat p<c^f$, and so Firm 1 exhausts low-cost capacity when $p\leq c^f$. In this case both left limits $\lim_{p\to c^{f-}} S_1(p)$ and $\lim_{p\to c^{f-}} S_2(p)$ equals the left limits of the expressions in (\ref{eq:system2}). It follows from Lemma \ref{monotone} that these left limits are monotonically increasing for $\kappa_1\in(0,\alpha)$.

By Lemma \ref{bertrand}, now the equilibrium can be pinned down by monotonically increasing $\kappa_1 \in (0, \alpha )$ until the first $\kappa_1$ that satisfies the boundary condition $\lim_{p\to c^{f-}} S_i(p) = \sum_\tau K_i^\tau $ is found for some $i\in\{1,2\}$. Such an equilibrium always exists due to Lemma \ref{monotone}.3. Moreover, any larger $\kappa_1$ will imply that $S_i$ exceeds $i$'s capacity as $p\to c^f$ from below due to the monotonicity. Therefore the equilibrium exists and is unique.  
\end{proof}

\subsubsection{Proof of Proposition \ref{prop:comp}.2.a \textit{(Under Abundance)}}\label{s:reallocation.abundance}

 \vspace{1ex}

\begin{proof}
It can be verified that in this case Firm~2 exhausts its capacity before Firm 1 exhausts low-cost capacity. From (\ref{eq:system1}), we have that $S_2^l(p) = \kappa_1 \cdot (p-c^h)$. In the limit $p\to c^{f-}$, $S_2^l(p)\to K_2^h$, pinpointing $\kappa_1 = \frac{K_2^h}{c^f-c^h}$. Since Firm 1 still has low-cost capacity, from the first equation in (\ref{eq:system1}), we have that $S_1(p) = S_1^l(p) = \frac{K_2^h}{c^f-c^h} \cdot (p - c^l)$. Hence as $p\to c^{f-}$, $K_1^l > \frac{p - c^l}{c^f-c^h}\big|_{p=c^f} \cdot K_2^h = S_1^l(c^f)$.
Given this value for $\kappa_1$ we can compute $\kappa_2$ and $\kappa_3$ from (\ref{coefficients}), thereby constructing $S_i(p)$ for $i=\{1,2\}$.
Joining the systems (\ref{eq:system0}), (\ref{eq:system1}), and (\ref{eq:system3}) yields
\[
    S_1 =\begin{cases}
        0, & \text{ when } p<c^h, \\
        \kappa_1 (p-c^l), & \text{ when } p \in [c^h,  c^f ),\\
        K_1^l + K_1^h,  & \text{ when } p = c^f, 
    \end{cases}
\text{ and } 
    S_2 =\begin{cases}
        0, & \text{ when } p<c^h, \\
        \kappa_1 (p-c^h), & \text{ when } p \in [c^h, c^f ].
    \end{cases}
\]

If we reduce $K_2^h$ to $K_2^h-\delta$ and increase $K_1^h$ to $K_1^h +\delta$ for a $\delta<K_2^h$, it will still hold that Firm~2 just exhausts its capacity at $p\to c^f$ and hence $\kappa_1 = \frac{K^h_2-\delta}{c^f-c^h}$ still holds. Therefore, $\kappa_1$ decreases and the market-wide production
\[
    S_1(p)+S_2(p) =\begin{cases}
        0 & \text{ when } p<c^h \\
        \frac{K^h_2-\delta}{c^f-c^h} (2p-c^l-c^h) & \text{ when } p \in [c^h, c^f )\\
        K_1^l + K_1^h +K_2^h  & \text{ when } p = c^f 
    \end{cases}
    \]
decreases at every price level as $\delta$ rises. Hence, the market clearing price rises. 
\end{proof}

\subsubsection{Proof of Proposition \ref{prop:comp}.2.b \textit{(Under Scarcity)}}\label{s:reallocation.scarcity}

\begin{proof}
It can be verified that for $K_1^h$ small enough (to be specified later), the SFE permits a switching price $\hat{p}$ such that, for $p \in [\hat{p}, c^f]$, $S_1^h(p)>0$ and $S_1^l(p)=K_1^l$  as Firm 1 exhausts its low-cost capacity at $p=\hat{p}$, and in addition, Firm 1 also exhausts its high-cost capacity in the limit as $p\to c^{f-}$. 

Suppose for contradiction, that for any $K_1^h$, Firm~2 exhausts its capacity as $p\to c^{f-}$. From (\ref{eq:system2}), Firm~2's production schedule at $p=c^f$ is 
$$S_2(c^f)=\kappa_2(c^f-c^h)-\frac{\kappa_3}{c^f-c^h}.$$ 
By Lemma \ref{monotone}, $S_2(c^f)$ is increasing in $\kappa_1$ and so there is a unique $\tilde \kappa_1$ such that $S_2(c^f)=K_2^h$. Since in this scenario Firm 1 produces also with high-cost capacity, and that $S_1^l(\hat p) =K_1^l$ for some switching price $\hat p< c^f$. At this price $K_1^l = S_1^l(\hat p) = \tilde{\kappa}_1 (\hat p - c^l)$, which means that $\tilde{\kappa}_1 = \frac{K_1^l}{\hat p-c^l} \geq \frac{K_1^l}{c^f-c^l}$ since $c^f \geq \hat p$. 
Denote by
$$\tilde S_1(c^f)=\tilde{\kappa}_2(c^f-c^h)+\frac{\tilde{\kappa}_3}{c^f-c^h},$$ 
where $\tilde{\kappa}_2$ and $\tilde{\kappa}_3$ are the corresponding coefficients evaluated at $\tilde \kappa_1$. Since Firm 1 is producing with the high-cost technology at high prices, $\tilde S_1(c^f) - K_1^l >0. $ To support this supply of $\tilde S_1(c^f)$, it has to hold that $ \tilde S_1(c^f) -K_1^l\leq K_1^h  $, where the left-hand side is a function of the primitives $\{K_2^h,K_1^l,c^h,c^l,c^f\}$ but not of $K_1^h$. Therefore, let $K_1^h$ be small enough so that $ \tilde S_1(c^f) - K_1^l> K_1^h  $ then $\tilde S_1(c^f) > K_1^l + K_1^h$ as $p\to c^{f-}$ is infeasible: Firm~2 does not exhaust its capacity.

Therefore by Lemma \ref{bertrand}, we must have an alternative coefficient $\kappa_1'$ such that Firm 1 just exhausts its capacity as $p\to c^{f-}$. Lemma \ref{monotone} then implies $\kappa_1'<\tilde{\kappa}_1$: Firm~2 has extra capacity in the limit $p\to c^f$. Hence the equilibrium parameter $\kappa_1'$ is uniquely solved for using $\lim_{p\to {c^f}^-} S_1(p) = K_1^h+K_1^l$, which gives
\begin{align*}
        \frac{\kappa_1'}{2} \left(2 + \frac{\kappa_1'}{\alpha  - \kappa_1'}  \right) ( c^f-c^h) + \frac{(c^h-c^l)^2}{2} \frac{ \alpha -\kappa_1'}{c^f-c^h} &=  K_1^h +K_1^l,
    \end{align*}
by plugging in $\kappa_2$ and $\kappa_3$ from (\ref{coefficients}). This unique $\kappa_1'$ can be found numerically.

Joining the systems of equations (\ref{eq:system0}), (\ref{eq:system1}), (\ref{eq:system2}), and (\ref{eq:system3}), and using $\kappa_2$ and $\kappa_3$ from (\ref{coefficients}) evaluated at this $\kappa_1'$, the equilibrium supply schedules are
\[
S_1 =\begin{cases}
    0, & \text{if } p<c^h, \\
    \kappa_1' (p- c^l), & \text{if } p \in [c^h, \hat{p}),\\
    \kappa_2 (p-c^h) + \dfrac{\kappa_3}{p-c^h}, & \text{if } p\in [\hat{p}, c^f],
\end{cases}
\text{ and }
S_2 =\begin{cases}
    0, & \text{if } p<c^h, \\
    \kappa_1' (p- c^h), & \text{if } p \in [c^h, \hat{p}),\\
    \kappa_2 (p-c^h) - \dfrac{\kappa_3}{p-c^h}, & \text{if } p\in [\hat{p}, c^f),\\
    K_2^h, & \text{if } p = c^f,
\end{cases}
\]
where $\hat{p}$ is the price at which $S_1^l(\hat p) = K_1^l$. Continuity of $S_2$ at $\hat{p}$ (Lemma~\ref{smooth}), together with the second line of (\ref{eq:system2}) and (\ref{eq:system1}) evaluated at $\hat{p}$, gives
$\hat{p} = c^h + \frac{\alpha-\kappa_1'}{\kappa_1'} \cdot \left(c^h-c^l\right)$.

Now if we reduce $K_2^h$ to $K_2^h-\delta$ and  increase $K_1^h$ by  $\delta>0$ small enough, it will still hold that Firm 1 just exhausts its capacity at $p\to c^f$. Hence,
\[\frac{\kappa_1'}{2} \left(2 + \frac{\kappa_1'}{\alpha  - \kappa_1'}  \right) (c^f-c^h) + \frac{(c^h-c^l)^2}{2} \frac{ \alpha -\kappa_1'}{c^f-c^h} =  K_1^h +K_1^l+\delta . \]
  By Lemma \ref{monotone} we have $\kappa_1'$ is increasing in $\delta$. Hence, the market-wide production
\[
    S_1+S_2 =\begin{cases}
        0, & \text{ if } p<c^h, \\
        \kappa_1' (2p- c^h -c^l), 
        & \text{ if } p \in [c^h, \hat{p} ),\\
         \kappa_1'   \left( 1+\frac{ \alpha }{\alpha  - \kappa_1'}\right)  (p-c^h),  
    & \text{ if }   p\in [\hat{p} , c^f),\\
    K_2^h +K_1^h+K_1^l, & \text{ if } p = c^f,
    \end{cases}
    \]
increases at every price level as $\delta$ increases: the market clearing price drops.
\end{proof}

\subsection{Numerical Continuation with Three Strategic Firms}\label{apndx:u_shape_n2n3}

Figure~\ref{fig:u_shape_n2n3} is computed by constructing the
supply-function equilibrium directly for each transfer share $\kappa$.
For $N=2$ the calibration is Example~\ref{ex:scarcity}. For $N=3$,
Firm~1 has $K_1^l=5$, the two strategic high-cost rivals have initial
capacities $K_i^h(0)=4$, demand is $D=8$, and the proportional transfer
rule is $  K_i^h(\kappa)=(1-\kappa)K_i^h(0)$ and
$ K_1^{\mathrm{tot}}(\kappa)=K_1^l+\kappa\sum_{i=2}^N K_i^h(0)$.

The construction reproduces the two ODE regimes of
Proposition~\ref{prop:comp}, separated by the switching price $\hat{p}$
at which Firm~1 just exhausts $K_1^l$. For $p < \hat{p}$, Firm~1
uses only its low-cost capacity, and by symmetry each rival of Firm~1
submits a common schedule on its high-cost technology, denoted
$S_{-1}^h(p)$, that solves
\[
    (N-1)(p-c^l)(p-c^h)(S_{-1}^h)''(p)
    +(2N-3)(p-c^h)(S_{-1}^h)'(p)-S_{-1}^h(p)=0,
\]
with boundary condition $S_{-1}^h(c^h)=0$, and Firm~1 satisfies
$S_1(p)=(p-c^l)(N-1)(S_{-1}^h)'(p)$. The ODE has a regular singular
point at $p=c^h$; by the Frobenius theorem, its vanishing solution
admits the local series
\(
S_{-1}^h(p)=(p-c^h)+\sum_{k\geq 1}b_k(p-c^h)^{k+1},
\)
with coefficients $b_k$ pinned by substitution into the ODE
(e.g., $b_1=-(N-2)/[(N-1)(c^h-c^l)]$). We start the numerical
integration at $c^h+\varepsilon$ using this series and integrate up
to $c^f$. For prices in $[\hat{p},c^f)$,
Firm~1 uses high-cost capacity in addition, and with $u=p-c^h$ the
schedules take the closed form
\[
    S_1(u)=Au^{1/(N-1)}+\frac{(N-1)C}{u}, \qquad
    S_{-1}^h(u)=Au^{1/(N-1)}-\frac{C}{u}.
\]
For each $\kappa$, the script tries the four cases that arise from
combining (i)~which firm is capacity-constrained at $c^f$---Firm~1
(scarcity) or a rival (abundance)---and (ii)~whether the switching
price $\hat{p}$ lies strictly below $c^f$ or coincides with it (so that
only the first regime is active). It accepts the unique case in which
the supply schedules are non-decreasing, capacity inequalities hold,
and at least one firm is capacity-constrained at $c^f$. 

For $N\geq4$, the smooth all-interior continuation is not sufficient.
The Frobenius expansion of the unconstrained system near $c^h$ implies
that $S_1'(c^{h+})$ is proportional to $(3-N)$, and is therefore
negative for $N\geq4$. This would imply a decreasing leader schedule.
This does not rule out an $N$-firm SFE; rather, it indicates that the
monotonicity constraints $S_i'(p)\geq0$ must bind locally. Imposing these
constraints directly would replace the unconstrained ODE with
Kuhn--Tucker conditions involving additional multipliers and corner
active sets. We therefore use the $N=3$ numerical continuation only to
illustrate that the mechanism extends beyond two strategic firms, leaving
the full constrained $N\geq4$ characterization for future work.

\subsection{Equilibrium Verification for Example~2}\label{apndx:verification}

We verify that the supply functions stated in Example~2 for the scarcity case ($K_1^l = 5$) after the transfer ($\delta = 0.5$) constitute an SFE. Specifically, we show that $\tilde{S}_1$ is a best response to $\tilde{S}_2$; the reverse follows by symmetry of the ODE system.

Given $\tilde{S}_2(p)$, Firm~1's residual demand is
\[
D^R_1(p,\epsilon) = 
\begin{cases}
D(\epsilon) & p \in (0, c^h), \\
\max\!\big\{0,\; D(\epsilon) - \tfrac{30}{11}(p - c^h)\big\} & p \in [c^h, \hat{p}),\\[4pt]
\max\!\big\{0,\; D(\epsilon) - \tfrac{48}{11}(p - c^h) + \tfrac{25}{22(p - c^h)}\big\} & p \in [\hat{p}, c^f),\\
\max\!\big\{0,\; D(\epsilon) - 3.5\big\} & p = c^f, \\
0 & p > c^f.
\end{cases}
\]
Firm~1 maximizes $D^R_1(p,\epsilon) \cdot p - C_1(D^R_1(p,\epsilon))$ over $p$, where $C_1(q) = c^l \cdot \min(q, K_1^l) + c^h \cdot \max(0, q - K_1^l)$.

\textit{Case 1}: $D(\epsilon) \in (0, \tfrac{30}{11}]$. Residual demand is $D(\epsilon)$ at $p = c^h$ and zero for $p$ slightly above $c^h$. The optimal price is $p = c^h$ with $\tilde{S}_1 = D(\epsilon) = \tfrac{30}{11}(p - c^l)$ at $p = c^h$, matching the supply function.

\textit{Case 2}: $D(\epsilon) \in (\tfrac{30}{11}, \tfrac{80}{11})$. The FOC of the ex post problem on the linear segment of $D^R_1$ yields an interior solution with market-clearing price $p^* = \frac{D(\epsilon) + \kappa_1 c^h}{2 \kappa_1}$ where $\kappa_1 = \frac{30}{11}$. The corresponding optimal quantity is $\kappa_1 \cdot p^* = \kappa_1(p^* - c^l)$, which traces out $\tilde{S}_1(p) = \frac{30}{11}(p - c^l)$ on $[c^h, \hat{p})$.

\textit{Case 3}: $D(\epsilon) \in (\tfrac{80}{11}, \tfrac{96}{11})$. The FOC on the second segment of $D^R_1$ (where Firm~1 uses both technologies) yields an interior solution tracing out $\tilde{S}_1(p) = \frac{48}{11}(p - c^h) + \frac{25}{22(p - c^h)}$ on $[\hat{p}, c^f)$.

\textit{Case 4}: $D(\epsilon) \geq \tfrac{96}{11}$. Residual demand at $c^f$ exceeds Firm~1's total capacity, so $\tilde{S}_1(c^f) = K_1^l + \delta = 5.5$.

In each case, the ex post optimal $(p, q)$ pair lies on $\tilde{S}_1(p)$, confirming it is a best response. The verification for $\tilde{S}_2$ best-responding to $\tilde{S}_1$ is analogous by the symmetric structure of the FOC system (\ref{eq:foc_main}).

\subsection{Proof of Proposition \ref{prop:symmetry} (\textit{Symmetric Case})}\label{apndx:proof:sym}

The proofs rely on the Lemmas and definitions listed in Appendix \ref{apndx:primitives}. 

 \begin{proof} Since $S_i(p)=0$ for all $p<c^l$, it follows from  Lemma \ref{smooth}.3 that supply functions are differentiable except at the price that one of the firms exhausts low-cost capacity or at $p=c^l$. By Lemma \ref{bertrand}, at least one of the firms exhausts all capacities as $p\to c^f$, and $S_i(p)=K^h+K^l$ for $p\geq c^f$. 
By the FOCs in (\ref{eq:focs}) and (\ref{eq:focs-simple}): 
\begin{equation}\label{eq:sym_system0}
S_i(p) = S_{-i}^\prime(p) \cdot (p-c^\tau).
\end{equation}
Lemma \ref{changepoint} implies $\tau=h$ is used only when the capacity for $\tau=l$ is exhausted.  Denote by $\hat{p}$ the lowest price at which one of the firms exhausts low-cost capacity, i.e.  $\hat p :=\min \{p|S_i(p)\geq K^l\text{ for some }i=1,2\}$. On the interval for $p\in [c^l, \hat{p})$ both firms compete using only $\tau=l$, so that  $c^\tau = c^l$.  Above this price $\hat p$, some firm competes using $\tau=h$ with  $c^\tau = c^h$.

Solving for the system of equations (\ref{eq:sym_system0}) for $p \in [c^l,\hat{p})$  gives
\begin{equation}
    \begin{cases}\label{eq:sym_systemlow}
        S_1(p) &= \frac{\kappa_7}{p-c^l} + \kappa_6 \cdot (p-c^l), \text{if }  p \in [c^l,\hat{p}), \\
        S_2(p) &= - \frac{\kappa_7}{p-c^l} + \kappa_6 \cdot (p-c^l) ,  \text{if }  p \in [c^l,\hat{p}).
    \end{cases}
\end{equation}
Since supply functions are non-decreasing, we have $\kappa_7=0$. Therefore $K_1^l=K_2^l=K^l>0$ implies both firms exhaust capacity at the same $\hat p$ and $\kappa_6>0$.

Now solve the system of equations (\ref{eq:sym_system0}) for $p \in [\hat{p},c^f)$. The solution is similarly,
\begin{equation}
    \begin{cases}\label{eq:sym_system}
        S_1(p) &= \frac{\kappa_4}{p-c^h} + \kappa_5 \cdot (p-c^h), \text{if }  p \in [\hat{p},c^f), \\
        S_2(p) &= - \frac{\kappa_4}{p-c^h} + \kappa_5 \cdot (p-c^h) ,  \text{if }  p \in [\hat{p},c^f).
    \end{cases}
\end{equation}  
Due to profit maximization, we have $\hat p\geq c^h$. If $\hat p=c^h$, then $\kappa_4=0$ by monotonicity of $S_i$. However this violates monotonicity of $S_i$ since $\lim_{p\to c^h+} S_i(p)=0<\lim_{p\to c^h-}S_i(p)=K^l$. Therefore $\hat p>c^h$, and hence Lemma \ref{smooth}.1 implies $S_i$'s are continuous at $\hat p$ with $K^l=\lim_{p\to \hat p}S_i(p)$ for both $i$, and so $\kappa_4=0$.
 To pin down $\kappa_5$, note that if Firm $i$ has a weakly smaller $K^h$, it exhausts $K^l+K^h$ exactly at limit $p\to c^f-$. Hence, $K^l+K^h=\kappa_5 \cdot (c^f-c^h)$, meaning that $\kappa_5=\frac{K^l+K^h}{c^f-c^h}$. We can then determine $\hat{p}$ as the solution of:
\begin{align*}
    K^l&= \lim_{p\to \hat{p}^+} S_i(p) \Rightarrow \hat{p} = \frac{c^h K^h  + c^f K^l }{K^l+K^h}.
\end{align*}
Therefore, the optimal response in this interval is:
\begin{equation}\label{eq:sym_system1}
        S_i(p) = \frac{K^l+K^h}{c^f-c^h} \cdot (p-c^h), \;\;\;\text{if }  p \in \Bigg[\frac{c^h K^h  + c^f K^l }{K^l+K^h}, c^f\Bigg).
\end{equation}

Turning back to the interval $p\in[c^l, \frac{c^h K^h  + c^f K^l }{K^l+K^h})$, consider the system  (\ref{eq:sym_systemlow}) with $\kappa_7=0$. To pin down $\kappa_6$, notice that $S_i(p) = K^l$ as $p \to \frac{c^h K^h  + c^f K^l }{K^l+K^h}$. Therefore,
\begin{align*}
    K^l&= \lim_{p\to \hat{p}^-} S_i(p) \Rightarrow \kappa_6 = \frac{K^l(K^l+K^h)}{(c^h-c^l)K^h+(c^f-c^l)K^l}.
\end{align*}
Therefore, the optimal supply in this interval is:
\begin{equation}\label{eq:sym_system2}
        S_i(p) = \frac{K^l(K^l+K^h)}{(c^h-c^l)K^h+(c^f-c^l)K^l} \cdot (p-c^l), \text{if }   p\in\Bigg[c^l, \frac{c^h K^h  + c^f K^l }{K^l+K^h}\Bigg).
\end{equation}

Joining all the intervals, we find that $i$ supplies:
 \[S_i(p)=\begin{cases}
  0 & p\in [0, c^l), \\
  \frac{K^l(K^l+K^h)}{(c^h-c^l)K^h+(c^f-c^l)K^l}(p-c^{l})
& p\in[c^l, \frac{c^h K^h  + c^f K^l }{K^l+K^h}),  \\
  \frac{K^l+K^h}{c^f -c^h} (p-c^{h}),
& p\in [\frac{K^h c^h + K^l c^f }{K^l+K^h}, c^f),\\
 K^l+K^h,
& p\in [c^f, \infty).
 \end{cases}
 \]

 \vspace{1ex}
 \noindent\textbf{Reallocation.} We move $\delta$ units of high-cost capacity from Firm~2 to Firm~1. The above solution holds with $K^h$ replaced by $K^h-\delta$ on $[0,c^f)$. When $p\geq c^f$, the supply of Firm~1 jumps to  $K^l+K^h+\delta$, while Firm~2 remains at $K^l+K^h-\delta$.
 The new supply functions are:
 \[S_1(p)=\begin{cases}
  0, & p\in [0, c^l), \\
	  \frac{K^l(K^l+K^h-\delta)}{(c^h-c^l)(K^h-\delta)+(c^f-c^l)K^l}(p-c^{l}),
& p\in[c^l, \frac{c^h (K^h-\delta)  + c^f K^l }{K^l+K^h-\delta}),  \\
  \frac{K^l+K^h-\delta}{c^f -c^h} (p-c^{h}),
& p\in [\frac{(K^h-\delta) c^h + K^l c^f }{K^l+K^h-\delta}, c^f),\\
 K^l+K^h+\delta,
& p\in [c^f, \infty),
 \end{cases}
 \]
 and
  \[S_2(p)=\begin{cases}
  0, & p\in [0, c^l), \\
	  \frac{K^l(K^l+K^h-\delta)}{(c^h-c^l)(K^h-\delta)+(c^f-c^l)K^l}(p-c^{l}),
& p\in[c^l, \frac{c^h (K^h-\delta)  + c^f K^l }{K^l+K^h-\delta}),  \\
  \frac{K^l+K^h-\delta}{c^f -c^h} (p-c^{h}),
& p\in [\frac{(K^h-\delta) c^h + K^l c^f }{K^l+K^h-\delta}, c^f),\\
 K^l+K^h-\delta,
& p\in [c^f, \infty).
 \end{cases}
 \]

It is clear that the solution is unique, and the supplies decrease at every price level when $\delta$ increases. This concludes the proof. \end{proof}

\subsection{A Multi-Technology Extension of Maggi (1996)}\label{apndx:maggi_extension}

Maggi's capacity-price model treats capacity as an imperfect constraint: a firm can produce at marginal cost $c$ up to capacity $k$ and at the higher marginal cost $c+\theta$ above capacity, where $\theta$ measures the importance of capacity constraints. Maggi assumes a common $\theta$ across firms but notes that the extension to asymmetric $\theta_A,\theta_B$ is straightforward \citepsec[][ p.~241]{maggi1996strategic}. We use the second-stage price subgame in his Section~I.A, where capacities are taken as given and firms choose prices, but allow asymmetric costs and transferred capacity that retains its original marginal cost. To make the recipient the capacity leader, let the differentiated products have different intercepts: $q_A=a_A-p_A+\rho p_B$ and $q_B=a_B-p_B+\rho p_A$, with $0<\rho<1$. Firm $A$ produces at cost $c_A$ up to $k_A$ and at $c_A+\theta_A$ thereafter; firm $B$ produces at cost $c_B$ up to $k_B$ and at $c_B+\theta_B$ thereafter. A transfer $\delta$ gives $A$ a block of $B$'s capacity at cost $c_B$ and reduces $B$'s low-cost capacity to $k_B-\delta$.

\paragraph{Profit maximization.} After the transfer, firm $A$'s marginal cost is $c_A$ for $q_A\in[0,k_A]$, $c_B$ for $q_A\in(k_A,k_A+\delta]$, and $c_A+\theta_A$ for $q_A>k_A+\delta$.\footnote{Since $c_A<c_B<c_A+\theta_A$, firm $A$ fills the cheapest block first, then the transferred block at $c_B$, and only then pays the overflow cost $c_A+\theta_A$. The transferred block retains its original marginal cost $c_B$ rather than being rebadged at $A$'s cost $c_A$.} Firm $B$'s marginal cost is $c_B$ for $q_B\in[0,k_B-\delta]$ and $c_B+\theta_B$ thereafter. Each firm chooses $p_i$ to maximize $\pi_i(p_i,p_{-i};\delta)=p_iq_i-C_i(q_i;\delta)$, taking $p_{-i}$ as given. Profit is concave in own price, so the Bertrand first-order condition is sufficient whenever the induced quantity lies in the stated cost tier.

\paragraph{Argument.} Let $H_A=c_A+\theta_A$ and fix active marginal costs $(m_A,m_B)$. The first-order conditions give
\[
p_A=\frac{2(a_A+m_A)+\rho(a_B+m_B)}{4-\rho^2},\qquad
p_B=\frac{\rho(a_A+m_A)+2(a_B+m_B)}{4-\rho^2},
\]
with $q_i=p_i-m_i$. Prices are increasing in both active costs. Hence prices fall when active costs move from $(H_A,c_B)$ to $(c_B,c_B)$, and, if $\theta_B>2(H_A-c_B)/\rho$, both prices under $(c_B,c_B+\theta_B)$ exceed those under $(H_A,c_B)$. It is therefore enough to ensure that the relevant cost tiers are active. A sufficient set of strict inequalities, with the recipient larger than the donor, is
\[
\begin{gathered}
q_B(H_A,c_B)<k_B<k_A<q_A(H_A,c_B),\\
0<q_A(c_B,c_B)-k_A<\delta_s<k_B-q_B(c_B,c_B),\\
\max\{\delta_s,\;q_A(c_B,c_B+\theta_B)-k_A,\;k_B-q_B(c_B,c_B+\theta_B)\}<\delta_\ell<k_B .
\end{gathered}
\]
Then at $\delta=0$ firm $A$ is constrained and firm $B$ is not; at $\delta_s$, $A$ uses the transferred block while $B$ remains unconstrained, so prices fall; at $\delta_\ell$, $B$ is constrained, so prices rise. These sufficient conditions are feasible. For example, take $\rho=3/4$, $(a_A,a_B)=(18,4)$, $(c_A,c_B,\theta_A,\theta_B)=(0,1,4,11)$, $(k_A,k_B)=(19/2,17/2)$, $\delta_s=2$, and $\delta_\ell=15/2$. The induced prices are $p^0=(764/55,424/55)$, $p^s=(668/55,388/55)$, and $p^\ell=(160/11,148/11)$; the corresponding quantities satisfy all three displayed inequalities, and $k_A>k_B$. The inequalities are strict, so the same pattern holds in a neighborhood of these primitives. In Maggi's full game with positive capacity costs, firms typically choose capacity to avoid overflow production or unused slack; our construction therefore uses the price subgame and applies directly to the full game when capacities are inherited, sunk, free, or supported by reserve obligations.

\subsection{Daily Ex Post Bellman Optimality}\label{apndx:bellman_expost}

This section provides sufficient conditions for the ex post Bellman optimality property in Section~\ref{s:model}. Absorbing the continuation value into a total economic cost reduces the dynamic problem to the static form of \citesec{klemperer1989supply}, and ex post optimality follows from typical concavity conditions.

The following notation simplifies Section~\ref{s:model}. Fix a firm $i$ and suppress the hour subscript $h$. The firm's supply function is $S_i(p)$, with production cost $C(S_i(p))$ and residual demand $D^R(p, \epsilon_t)  $, where $\epsilon_t$ is an additive demand shock with full support. The market clears at price $p(\epsilon_t)$, and the water stock evolves as $w_{i,t+1} = w_{i,t} - S_i(p(\epsilon_t)) + \xi_{i,t}$, where $\xi_{i,t} \sim f(\cdot|\mathbf Z_t)$ denotes random inflows. Throughout, firm $i$ take opponents' strategies and its own value function $V_i$   as given.

\noindent\textbf{Value function and economic cost.}
By (\ref{eq:inter.opt}),   firm $i$ chooses $S_i(\cdot)$ to maximize
\begin{align*}
	V_i(w_t) = \max_{S_i(\cdot)}\;
	\mathbb E_\epsilon \Big[  S_i(p(\epsilon_t))\, p(\epsilon_t) - C(S_i (p(\epsilon_t)))
	+\beta \int_{\mathbb W} V_i(w_{t}-S_i(p(\epsilon_t))+\xi)\, f( \xi|\mathbf Z_t)\, \mathrm d\xi \Big].
\end{align*}
Following   \citesec{klemperer1989supply},  market must clear at $S_i(p(\epsilon_t))= D^R(p, \epsilon_t)$. Hence the ex post optimization is equivalent to choosing an optimal  $p$ for each realization $\epsilon_t$. Define the \emph{total economic cost}  that captures both the production cost $C(q)$ and the change in continuation value from producing $q$ today:
\begin{equation}\label{eq:total_econ_cost}
	\tilde C(q;\, w_t, \mathbf Z_t )
	\equiv C(q)-\beta \int_{\mathbb W} V_i(w_{t} - q+\xi)\, f( \delta|\mathbf Z_t)\, \mathrm d\xi.
\end{equation}
At market clearing, $q = D^R(p, \epsilon_t)$, the realized daily payoff at price $p$ becomes
\begin{equation}\label{eq:daily_payoff}
	\Psi(p;\, \epsilon_t, w_t, \mathbf Z_t) = D^R(p,\epsilon_t) \cdot p - \tilde C(D^R(p,\epsilon_t);\, w_t, \mathbf Z_t ),
\end{equation}
which takes the same form as the static payoff in \citesec{klemperer1989supply}. Assuming all integrals are well-defined, the firm's ex ante problem is then
\begin{equation}\label{eq:ex_ante}
	V_i(w_t) = \max_{S_i(\cdot)} \; \mathbb{E}_\epsilon\!\left[\Psi(p(\epsilon_t);\, \epsilon_t, w_t, \mathbf Z_t)\right].
\end{equation}

\begin{assumption}\label{ass:concavity}
	For all $\epsilon_t, w_t, \mathbf Z_t$, it holds that \emph{(i)} the residual demand $D^R(p, \epsilon_t)$ is twice differentiable, strictly decreasing, and weakly concave in $p$; \emph{(ii)} the total economic cost $\tilde C(D^R(p,\epsilon_t);\, w_t, \mathbf Z_t)$, viewed as a function of $p$, is twice differentiable and convex; \emph{(iii)} the realized daily payoff $\Psi(p;\, \epsilon_t, w_t, \mathbf Z_t)$ satisfies Inada conditions.
\end{assumption}

\noindent This assumption is sufficient for Bellman ex post optimality for the focal firm:

\begin{proposition}\label{prop:bellman_expost}
	Under Assumption~\ref{ass:concavity}, given $D^R$ and $V_i$, there exists an increasing supply function $S_i(p^*) = q^*$, depending on $(w_t, \mathbf Z_t)$, that is an ex post optimal solution to~(\ref{eq:ex_ante}).
\end{proposition}

\begin{proof}
	Fix $(w_t, \mathbf Z_t)$ and suppress them from the notation. Since $\epsilon_t$ is additive, write $D^R(p,\epsilon_t) = D^R(p) + \epsilon_t$, so $\partial D^R / \partial \epsilon_t = 1$.
	
	\emph{Step 1: unique $p^*$, increasing in $\epsilon_t$.} Uniqueness is immediate. Differentiability implies the maximizer $p^*(\epsilon_t)$ satisfies
	\begin{equation}\label{eq:FOC_app}
		\frac{\partial \Psi}{\partial p} \;=\; D^{R\prime}(p)\!\left[p - \tilde C'(D^R + \epsilon_t)\right] + (D^R(p) + \epsilon_t) \;=\; 0.
	\end{equation}
	Holding $p$ fixed and differentiating in $\epsilon_t$,
	\begin{equation}\label{eq:cross_partial}
		\frac{\partial^2 \Psi}{\partial p\, \partial \epsilon_t} \;=\; -D^{R\prime}(p) \cdot \tilde C''(D^R + \epsilon_t) + 1 \;\geq\; 1 \;>\; 0,
	\end{equation}
	so $p^*(\epsilon_t)$ is strictly increasing.
	
	\emph{Step 2: $q^*$ increasing in $\epsilon_t$.}
	Totally differentiating~(\ref{eq:FOC_app}) along the optimal path $p = p^*(\epsilon_t)$, $q^* = D^R(p^*) + \epsilon_t$,
	\begin{equation}\label{eq:total_diff}
		\left[D^{R\prime\prime}(p^* - \tilde C') + D^{R\prime}\right]\frac{dp^*}{d\epsilon_t} \;+\; \left[1 - D^{R\prime}\, \tilde C''\right]\frac{dq^*}{d\epsilon_t} \;=\; 0.
	\end{equation}
	By Assumption~\ref{ass:concavity}, $1 - D^{R\prime}\tilde C'' > 0$ and $D^{R\prime\prime}(p^* - \tilde C') + D^{R\prime} < 0$, so $dq^*/d\epsilon_t > 0$.
	
Since both $p^*$ and $q^*$ are increasing in $\epsilon_t$, the locus $(p^*(\epsilon_t), q^*(\epsilon_t))$ defines an increasing supply  $S_i(p^*) = q^*$. This supply function clears the market at the ex post optimal $p^*(\epsilon_t)$ for every $\epsilon_t$, so it also solves the ex ante problem~\eqref{eq:ex_ante}.
\end{proof}

\FloatBarrier
\setcounter{table}{0}
\renewcommand{\thetable}{B\arabic{table}}
\setcounter{figure}{0}
\renewcommand{\thefigure}{B\arabic{figure}}
\setcounter{equation}{0}
\renewcommand{\theequation}{B\arabic{equation}}
\section{Omitted Figures}

\begin{figure}[!htbp]
	\caption{Location of dams}\label{fig:omitted}
	\centering
	\includegraphics[width=0.6\textwidth]{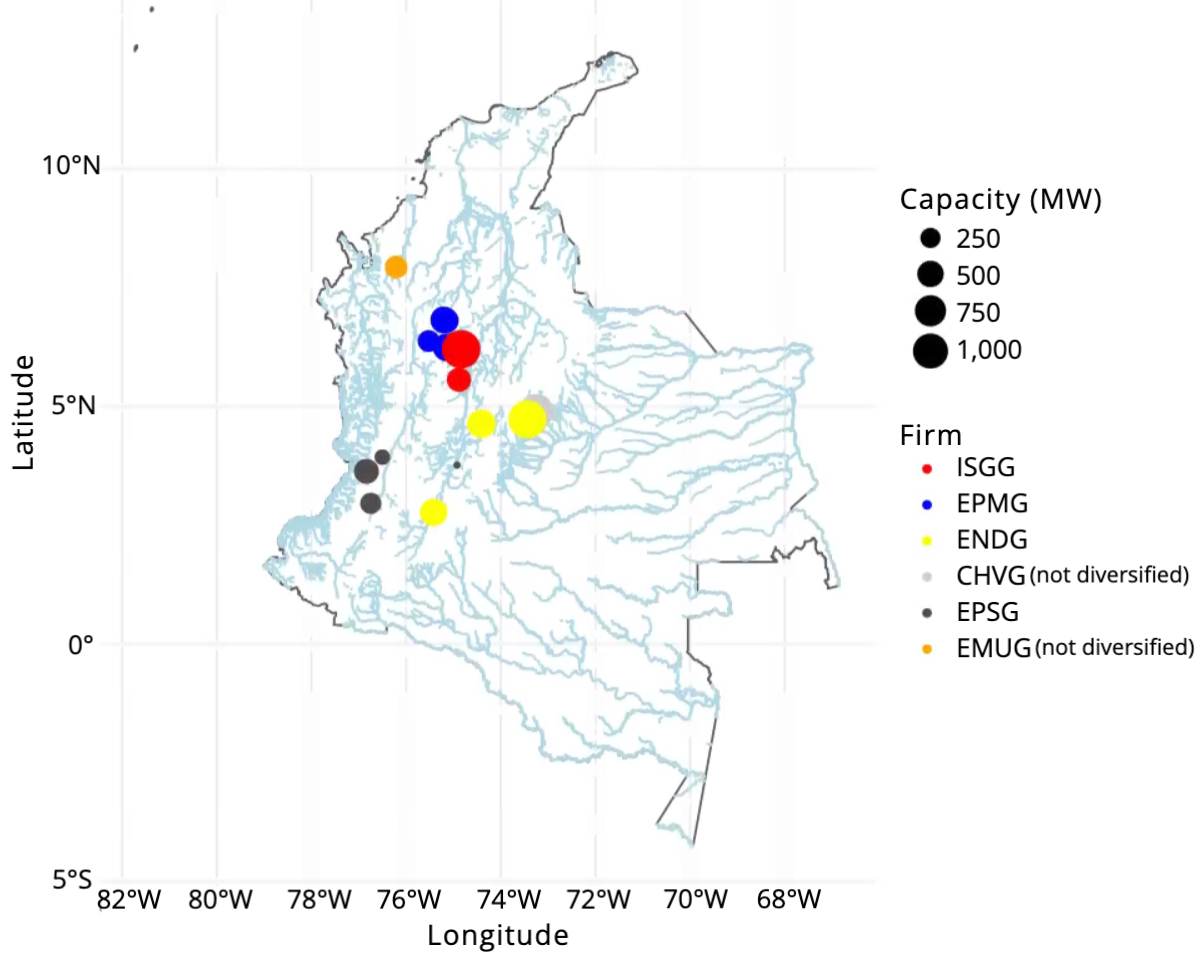}
	\label{fig:dam_locations}
	\begin{minipage}{1 \textwidth}
		{\footnotesize Notes: Location of Colombia dams by firm (color) and capacity (size). Colombia’s West border is
			with the Pacific Ocean while rivers streaming East continue through Brazil and Venezuela. To give a
			sense of the extension of Colombia, its size is approximately that of Texas and New Mexico combined}
	\end{minipage}
\end{figure}

\begin{figure}[!htbp]
	\caption{Supplies of Firm~2 and fringe firms before and after the transfer}
	\label{apndx:fig:cases_competitors}
	\centering
	\begin{subfigure}{0.44\textwidth}
		\centering
		\includegraphics[width=\textwidth]{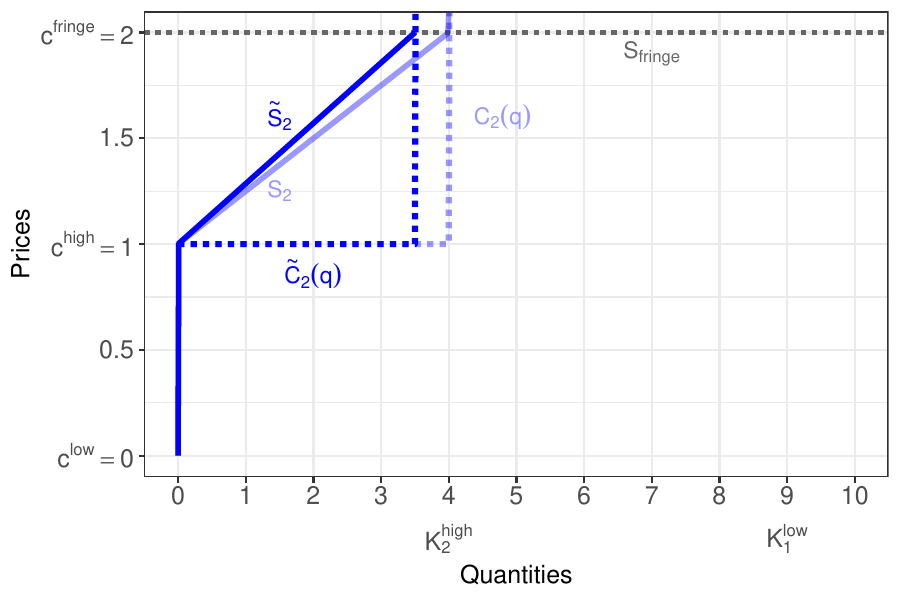}
		\captionsetup{width=0.9\linewidth, justification=centering}
		\caption{Abundance scenario \\ $K_1 = (9, 0, 0) \, \& \,  K_2 = (0, 4, 0) $}
	\end{subfigure}\hfill
	\begin{subfigure}{0.44\textwidth}
		\centering
		\includegraphics[width=\textwidth]{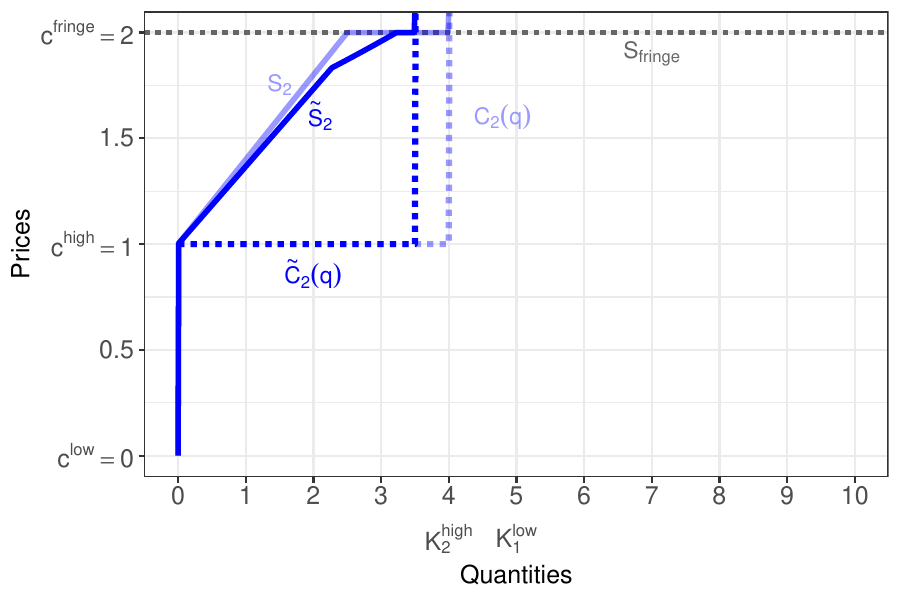}
		\captionsetup{width=0.9\linewidth, justification=centering}
		\caption{Scarcity scenario \\ $K_1 = (5, 0, 0) \, \& \,  K_2 = (0, 4, 0)$}
	\end{subfigure}
	\begin{minipage}{1 \textwidth}
		{\footnotesize Notes: Each panel illustrates the equilibrium supply of Firm~2 and the supply of fringe firms (gray dotted lines) under abundance in Panel (a) and scarcity in Panel (b) using the parameters as in Figures \ref{fig:cases} and \ref{fig:transfer}, respectively. The cost of Firm~2 is the dotted blue line. Solid (shaded) lines refer to Firm~2's costs and supply after (before) the capacity transfer of 0.5 units from Firm~2 to Firm 1.  }
	\end{minipage}
\end{figure}

\setcounter{table}{0}
\renewcommand{\thetable}{C\arabic{table}}
\setcounter{figure}{0}
\renewcommand{\thefigure}{C\arabic{figure}}
\setcounter{equation}{0}
\renewcommand{\theequation}{C\arabic{equation}}

\section{Inflow Forecasts}\label{apndx:forecast}
First, we run the following ARDL model using the weekly inflows of each unit $j$ as dependent variable,
\begin{equation}\label{eq:ARDL}
\delta_{j,t}	=	\mu_{0}+\sum_{1\leq p\leq t}^P\alpha_{p}\xi_{j,t-p}+\sum_{1\leq q\leq t}^Q\mathbf{\beta}_{q}\mathbf{x}_{j,t-q}+\epsilon_{j,t} \, \, \forall j
\end{equation}
We denote by $\xi_{j,t}$ the inflow to the focal dam in week $t$. $\mathbf{x}_{j,t}$ is a vector that includes the average maximum temperature and rainfalls in the past week at dam $j$, and information about the future probabilities of el ni\~no. We average the data at the weekly level to reduce the extent of autocorrelation in the error term. Importantly for forecasting, the model include only lagged regressors.

For forecasting, we first determine the optimal number of lags for $P$ and $Q$ for each dam $j$ using the BIC criterion. As standard, we set $Q=P$ in (\ref{eq:ARDL}) to reduce the computation burden. For an h-ahead week forecast, we estimate:
\begin{equation}\label{eq:forecast}
\xi_{j,t+h}	=	\hat{{\mu}}_{0}+\hat{\alpha}_{1}\xi_{t}+\cdots+\hat{\alpha}_{P}\xi_{j,t-P+1}+\sum_{k=1}^{K}\hat{\beta}_{1,k}x_{j,t,k}+\cdots+\hat{\beta}_{q,k}x_{j,t-Q+1,k}+\epsilon_{t}, 
\end{equation}
where $K$ denotes the number of control variables in $\textbf{x}_{j,t-q}$. 

For each week $t$ of time series of dam $j$, we estimate (\ref{eq:forecast}) for $h\in\{4,8,12,16,20\}$ weeks ahead (i.e., for each month up to five months ahead)  using only data for the 104 weeks (2 years) before week $t$. In the analysis, we only keep dams for which we have at least 2 years of data to perform the forecast. Dropping this requirement does not affect the results.

We verify the quality of the ARDL forecasts through standard diagnostics. Autocorrelation function plots and Ljung-Box tests for the largest dams in Colombia confirm that the residuals $\epsilon_t$ in (\ref{eq:forecast}) exhibit no significant autocorrelation. At the firm level, we estimate a similar ARDL model with explanatory variables averaged over months to capture seasonal heterogeneity, controlling for El Ni\~no probabilities. The error term $\epsilon_{j,t}$ is modeled through a Pearson Type IV distribution, which accommodates the asymmetry between dry and wet seasons common in hydrological data. The Pearson IV distribution fits well for all four largest diversified firms. Diagnostic plots are available upon request.


\setcounter{table}{0}
\renewcommand{\thetable}{D\arabic{table}}
\setcounter{equation}{0}
\renewcommand{\theequation}{D\arabic{equation}}
\setcounter{figure}{0}
\renewcommand{\thefigure}{D\arabic{figure}}
\section{Additional Empirical Results}\label{apndx:additional_results}

\begin{figure}[!htbp]
    \caption{Robustness of the U-shape}
     \label{fig:empiricalU_robustness}
    \centering
    \begin{subfigure}{0.44\textwidth}
        \centering
        \includegraphics[width=\textwidth]{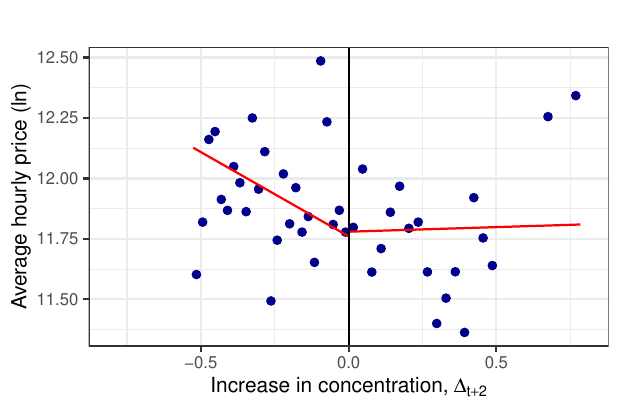}
        \captionsetup{width=0.9\linewidth, justification=centering}
        \caption{Two-month-ahead forecasts ($\Delta_{t+2}$), daily}
    \end{subfigure}\hfill
    \begin{subfigure}{0.44\textwidth}
        \centering
        \includegraphics[width=\textwidth]{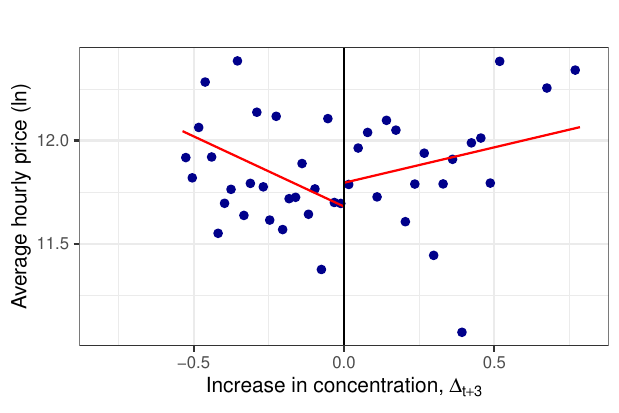}
        \captionsetup{width=0.9\linewidth, justification=centering}
        \caption{Three-month-ahead forecasts ($\Delta_{t+3}$), daily}
    \end{subfigure}
\begin{minipage}{1 \textwidth}
{\footnotesize Notes: Binned scatter plots of log market prices against $\Delta_{t+l}$, as in Figure~\ref{fig:empiricalU}. Both panels use daily data and residualize month-by-hour, year, and day-of-week-by-hour fixed effects. Bootstrap 95\% CIs (300 repetitions). Panel~(a): left slope $[-0.727,-0.660]$, right slope $[0.012,0.063]$, slope change $[0.678,0.779]$. Panel~(b): left slope $[-0.703,-0.647]$, right slope $[0.320,0.372]$, slope change $[0.982,1.056]$. \par}
\end{minipage}
\end{figure}

 Figure~\ref{fig:empiricalU_robustness} replicates the main result from Figure~\ref{fig:empiricalU} in daily data. Panel~(a) uses two-month-ahead forecasts ($\Delta_{t+2}$) instead of three-month-ahead. Panel~(b) uses three-month-ahead forecasts ($\Delta_{t+3}$), matching the horizon in the main figure.

\begin{figure}[!htbp]
    \caption{U-shape with forecast-continuous running variable}
    \label{fig:empiricalU_continuous}
    \centering
    \begin{subfigure}{0.44\textwidth}
        \centering
        \includegraphics[width=\textwidth]{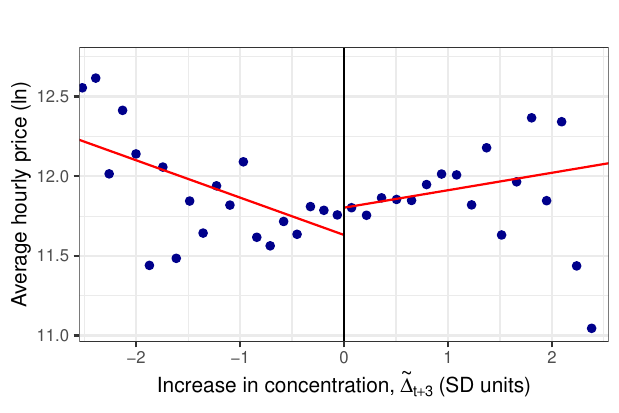}
        \captionsetup{width=0.9\linewidth, justification=centering}
        \caption{Three-month-ahead forecasts ($\widetilde{\Delta}_{t+3}$), daily}
    \end{subfigure}\hfill
    \begin{subfigure}{0.44\textwidth}
        \centering
        \includegraphics[width=\textwidth]{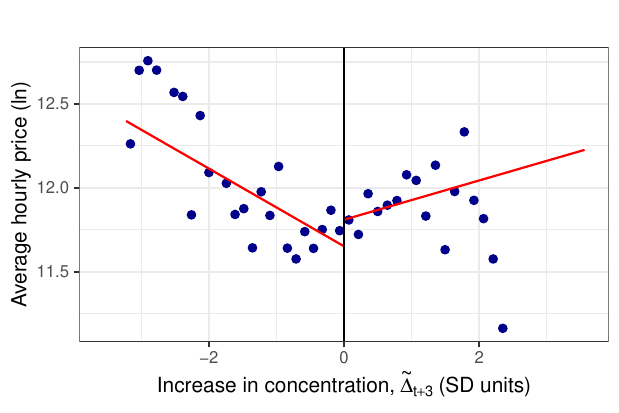}
        \captionsetup{width=0.9\linewidth, justification=centering}
        \caption{Three-month-ahead forecasts ($\widetilde{\Delta}_{t+3}$), weekly}
    \end{subfigure}
\begin{minipage}{1 \textwidth}
{\footnotesize Notes: Binned scatter plots of log market prices against $\widetilde{\Delta}_{t+3}$, as defined in \eqref{eq:tilde_delta}. The horizontal axis is in units of dam-level forecast standard deviations, capacity-share-squared weighted and summed across the four focal firms. Panel~(a): daily data, FEs are month-by-hour, year, and day-of-week-by-hour. Bootstrap 95\% CIs (200 repetitions): left slope $[-0.240,-0.226]$, right slope $[0.100,0.116]$, slope change $[0.328,0.350]$. Panel~(b): weekly data, FEs are month-by-hour and year. Bootstrap 95\% CIs: left slope $[-0.238,-0.206]$, right slope $[0.095,0.130]$, slope change $[0.308,0.362]$. \par}
\end{minipage}
\end{figure}

\paragraph{Using continuous forecasts.} Figure~\ref{fig:empiricalU_continuous} replicates Figure~\ref{fig:empiricalU} using the standardized dam-level forecast directly in place of the $\pm 1$ standard deviation indicator. This makes the running variable the continuous counterpart of $\Delta_{t+l}$ in the first-order expansion of the footnote to equation~(\ref{eq:delta}): there, a standardized adverse forecast lowers a firm's effective capacity in proportion to its size, so to first order the capacity-based $\text{HHI}$ moves with the $s_i^2$-weighted sum of standardized forecasts. The main-text $\Delta_{t+l}$ coarsens each dam's standardized forecast to its $\pm 1$ standard deviation sign; here we keep the standardized magnitude itself. Let $\tilde{\xi}_{j,t+l}$ denote the $l$-month-ahead forecast deviation at dam $j$, standardized to mean zero and unit variance over the dam's own time series. We aggregate to the firm level by summing across the $n_i$ dams of firm $i$,
$\tilde{\xi}^i_{t+l} \;\equiv\; \sum_{j \in i} \tilde{\xi}_{j,t+l}$,
and to the market level with the same capacity-share-squared weight as in (\ref{eq:delta}):
\begin{equation}\label{eq:tilde_delta}
	\widetilde{\Delta}_{t+l} \;\equiv\; \sum_{i} \tilde{\xi}^i_{t+l} \times s_{i,t-1}^2.
\end{equation}
Under this specification, each dam contributes a continuous value in units of its own forecast standard deviation, rather than a $\{-1, 0, +1\}$ indicator as in the main text. The firm-level sum $\tilde{\xi}^i_{t+l}$ has mean zero by construction, and variance between $n_i$ (independent dams) and $n_i^2$ (perfectly correlated dams). Empirically, across the four focal firms, within-firm correlations range from near zero to roughly $0.5$, placing the firm-level SD closer to $\sqrt{n_i}$ than to $n_i$; a firm with more dams nonetheless contributes a larger absolute signal. This scaling changes the units of the slope coefficients, but not the qualitative comparison of left and right slopes at the cutoff. 

The U-shape is preserved; Panel~(a) uses daily data, which are noisier than the weekly data in Panel~(b). Table~\ref{tab:empiricalU_piecewise} reports the daily estimates of equation~(\ref{eq:rd_piecewise}) using both the binary running variable (Columns~1 and~3) and the continuous running variable (Columns~2 and~4). 

\begin{figure}[!htbp]
    \caption{Responses to forecast errors}
     \label{fig:forecast_errors}
    \centering
        \includegraphics[width=0.48\textwidth]{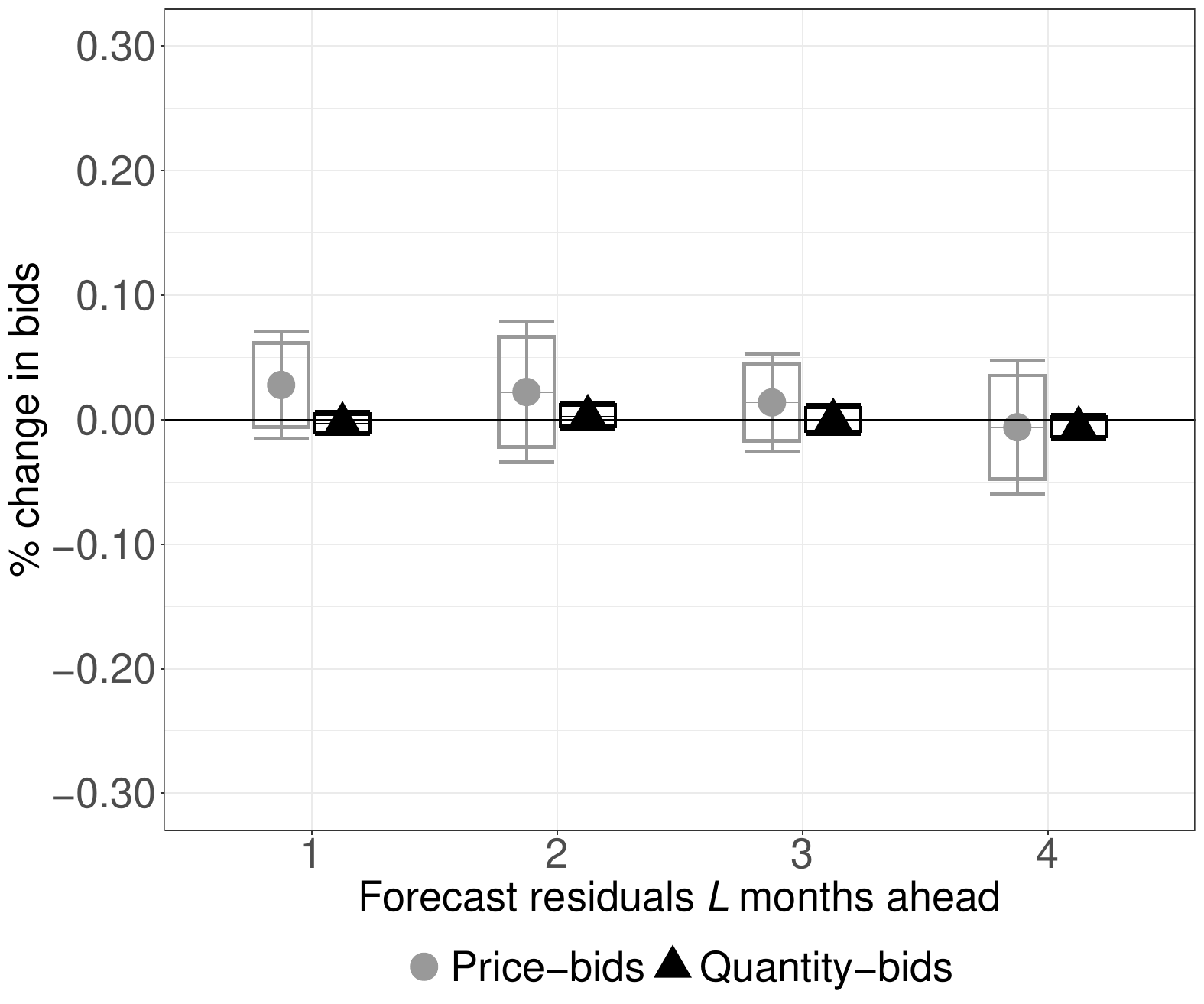}
\begin{minipage}{1 \textwidth}
{\footnotesize Notes: Estimates of $\{\beta_\ell\}_\ell$ from separate regressions using forecast residuals as regressors, one for each month-ahead forecast ($\ell=1,2,3,4$), with price bids (gray) and quantity bids (black) as dependent variables. Error bars (boxes) report 95\% (90\%) CIs. FEs: unit-by-month, firm-by-year, hour, week-by-year. SEs clustered by unit, month, and year. \par}
\end{minipage}
\end{figure}

\begin{table}[!htbp]
	\centering
	\caption{Piecewise-linear regression estimates at $l=3$}
	\label{tab:empiricalU_piecewise}
	\small
		\input{Tables/empirical_evidence/table_D1.tex}
	\begin{minipage}{0.95\textwidth}
		\vspace{0.3ex}
			\footnotesize Notes: Daily estimates of \eqref{eq:rd_piecewise} at $l=3$; Newey--West standard errors in parentheses (lag length = 96 hourly observations, or 4 days). Columns~(1) and~(3) use the $\{-1,0,1\}$ net-adverse running variable \eqref{eq:delta}, while Columns~(2) and~(4) use the standardized forecast \eqref{eq:tilde_delta}. All specifications control for $\text{HHI}_{t-1}$, the aggregate forecast (either through  $\sum_i \text{net adverse}_{i,t+3}$ or equation \ref{eq:tilde_delta}), and the lagged vector $\mathbf{x}_{t-1}$ (log demand, log forward contracts, log price, and log total available capacity) interacted with $\mathds{1}[\Delta_{t+3}>0]$. At the weekly level, excluding Day-of-the-week FEs, the slope change is $+0.270$ $(0.115)$ and $+0.064$ $(0.032)$ for the indicator and continuous specifications, and $+0.256$ $(0.117)$ and $+0.042$ $(0.033)$ with the $\text{HHI}_{t-1}\times\text{forecast}_{t+l}$ control (slope coefficient not reported due to space constraints). 
		
	\end{minipage}
\end{table}
\paragraph{Information content of inflow forecasts. }
To verify that our forecasts capture information material to firms, we regress bids on forecast \textit{residuals} ($inflow_{ij,t+\ell} - \widehat{inflow}_{ij,t+\ell}$) rather than on forecasts. If bids responded to forecast errors, our model would miss information that firms act on. Figure \ref{fig:forecast_errors} shows that this is not the case.

\begin{figure}[!htbp]
    \caption{Responses to competitors’ inflow forecasts}
     \label{fig:comp_responses}
    \centering
    \begin{subfigure}{0.44\textwidth}
        \centering
        \includegraphics[width=\textwidth]{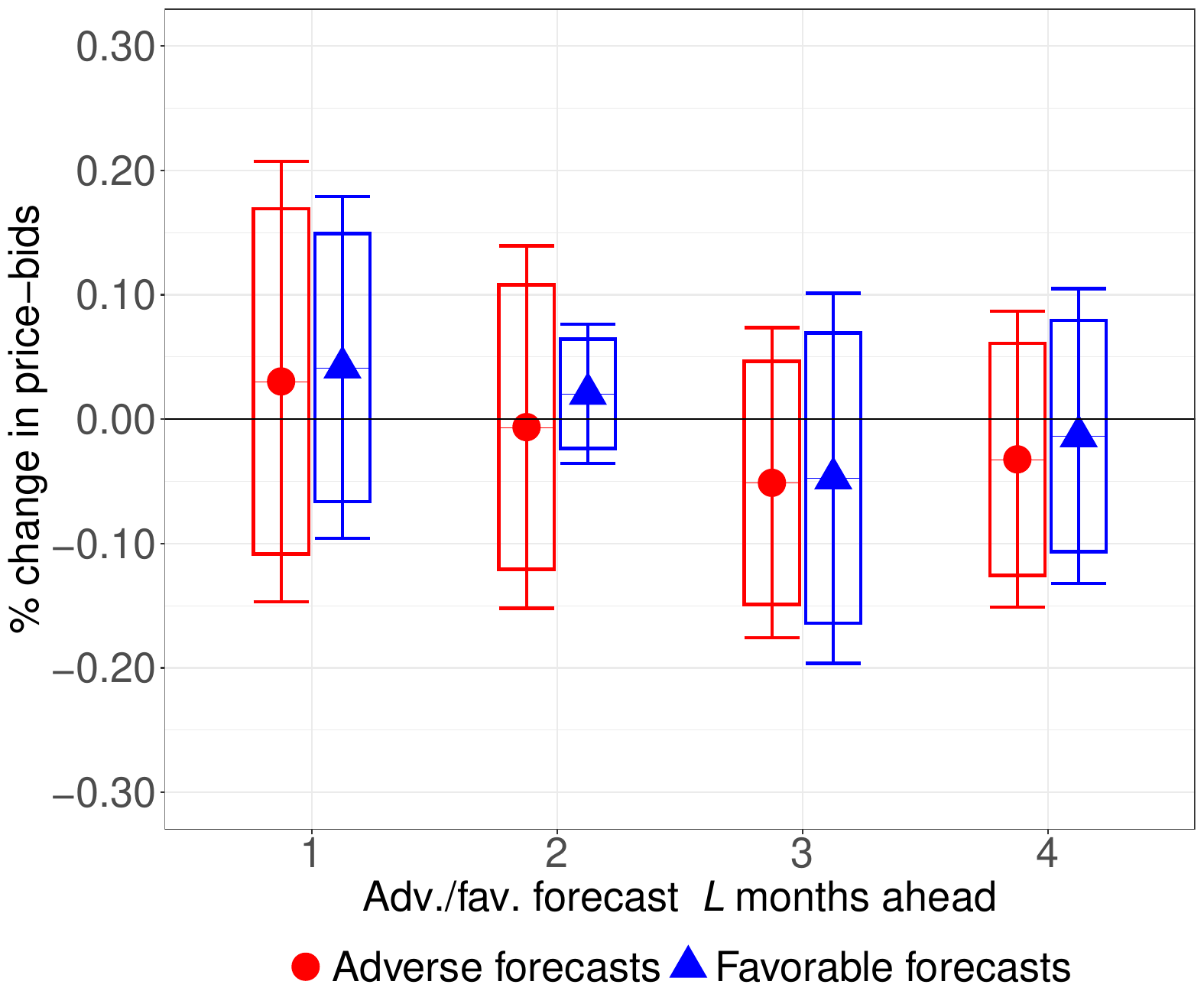}
        \captionsetup{width=0.9\linewidth, justification=centering}
        \caption{Price bids}
    \end{subfigure}\hfill
    \begin{subfigure}{0.44\textwidth}
        \centering
        \includegraphics[width=\textwidth]{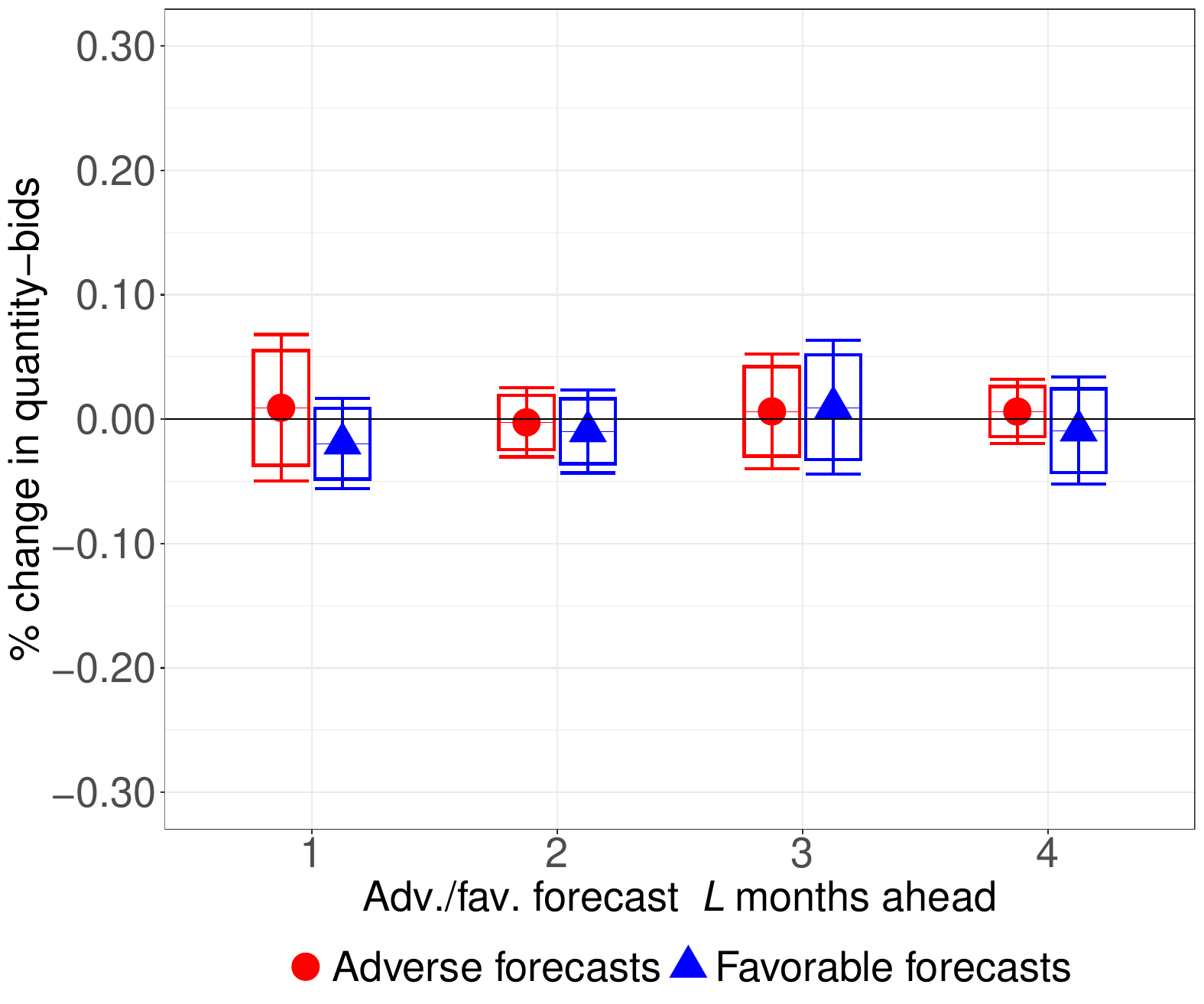}
        \captionsetup{width=0.9\linewidth, justification=centering}
        \caption{Quantity bids}
    \end{subfigure}
\begin{minipage}{1 \textwidth}
{\footnotesize Notes: Estimates of $\{\beta^{low}_\ell\}$ (red) and $\{\beta^{high}_\ell\}$ (blue) from (\ref{eq:as_inflow}) using competitors’ forecast indicators, at one to four months ahead. Error bars (boxes) report 95\% (90\%) CIs. FEs: unit-by-month, firm-by-year, hour, week-by-year. SEs clustered by unit, month, and year. \par}
\vspace{1cm}
\end{minipage}
\end{figure}

\paragraph{Responses to competitors’ inflow forecasts.} 
To test whether units react to competitors’ forecasts, we replace own-forecast indicators in (\ref{eq:as_inflow}) with aggregate competitors’ firm-level forecast indicators. Figure~\ref{fig:comp_responses} shows no significant response: magnitude changes are within $\pm 1\%$ and joint significance tests do not reject zero at standard levels.

\FloatBarrier

\setcounter{table}{0}
\renewcommand{\thetable}{E\arabic{table}}
\setcounter{figure}{0}
\renewcommand{\thefigure}{E\arabic{figure}}
\section{Model Estimation and Counterfactuals}\label{apndx:tabs}

\paragraph{Smoothing variables.}\label{apndx:smooth}
Following \citesec{wolak2007quantifying} and \citesec{ryan2021competitive}, we smooth supply schedules using the standard normal CDF as kernel $\mathcal{K}$, with bandwidth $bw$ equal to 10\% of the expected price. Let $\mathcal{J}_i=\mathcal{D}_i\cup\mathcal{T}_i$ denote firm $i$'s set of generation units. The smoothed supply of firm $i$'s competitors is $\tilde{S}_{-iht}(p) = \sum_{m \neq i} \sum_{j\in\mathcal{J}_m} q_{mjht}\, \mathcal{K}\!\left(\frac{p - b_{mjt}}{bw}\right)$, and the smoothed residual demand is $\tilde{D}^R_{iht}(p, \epsilon) = D_{ht}(\epsilon) - \tilde{S}_{-iht}(p)$. Firm $i$'s own smoothed supply is $\tilde{S}_{iht}(p) = \sum_{j\in\mathcal{J}_i} q_{ijht}\, \mathcal{K}\!\left(\frac{p - b_{ijt}}{bw}\right)$. The key derivatives entering the FOC (\ref{eq:FOC.q}) are:
\begin{equation*}
	\frac{\partial \tilde{S}_{iht}}{\partial q_{ijht}} = \mathcal{K}\!\left(\frac{p - b_{ijt}}{bw}\right), \qquad
	\frac{\partial \tilde{S}_{iht}}{\partial b_{ijt}} = -q_{ijht}\, \mathcal{K}'\!\left(\frac{p - b_{ijt}}{bw}\right)\frac{1}{bw}, \qquad
	\frac{\partial p_{ht}}{\partial q_{ijht}} = \frac{\mathcal{K}\!\left(\frac{p - b_{ijt}}{bw}\right)}{\frac{\partial \tilde{D}^R_{iht}}{\partial p_{ht}} - \frac{\partial \tilde{S}_{iht}}{\partial p_{ht}}}.
\end{equation*}
Technology-specific derivatives $\partial \tilde{S}_{iht}^\tau / \partial q_{ijht}$ and $\partial \tilde{S}_{iht}^\tau / \partial b_{ijt}$ are defined analogously, restricting the sums to units of technology $\tau$.

\input{Tables/structural_model/res2_p4_rob}

\input{Tables/structural_model/res2_p5_rob_normal}

\begingroup
\setlength{\intextsep}{1pt plus 1pt minus 1pt}%
\setlength{\textfloatsep}{0pt plus 1pt minus 1pt}%
\setlength{\abovecaptionskip}{2pt}%
\setlength{\belowcaptionskip}{2pt}%

\input{Figures/Model_fit/Table_Price_Diff_Wide}

\input{Figures/Model_fit/Regressions_Fit_Levels}

\FloatBarrier

\vspace{7pt}

\refstepcounter{figure}%
\label{fig:counter_price_myopic}%
\begin{center}
\captionsetup{type=figure}%
\captionof{figure}{Counterfactual capacity transfers using observed water stocks}
\begin{minipage}{0.44\textwidth}
	\centering
	\includegraphics[width=\textwidth]{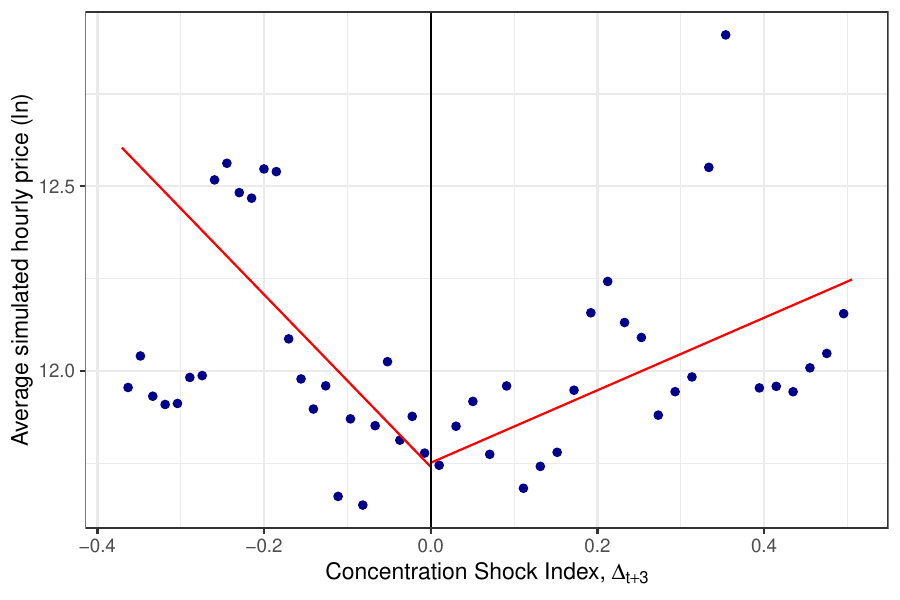}
	\subcaption{Simulated U-shape (cf.\ Figure~\ref{fig:empiricalU})}
\end{minipage}\hfill
\begin{minipage}{0.44\textwidth}
	\centering
	\includegraphics[width=\textwidth]{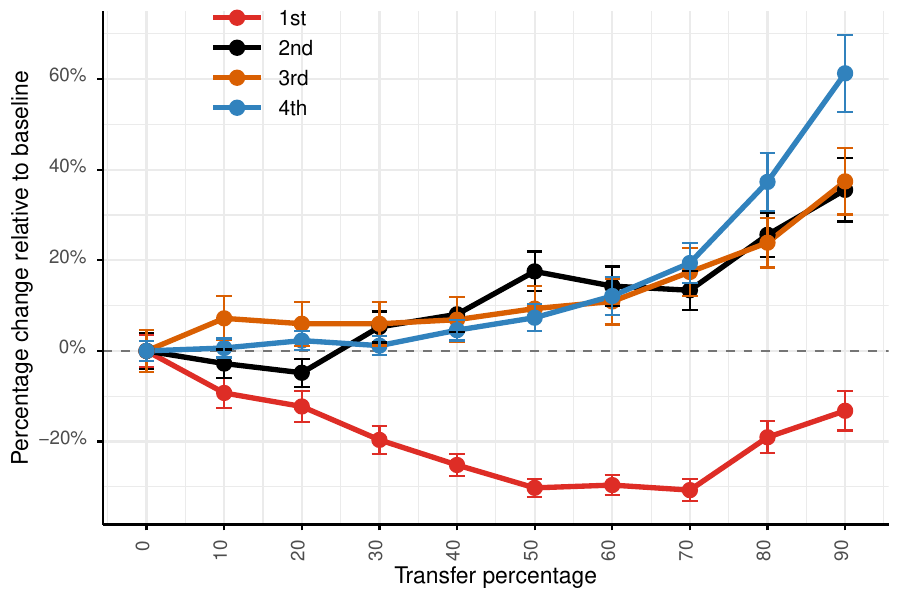}
	\subcaption{Price changes by HHI quartile}
\end{minipage}\\[2pt]
\begin{minipage}{1\textwidth}
	{\footnotesize Notes: Same exercise as Figure~\ref{fig:counter_price}, except that each period's water stock is reset to observed data values rather than carried forward from the previous simulation. This removes cumulative errors in the value-function approximation. Panel~(a): RD plot of log simulated prices on $\Delta_{t+3}$. Panel~(b): percentage price change by HHI quartile as a function of the transfer share $\kappa$. The U-shape is robust to this specification.\par}
\end{minipage}
\end{center}

\endgroup\vspace{1ex}
\begin{spacing}{0.95}
	\bibliographystylesec{ecca}
	\bibliographysec{sample.bib}
\end{spacing}
\end{document}

%% file: Tables/structural_model/res2_p5_rob.tex
\begin{table}[htb!]\caption{Estimated model primitives \label{tab:estimates}}
\centering
\begin{adjustbox}{width=.85\textwidth,center}
\begin{tabular}{l*{4}{r @{} l}}
\toprule
                &         (1)&        &         (2)&        &         (3)&        &         (4)&        \\
\midrule
\multicolumn{9}{c}{\textbf{Marginal Costs (COP/MWh)}} \\
Thermal $(c^{T})$&    203,677.62&$^{***}$&    141,668.46&$^{***}$&    221,304.18&$^{***}$&    144,744.21&$^{***}$\\
                &   (1,711.10)&        &   (1,875.29)&        &   (1,665.82)&        &   (1,561.62)&        \\
Hydropower $(c^{H})$&    64,258.02&$^{***}$&    20,123.07&$^{***}$&    29,187.79&$^{***}$&    52,755.37&$^{***}$\\
                &   (6,692.81)&        &   (5,309.73)&        &   (3,931.92)&        &   (3,731.31)&        \\
\multicolumn{9}{c}{\textbf{Intertemporal Value of Water (COP/MWh)}} \\

Spline 1 $(\gamma_1)$&   --2,950.20&$^{***}$&   --6,812.77&$^{***}$&   --11664.64&$^{***}$&   --3,812.18&$^{***}$\\
                &     (908.25)&        &     (528.12)&        &     (526.66)&        &     (385.60)&        \\
Spline 2 $(\gamma_2)$&  --2.301e-03&$^{***}$&  --1.546e-04&        &    6.286e-04&$^{***}$&  --8.402e-04&$^{***}$\\
                &  (3.213e-04)&        &  (1.456e-04)&        &  (1.819e-04)&        &  (1.036e-04)&        \\
Spline 3 $(\gamma_3)$&  --3.527e-09&$^{***}$&    1.919e-08&$^{***}$&  --1.932e-08&$^{***}$&    1.712e-08&$^{***}$\\
                &  (1.282e-09)&        &  (1.041e-09)&        &  (1.174e-09)&        &  (8.106e-10)&        \\
Spline 4 $(\gamma_4)$&    3.246e-08&$^{***}$&  --3.119e-08&$^{***}$&    4.536e-08&$^{***}$&  --2.729e-08&$^{***}$\\
                &  (2.422e-09)&        &  (1.894e-09)&        &  (2.057e-09)&        &  (1.492e-09)&        \\
Spline 5 $(\gamma_5)$&  --1.414e-08&$^{***}$&    9.357e-08&$^{***}$&    5.167e-08&$^{***}$&    8.566e-08&$^{***}$\\
                &  (3.053e-09)&        &  (2.950e-09)&        &  (2.480e-09)&        &  (2.568e-09)&        \\
\multicolumn{9}{c}{\textbf{Fixed Effects}} \\
Firm        & $\checkmark$&        & $\checkmark$&        & $\checkmark$&        & $\checkmark$&        \\
Unit   & $\checkmark$&        & $\checkmark$&        & $\checkmark$&        & $\checkmark$&        \\
Month-by-technology&             &        & $\checkmark$&        &             &        & $\checkmark$&        \\
Hour        & $\checkmark$&        & $\checkmark$&        & $\checkmark$&        & $\checkmark$&        \\
Week-by-year& $\checkmark$&        & $\checkmark$&        &             &        &             &        \\
Date        &             &        &             &        & $\checkmark$&        & $\checkmark$&        \\
\midrule
SW F $(c^{T})$&     3,300.46&        &     1,170.68&        &     2,889.26&        &     1,000.69&        \\
SW F $(c^{H})$&       458.72&        &       267.20&        &       877.47&        &       360.16&        \\
SW F $(c^{\gamma_1})$&       272.15&        &       201.92&        &       274.58&        &       214.83&        \\
SW F $(c^{\gamma_2})$&       220.24&        &       266.10&        &       266.58&        &       292.22&        \\
SW F $(c^{\gamma_3})$&       466.55&        &       494.96&        &       291.98&        &       491.58&        \\
SW F $(c^{\gamma_4})$&       570.57&        &       577.42&        &       292.40&        &       566.81&        \\
SW F $(c^{\gamma_5})$&       419.44&        &     1,156.74&        &       432.57&        &       920.39&        \\
Anderson Rubin F&     1,213.31&        &     1,395.05&        &     1,527.77&        &     1,539.08&        \\
KP Wald         &       143.70&        &       142.85&        &       138.91&        &       147.23&        \\
N               &    1,451,592&        &    1,451,592&        &    1,451,592&        &    1,451,592&        \\
\bottomrule
\multicolumn{9}{l}{* -- $p < 0.1$; ** -- $p < 0.05$; *** -- $p < 0.01$}
\end{tabular}%
\end{adjustbox}
\begin{tablenotes} \footnotesize
\item Notes: This table presents the coefficients obtained estimating (\ref{eq:2sls}) by two-stage least squares on daily data between January 1, 2010, and December 31, 2015. The top panels separate the marginal cost estimates and the value function parameters from the fixed effects used in estimation, which vary across columns. Our favorite specification is in Column (4), which includes day and month-by-technology fixed effects. The bottom panel provides diagnostic tests for the first stage. 2,900 COP $\simeq$ 1 US\$.
\end{tablenotes}
\end{table}

%% file: Tables/empirical_evidence/table_D1.tex
\resizebox{\columnwidth}{!}{%
\begin{tabular}{lcccc}
	\toprule
	 & (1) & (2) & (3) & (4) \\
	\midrule
	Left slope ($\gamma_1^-$)                & $-0.080^{***}$ & $-0.035^{***}$ & $-0.081^{***}$ & $-0.041^{***}$ \\
	                                         & $(0.029)$ & $(0.009)$ & $(0.029)$ & $(0.009)$ \\
	Right slope ($\gamma_1^+$)               & $+0.026$ & $+0.003$ & $+0.026$ & $-0.014$ \\
	                                         & $(0.027)$ & $(0.010)$ & $(0.028)$ & $(0.010)$ \\
	\multicolumn{5}{c}{ }\\
	Slope change ($\gamma_1^+ - \gamma_1^-$) & $+0.106^{***}$ & $+0.038^{***}$ & $+0.107^{***}$ & $+0.028^{***}$ \\
	                                         & $(0.038)$ & $(0.011)$ & $(0.039)$ & $(0.011)$ \\
	\midrule
	Forecast measure:& Indicator & Continuous & Indicator & Continuous \\
	Month-by-hour and year FEs &\checkmark  &\checkmark  & \checkmark & \checkmark \\
	Day-of-the-week-by-hour FEs & \checkmark & \checkmark & \checkmark & \checkmark \\
	Include $\text{HHI}_{t-1}\!\times\!\sum \text{forecast}_{t+3}$ &  &  & \checkmark & \checkmark \\
	Observations                             & 52{,}416 & 52{,}416 & 52{,}416 & 52{,}416 \\
	Newey--West lag length                   & 4 days & 4 days & 4 days & 4 days \\
	\bottomrule
	\multicolumn{5}{l}{* -- $p < 0.1$; ** -- $p < 0.05$; *** -- $p < 0.01$}\\
\end{tabular}}

%% file: Tables/structural_model/res2_p4_rob.tex
\begin{table}[!htbp]\caption{Estimated model primitives---employing 4 polynomials for $V(\cdot)$\label{tab:estimates_p4} }
\renewcommand{\arraystretch}{0.9}%
\begin{adjustbox}{width=.78\textwidth,center}
\begin{tabular}{l*{4}{r @{} l}}
\toprule
                 &         (1)&        &         (2)&        &         (3)&        &         (4)&        \\
\midrule
\multicolumn{9}{c}{\textbf{Marginal costs (COP/MWh)}} \\
Thermal $(c^{T})$&    204460.10&$^{***}$&    151965.08&$^{***}$&    213177.19&$^{***}$&    149699.10&$^{***}$\\
                &   (1,880.65)&        &   (1,840.22)&        &   (1,626.95)&        &   (1,624.97)&        \\
Hydropower $(c^{H})$&    76,022.12&$^{***}$&    28,820.29&$^{***}$&    44,941.37&$^{***}$&    51,297.15&$^{***}$\\
                &   (6,601.03)&        &   (6,290.74)&        &   (3,368.43)&        &   (4,638.29)&        \\
\multicolumn{9}{c}{\textbf{Intertemporal value of water (COP/MWh)}} \\                
Spline 1 $(\gamma_1)$&   --2,216.01&$^{***}$&     6,992.47&$^{***}$&     2,297.60&$^{***}$&    10,797.46&$^{***}$\\
                &     (710.63)&        &     (524.16)&        &     (372.92)&        &     (474.35)&        \\
Spline 2 $(\gamma_2)$&  --2.773e-03&$^{***}$&  --2.668e-03&$^{***}$&  --3.672e-03&$^{***}$&  --3.576e-03&$^{***}$\\
                &  (2.612e-04)&        &  (1.550e-04)&        &  (1.398e-04)&        &  (1.402e-04)&        \\
Spline 3 $(\gamma_3)$&    5.359e-09&$^{***}$&    1.386e-08&$^{***}$&    5.512e-09&$^{***}$&    1.382e-08&$^{***}$\\
                &  (6.862e-10)&        &  (6.043e-10)&        &  (4.654e-10)&        &  (5.307e-10)&        \\
Spline 4 $(\gamma_4)$&    2.364e-08&$^{***}$&  --1.893e-08&$^{***}$&    1.220e-08&$^{***}$&  --1.996e-08&$^{***}$\\
                &  (2.347e-09)&        &  (1.773e-09)&        &  (1.382e-09)&        &  (1.536e-09)&        \\
\multicolumn{9}{c}{\textbf{Fixed Effects}} \\
Firm        & $\checkmark$&        & $\checkmark$&        & $\checkmark$&        & $\checkmark$&        \\
Unit   &             &        &             &        &             &        & $\checkmark$&        \\
Month-by-technology&             &        & $\checkmark$&        &             &        & $\checkmark$&        \\
Hour        & $\checkmark$&        & $\checkmark$&        & $\checkmark$&        & $\checkmark$&        \\
Week-by-year& $\checkmark$&        & $\checkmark$&        &             &        &             &        \\
Date        &             &        &             &        & $\checkmark$&        & $\checkmark$&        \\
\midrule
SW F $(c^{thermal})$&     2,314.39&        &       760.89&        &     3,458.28&        &       700.41&        \\
SW F $(c^{hydro})$&       425.23&        &       245.77&        &       866.70&        &       333.94&        \\
SW F $(\gamma_1)$&       318.26&        &       242.54&        &       392.31&        &       269.99&        \\
SW F $(\gamma_2)$&       244.94&        &       275.53&        &       366.43&        &       323.50&        \\
SW F $(\gamma_3)$&       623.17&        &       484.66&        &       519.66&        &       491.27&        \\
SW F $(\gamma_4)$&       394.17&        &       448.36&        &       445.26&        &       446.16&        \\
Anderson Rubin F&     1,213.31&        &     1,395.05&        &     1,527.77&        &     1,539.08&        \\
N               &    1,451,592&        &    1,451,592&        &    1,451,592&        &    1,451,592&        \\
\bottomrule
\multicolumn{9}{l}{* -- $p < 0.1$; ** -- $p < 0.05$; *** -- $p < 0.01$}
\end{tabular}
\end{adjustbox}
\begin{tablenotes}\footnotesize
\item Notes: This table presents the coefficients obtained estimating (\ref{eq:2sls}) by two-stage least squares on daily data between January 1, 2010, and December 31, 2015. Unlike the results presented in the main text (Table \ref{tab:estimates}), these estimates are based on an approximation of the value function over four knots instead of five, meaning that we estimate only four $\{\gamma\}_{r=1}^4$. The top panels separate the marginal cost estimates and the value function parameters from the fixed effects used in estimation, which vary across columns. Our favorite specification is in Column (4), which includes day-fixed effects. The bottom panel provides diagnostic tests in the first stage. 2,900 COP $\simeq$ 1 US\$
\end{tablenotes}
\end{table}

%% file: Tables/structural_model/res2_p5_rob_normal.tex
\begin{table}[!htbp]\caption{Estimated model primitives---employing a normal density for the transition matrix and 5 polynomials for $V(\cdot)$ \label{tab:estimates_normal} }
\renewcommand{\arraystretch}{0.9}%
\begin{adjustbox}{width=.78\textwidth,center}
\begin{tabular}{l*{4}{r @{} l}}
\toprule
                 &         (1)&        &         (2)&        &         (3)&        &         (4)&        \\
\midrule
\multicolumn{9}{c}{\textbf{Marginal costs (COP/MWh)}} \\
Thermal $(c^{T})$&    204727.14&$^{***}$&    143319.87&$^{***}$&    220441.60&$^{***}$&    146635.86&$^{***}$\\
                &   (1,803.36)&        &   (1,843.71)&        &   (1,644.41)&        &   (1,529.32)&        \\
Hydropower $(c^{H})$&    46,491.28&$^{***}$&    28,163.59&$^{***}$&    28,458.10&$^{***}$&    60,353.00&$^{***}$\\
                &   (7,097.17)&        &   (5,026.09)&        &   (3,774.68)&        &   (3,616.79)&        \\
\multicolumn{9}{c}{\textbf{Intertemporal value of water (COP/MWh)}} \\
Spline 1 $(\gamma_1)$&     --797.45&        &   --6,751.10&$^{***}$&   --9,712.11&$^{***}$&   --3,744.90&$^{***}$\\
                &   (1,018.26)&        &     (504.77)&        &     (526.92)&        &     (364.28)&        \\
Spline 2 $(\gamma_2)$&  --3.346e-03&$^{***}$&  --3.154e-04&$^{**}$ &  --2.173e-04&        &  --1.064e-03&$^{***}$\\
                &  (3.548e-04)&        &  (1.421e-04)&        &  (1.806e-04)&        &  (1.016e-04)&        \\
Spline 3 $(\gamma_3)$&  --4.894e-09&$^{***}$&    2.009e-08&$^{***}$&  --1.621e-08&$^{***}$&    1.837e-08&$^{***}$\\
                &  (1.497e-09)&        &  (1.055e-09)&        &  (1.093e-09)&        &  (8.171e-10)&        \\
Spline 4 $(\gamma_4)$&    4.070e-08&$^{***}$&  --3.179e-08&$^{***}$&    4.205e-08&$^{***}$&  --2.848e-08&$^{***}$\\
                &  (2.795e-09)&        &  (1.931e-09)&        &  (1.926e-09)&        &  (1.508e-09)&        \\
Spline 5 $(\gamma_5)$&  --2.216e-08&$^{***}$&    8.949e-08&$^{***}$&    4.422e-08&$^{***}$&    8.251e-08&$^{***}$\\
                &  (3.405e-09)&        &  (2.893e-09)&        &  (2.274e-09)&        &  (2.500e-09)&        \\
\multicolumn{9}{c}{\textbf{Fixed Effects}} \\ Firm        & $\checkmark$&        & $\checkmark$&        & $\checkmark$&        & $\checkmark$&        \\
Unit   & $\checkmark$&        & $\checkmark$&        & $\checkmark$&        & $\checkmark$&        \\
Month-by-technology&             &        & $\checkmark$&        &             &        & $\checkmark$&        \\
Hour        & $\checkmark$&        & $\checkmark$&        & $\checkmark$&        & $\checkmark$&        \\
Week-by-year& $\checkmark$&        & $\checkmark$&        &             &        &             &        \\
FE: Date        &             &        &             &        & $\checkmark$&        & $\checkmark$&        \\               
\midrule
SW F $(c^{T})$&     3,129.14&        &     1,257.31&        &     2,991.60&        &     1,097.29&        \\
SW F $(c^{H})$&       443.62&        &       272.74&        &       883.91&        &       367.55&        \\
SW F $(c^{\gamma_1})$&       251.73&        &       213.67&        &       285.62&        &       225.96&        \\
SW F $(c^{\gamma_2})$&       219.64&        &       273.32&        &       270.25&        &       300.83&        \\
SW F $(c^{\gamma_3})$&       441.27&        &       476.05&        &       297.84&        &       482.88&        \\
SW F $(c^{\gamma_4})$&       522.38&        &       550.59&        &       296.71&        &       553.52&        \\
SW F $(c^{\gamma_5})$&       403.80&        &     1,255.92&        &       485.36&        &     1,018.15&        \\
Anderson Rubin F&     1,213.31&        &     1,395.05&        &     1,527.77&        &     1,539.08&        \\
N               &    1,451,592&        &    1,451,592&        &    1,451,592&        &    1,451,592&        \\
\bottomrule
\multicolumn{9}{l}{* -- $p < 0.1$; ** -- $p < 0.05$; *** -- $p < 0.01$}
\end{tabular}
\end{adjustbox}
\begin{tablenotes}\footnotesize
\item Notes: This table presents the coefficients obtained estimating (\ref{eq:2sls}) by two-stage least squares on daily data between January 1, 2010, and December 31, 2015. Unlike the results presented in the main text (Table \ref{tab:estimates}), these estimates assume that the transition matrix is normally distributed. The top panels separate the marginal cost estimates and the value function parameters from the fixed effects used in estimation, which vary across columns. Our favorite specification is in Column (4), which includes day-fixed effects. The bottom panel provides diagnostic tests in the first stage. 2,900 COP $\simeq$ 1 US\$
\end{tablenotes}
\end{table}

%% file: Figures/Model_fit/Table_Price_Diff_Wide.tex
\begin{table}[!htbp] \centering
\caption{Average absolute percentage price difference by hour}
\label{tab:model_fit_hp}
\begin{adjustbox}{width=.9\textwidth,center}
\begin{tabular}{lccccccccccccc}
\toprule
Hour & 0 & 1 & 2 & 3 & 4 & 5 & 6 & 7 & 8 & 9 & 10 & 11 & \\
\% Diff & 16.6 & 16.2 & 17.3 & 17.2 & 16.2 & 15.3 & 14.2 & 13.9 & 12.7 & 12.8 & 12.5 & 13.1 & \\
\midrule
Hour & 12 & 13 & 14 & 15 & 16 & 17 & 18 & 19 & 20 & 21 & 22 & 23 & All \\
\% Diff & 12.3 & 12.8 & 13 & 12.7 & 12.2 & 12.4 & 13.9 & 17.4 & 15.5 & 12.9 & 13.1 & 15.1 & 14.2 \\
\bottomrule
\end{tabular}
\end{adjustbox}
\begin{minipage}{1\textwidth}
\vspace{0.5em}
{\footnotesize Notes: Average absolute percentage difference between simulated and observed hourly market-clearing prices, $|p^{sim}_{ht} - p_{ht}|/p_{ht} \times 100$, pooled across the four largest firms (ENDG, EPMG, EPSG, ISGG) and all markets in the sample. The simulation uses ten discretization steps for demand, supply, and the value function ($J=K=M=10$). 2,900 COP $\simeq$ 1 US\$. \par}
\end{minipage}
\end{table}

%% file: Figures/Model_fit/Regressions_Fit_Levels.tex
\begin{table} \vspace{-1.5cm}\centering
\caption{Model fit: simulated vs.\ observed outcomes}
\label{tab:model_fit_reg}
\begin{adjustbox}{width=.55\textwidth,center}
\begin{tabular}{lccc}
\toprule
 & (1) & (2) & (3) \\
 & Price & Quantity & Water \\
\midrule
Simulated Price & 1.049$^{***}$ & & \\
 & (0.022) & & \\
Simulated Quantity & & 0.795$^{***}$ & \\
 & & (0.006) & \\
Simulated Water & & & 0.722$^{***}$ \\
 & & & (0.011) \\
\midrule
Observations & 29,952 & 26,672 & 1,103 \\
\bottomrule
\multicolumn{4}{l}{* -- $p < 0.1$; ** -- $p < 0.05$; *** -- $p < 0.01$}
\end{tabular}
\end{adjustbox}
\begin{minipage}{1\textwidth}
{\footnotesize Notes: OLS regressions of observed on simulated outcomes without intercept. Columns differ in unit of observation: (1) market-hour prices (COP/MWh), (2) unit-hour quantities (MWh, excluding zeros), (3) firm-week water usage (GWh). All columns pool the four focal firms. Standard errors clustered by week.\par}
\end{minipage}
\end{table}